\documentclass[11pt,english]{article}
\usepackage{graphicx}
\usepackage{xcolor}
\usepackage{amstext}
\usepackage{caption}
\usepackage{etoolbox}
\usepackage{makeidx}
\usepackage{amsfonts}
\usepackage{authblk} 
\usepackage{sectsty}
\usepackage{amsmath,amssymb,epsfig}
\usepackage{amscd}
\usepackage{amsthm}
\usepackage{mathrsfs}
\usepackage{dsfont}
\usepackage[applemac]{inputenc}
\usepackage[english]{babel}
\usepackage{enumitem} 
\usepackage[]{latexsym}
\usepackage{subfigure}
\usepackage{hyperref}
\usepackage{blindtext}
\usepackage{verbatim}
\usepackage{cancel}
\usepackage{cases}
\usepackage[notext]{stix}
\usepackage{cite}

\hypersetup{
pdftitle={},%
pdfauthor={},%
pdfsubject={},%
pdfkeywords={},%
colorlinks=true,%
linkcolor=blue,%
citecolor=crimson,%
linktocpage=true,%
hyperfootnotes=true,%
pageanchor=true
}

\definecolor{crimson}{rgb}{0.7, 0.08, 0.24}

\newcommand*{\affaddr}[1]{#1}
\newcommand*{\affmark}[1][*]{\textsuperscript{#1}}

\newtheorem*{proof*}{Proof}

\newcommand{\be}{\begin{equation}}

\newcommand{\ee}{\end{equation}}
\def\beqa{\begin{eqnarray}}
\def\eeqa{\end{eqnarray}}
\def\bean{\begin{eqnarray*}}
\def\eean{\end{eqnarray*}}

\newcommand{\dd}{\mathrm{d}}

\renewenvironment{thebibliography}[1]
         {\section*{References}\frenchspacing\small
          \begin{list}{[\arabic{enumi}]}
         {\usecounter{enumi}\parsep=2pt\topsep 0pt
         \settowidth{\labelwidth}{[#1]}
         \leftmargin=\labelwidth\advance\leftmargin\labelsep
         \rightmargin=0pt\itemsep=1pt\sloppy}}{\end{list}}

 \numberwithin{equation}{section}

\input xy
\xyoption{all}

\usepackage[normalem]{ulem}

\title{\textbf{\textsf{Mass and Horizon Dirac Observables in Effective Models of Quantum Black-to-White Hole Transition}}\vspace{0.35cm}}

\author{
\textsf{Norbert Bodendorfer\affmark[1]\footnote{\texttt{norbert.bodendorfer@physik.uni-r.de}}, Fabio M. Mele\affmark[1]\footnote{\texttt{fabio.mele@physik.uni-r.de}}, and Johannes M\"unch\affmark[1]\footnote{\texttt{johannes.muench@physik.uni-r.de}}}\\
\affaddr{\affmark[1]\textsf{Institute for Theoretical Physics, University of Regensburg,}}\\
\affaddr{\textsf{93040 Regensburg, Germany}}\vspace{-0.5cm}
}

\begin{document}

\maketitle

\begin{abstract}
	\textsf{In the past years, black holes and the fate of their singularity have been heavily studied within loop quantum gravity. Effective spacetime descriptions incorporating quantum geometry corrections are provided by the so-called \textit{polymer models}. Despite the technical differences, the main common feature shared by these models is that the classical singularity is resolved by a black-to-white hole transition. In a recent paper \cite{BodendorferEffectiveQuantumExtended}, we discussed the existence of two Dirac observables in the effective quantum theory respectively corresponding to the black and white hole mass. Physical requirements about the onset of quantum effects then fix the relation between these observables after the bounce, which in turn corresponds to a restriction on the admissible initial conditions for the model. In the present paper, we discuss in detail the role of such observables in black hole polymer models. First, we revisit previous models and analyse the existence of the Dirac observables there. Observables for the horizons or the masses are explicitly constructed. In the classical theory, only one Dirac observable has physical relevance. In the quantum theory, we find a relation between the existence of two physically relevant observables and the scaling behaviour of the polymerisation scales under fiducial cell rescaling. We present then a new model based on polymerisation of new variables which allows to overcome previous restrictions on initial conditions. Quantum effects cause a bound of a unique Kretschmann curvature scale, independently of the relation between the two masses.}   
\end{abstract}

\section{Introduction}

Understanding the fate of classical gravitational singularities is one of the key questions that any quantum theory of gravity needs to address. In this respect, symmetry reduced spacetimes in which such singularities occur classically offer on the one hand a simplified setting where explicit calculations are possible and, on the other hand, they play a crucial role in the attempt to identify possible observational signatures of quantum gravitational effects. The application of quantisation techniques inspired by loop quantum gravity (LQG) in symmetry-reduced situations has proven very successful. In the cosmological setting, this has lead to the wide and active field of loop quantum cosmology (LQC) \cite{AshtekarLoopQuantumCosmology,OritiBouncingCosmologiesFrom,AgulloLoopQuantumCosmology,AshtekarQuantumNatureOf,AshtekarLoopQuantumCosmologyFrom} (see also \cite{DienerNumericalSimulationsOf,AshtekarLoopQuantumCosmologyBianchiI} for results in non-isotropic cosmology). At the semi-classical level, some of the relevant quantum corrections are captured by a phase space regularisation called \textit{polymerisation} according to which the canonical momenta are replaced by (combinations of) their exponentiated versions (point holonomies). These are the so-called holonomy corrections, and essentially they are the analogue of approximating the field strength by holonomies of the gauge connection along plaquettes in lattice gauge theory. The structure of such modifications is motivated by a mini-superspace polymer-like quantisation \cite{CorichiPolymerQuantumMechanics, AshtekarQuantumGravityShadow, AshtekarMathematicalStructureOf} inspired by LQG, which in turn can be thought of as a diffeomorphism invariant extension of lattice gauge theory where the dynamical lattice itself encodes the quantum properties of spacetime geometry \cite{ThiemannModernCanonicalQuantum, RovelliQuantumGravity, RovelliBook2,BodendorferAnElementaryIntroduction}. In the resulting effective quantum corrected cosmological spacetime, quantum geometry effects induce a natural cutoff for spacetime curvature invariants and the initial big bang singularity is resolved by a quantum bounce interpolating between a contracting and an expanding branch well approximated by classical geometries far from the Planck regime \cite{AshtekarLoopQuantumCosmology,AshtekarQuantumNatureOf}. Remarkably, the effective dynamics can be derived from the LQC quantum theory by considering expectation values on suitable semi-classical states peaked on classical phase space points for large volumes \cite{TaverasCorrectionstothe,DienerNumericalSimulationsOf,RovelliWhyAreThe}, thus showing that the polymerisation procedure is able to capture (some of) the relevant features of the quantum theory.

The application of LQG techniques to other spacetime singularities such as those occurring inside a black hole (BH) is however still limited. Despite of the large effort, no definite consensus has been reached so far and several effective models have been proposed \cite{AshtekarQuantumGeometryand,ModestoLoopquantumblack,GambiniQuantumblackholes,BianchiWhiteHolesasRemnants,CorichiLoopquantizationof,ModestoSemiclassicalLoopQuantum,BoehmerLoopquantumdynamics,ChiouPhenomenologicalloopquantum,OlmedoFromblackholesto,AshtekarQuantumTransfigurarationof,AshtekarQuantumExtensionof,OritiBlackHolesas,BenAchourPolymerSchwarzschildBlack,BojowaldEffectivelineelements,LoboRainbow-likeBlackHole,Morales-TecotlEffectivedynamicsof}. The starting point of these models is the observation that the interior region of Schwarzschild black holes, foliated with respect to the radial time-like coordinate, can be modelled as a Kantowski-Sachs cosmological spacetime so that techniques from homogeneous and non-isotropic LQC can be applied to construct the effective quantum theory. Besides of the technical differences, these polymer black hole effective models share common qualitative features such as the resolution of the central singularity, which is then replaced by a black-to-white hole quantum bounce. Undesirable outcomes concerning the onset of quantum effects and the curvature upper bound can however emerge depending on the details of the model. Recently, some advances in the attempt to overcome previous limitations were taken by the authors in \cite{BodendorferEffectiveQuantumExtended}. There, inspired by the construction of the variables successful for LQC, we introduced canonical variables for Schwarzschild black holes adapted to physical considerations about the onset of quantum effects in such a way that the simplest polymerisation scheme can be used to construct an effective model with satisfactory physical predictions. The main idea is to construct canonical momenta which are related to spacetime curvature invariants so that the resulting polymerisation induces a natural curvature bound in the Planck regime and quantum effects become negligible in the low curvature regime. In particular, in analogy to the so-called $(b,v)$-variables in LQC, where the canonical momentum $b$ is the Hubble rate which in turn is related to the Ricci scalar ($R\propto b^2$), the on-shell value of one of the momenta of the model is constructed to be proportional to the square root of the Kretschmann scalar. The main novel feature of our analysis was the observation that in the effective quantum theory there exist two independent Dirac observables corresponding to the black and white hole masses, respectively. However, as shown by the detailed analysis of the Dirac observables \cite{BodendorferEffectiveQuantumExtended}, in order to achieve physical reliable predictions such as a unique mass independent curvature upper bound, certain initial conditions and in turn certain relations between the black hole and white hole masses have to be selected. The source of such limitation is rooted in the fact that the on-shell canonical momentum is not exactly proportional to (the square root of) the Kretschmann scalar unless the integration constant entering the proportionality factor is selected to be independent of the mass. Thus, the canonical momentum comes to be proportional to the Kretschmann scalar only after restricting to a certain subset of initial conditions.

Given the above situation, the purpose of the present paper is twofold. On the one hand, we want to further investigate the role of the Dirac observables in effective LQG models for black-to-white hole transition, on the other hand previous limitations to achieve physically reliable effects need to be solved. Therefore, in the first part of the paper we focus on the question of whether such previously unnoticed observables exist also in other models. We thus scan the previous literature and show how the study of the mass and horizon Dirac observables leads to similar restrictions on the initial conditions and re-analyse previous results in this new light. The second part of the paper is devoted to introduce a new effective model for polymer Schwarzschild black holes in which such limitations are resolved and all criteria of physical viability (mass independent Planckian curvature upper bound, see \cite{Bodendorferbvtypevariables}) can be achieved for a large class of initial conditions independently of the relation between the black and white hole masses. The key insight of the new model is the construction of canonical variables in which one of the (on-shell) momenta is now directly related to the Kretschmann scalar with no restrictions on the allowed initial conditions. In the resulting effective quantum corrected spacetime, the central singularity is again resolved by a 3-dimensional space-like transition surface smoothly connecting a trapped (black hole) and a anti-trapped (white hole) interior region. Quantum effects become relevant in the high curvature regime close to the Planck scale and rapidly decay far from it so that classical Schwarzschild solution is recovered in the low curvature regime. By analysing the onset of quantum effects, we also argue that, among all possible relations between the masses, the symmetric bounce scenario is preferred as it would correspond to the case in which both types of quantum corrections coming from the polymerisation of the canonical momenta align, thus making them both appearing at high curvatures. Moreover, the simple form of the effective Hamiltonian characterising our previous model is remarkably unaffected by this canonical transformation and the model can still be solved analytically. In particular, as already discussed in \cite{BodendorferEffectiveQuantumExtended}, the resulting quantum theory can be constructed by means of standard techniques and the kernel of the corresponding Hamiltonian constraint operator can be explicitly computed.

Finally, we further explore the relation between the mass Dirac observables and the scaling properties of the polymerisation scales under a rescaling of the fiducial cell. As a concrete example, we discuss a class of canonical variables for which both canonical momenta (and hence the corresponding polymerisation scales) are independent of fiducial cell rescaling, while keeping one of them to be the (square root of the) Kretchmann scalar. In this case, there is no second fiducial cell independent Dirac observable which can be related with the white hole mass and the relation between the masses is determined as an outcome of the effective dynamics.  

The paper is organised as follows. In Sec. \ref{sec:Classical} we briefly recall the Hamiltonian framework for classical Schwarzschild black holes by focusing on how to fix the integration constants in a coordinate-free way by means of Dirac observables. As already pointed out in \cite{BodendorferEffectiveQuantumExtended}, in the classical theory there is only one fiducial cell independent Dirac observable corresponding to the black hole mass. On the contrary, in the effective quantum theory it is possible to exhibit two fiducial cell independent Dirac observables whose on-shell values can be interpreted as the black and white hole masses, respectively. Therefore, in order to emphasise on the role of such observables in properly fixing the integration constants for effective models, in Sec. \ref{sec:effectivequantumtheory} we first review the construction of these Dirac observables in our previous model \cite{BodendorferEffectiveQuantumExtended}, and then analyse in detail previous effective polymer black hole models in the LQG literature. In particular, we show that the analysis of the mass observables leads to similar restrictions on the admissible initial conditions of the model, thus reinterpreting previous results for the proposed relation between the black and white hole masses accordingly. In Sec. \ref{sec:newvariables}, we then introduce our new model based on adapted canonical variables which allow us to overcome previous limitations. The resulting quantum corrected effective spacetime and its causal structure is studied in Sec. \ref{sec:spacetimestructure}. The relation of the new variables with connection variables usually adopted in LQG-based investigations and the corresponding polymerisation scheme is discussed in Sec. \ref{sec:comparison}. Finally, in Sec. \ref{sec:noscalingmodels} we focus on th above mentioned relation between the existence of two independent Dirac observables and the scaling properties of the polymerisation scales. A summary of the results and some future perspectives are reported in Sec. \ref{sec:conclusions}.

\section{Integration constants in the classical theory}\label{sec:Classical}

Let us start by studying the integration constants appearing in the classical setting of black holes, more precisely static and spherically symmetric solutions of Einstein equations. The most general ansatz for the metric is given by \cite{CampigliaLoopquantizationof,VakiliClassicalpolymerizationof}

\begin{equation}\label{eq:metricansatz}
\dd s^2 = - \bar{a}(r) \dd t^2 + N(r) \dd r^2 + 2 \bar{B}(r) \dd r \dd t + \mathscr b(r)^2 \dd \Omega_2^2 \;,
\end{equation} 
\noindent
where $\dd \Omega_2$ denotes the metric on the $r,t = const.$ round 2-sphere.
The dynamics of the system is then described by the source-less (in fact there is a matter source of the form $\rho(r) \propto M \delta(r)$) Einstein-Hilbert action, leading to the Schwarzschild solution. For later use let us define the integrated quantities

$$
\sqrt{a} = \int_{0}^{L_o} \sqrt{\bar{a}} \;\dd t = L_o\sqrt{\bar{a}} \;, \quad B = \int_{0}^{L_o} \bar{B} \;\dd t = L_o \bar{B} \; , \quad n = Na + B^2\;,
$$
\noindent
where $L_o$ is the coordinate size of a fiducial cell in the non-compact $t$-direction, which is necessary to regularise the otherwise divergent integrals in the canonical analysis. We further define $\mathscr L_o = \int_{0}^{L_o} \left.\sqrt{\bar{a}}\right|_{r=r_{\text{ref}}}\dd t$ to be the physical size of the fiducial length at the reference point $r_{\text{ref}}$. 

As it is well-known, in the interior of the black hole $\bar{a}(r), N(r) < 0$, i.e. the coordinate $r$ becomes time-like and $t$ spacelike, thus leading to a homogeneous spherically symmetric cosmological model, namely the Kantowski-Sachs cosmology of topology $\mathbb{R} \times \mathbb{R} \times \mathbb{S}^2$. Concluding, the interior of a black hole is actually isometric to a cosmological spacetime which is well-suited for the framework of loop quantum cosmology (LQC). In this case $N(r)$ can be interpreted as the lapse of time-evolution and $\bar{B}(r)$ as the shift. The Hamiltonian analysis shows (as known from the ADM formalism) that both are purely gauge. As done by several authors (see e.g. \cite{CampigliaLoopquantizationof,AshtekarQuantumGeometryand,ModestoLoopquantumblack,CorichiLoopquantizationof,ModestoSemiclassicalLoopQuantum,BoehmerLoopquantumdynamics,ChiouPhenomenologicalloopquantum,OlmedoFromblackholesto,AshtekarQuantumTransfigurarationof,AshtekarQuantumExtensionof}), the metric can also be rewritten in connection variables as

\begin{equation}\label{eq:metricansatzconn}
ds^2 = -N_T^2(T) \,\dd T^2 + \frac{p_b^2(T)}{L_o^2 |p_c(T)|}\, \dd x^2 + |p_c(T)| \,\dd \Omega_2^2 \;,
\end{equation}
\noindent
where $p_b, p_c$ here correspond to the independent components of the triads $E_j^a$ in the symmetry reduced setting conjugate to the independent components $c,b$ of the Ashtekar-Barbero connection $A_a^j$. The latter is playing the role of configuration space variables, i.e.
\begin{align}
A_a^j\tau_j\dd x^a&=\frac{c}{L_o}\tau_3\dd x+b\tau_2\dd\theta-b\tau_1\sin\theta\dd\phi+\tau_3\cos\theta\dd\phi\;,\\
E^a_j\tau^j\partial_a&=p_c\tau_3\sin\theta\partial_x+\frac{p_b}{L_o}\tau_2\sin\theta\partial_\theta-\frac{p_b}{L_o}\tau_1\partial_\phi\;,
\end{align}
with $\tau_j=-i\sigma_j/2$, $j=1,2,3$, denoting the standard basis of the $\mathfrak{su}(2)$ Lie algebra with $\sigma_j$ being the Pauli matrices.
The metric \eqref{eq:metricansatzconn} describes the interior region of a black hole and is of course identical to \eqref{eq:metricansatz} by identifying 

\begin{equation}\label{eq:metricmap}
T = r\quad,\quad x = t\quad ,\quad   |p_c| = \mathscr b^2\quad ,\quad p_b^2 = -a \mathscr b^2\quad,\quad N = -N^2_T
\end{equation} 
\noindent
and the gauge $\bar{B} = 0$. The dynamics of this metric is described in the Hamiltonian framework within a phase space spanned by $\left( b, p_b\right)$, $\left( c, p_c\right)$,
and equipped with the Possion brackets

$$
\left\{b, p_b\right\} = G \gamma \quad , \quad \left\{c, p_c\right\} = 2 G \gamma \;,
$$
\noindent
where $\gamma$ is the Barbero-Immizri parameter, $G$ is the gravitational constant.
The Hamiltonian constraint reads 

\begin{equation}\label{eq:classHamconn}
H = N_T \mathcal{H} \quad , \quad \mathcal{H} = - \frac{b}{2 G \gamma^2 \text{sign}(p_c) \sqrt{|p_c|}} \left( 2 c p_c + \left(b+ \frac{\gamma^2}{b} \right) p_b \right) \approx 0 \;,
\end{equation}
\noindent
which in the following we set $G = 1$ as well as we already assumed $c=1$.
Note that to arrive at this result, a detailed discussion of Ashtekar-Barbero connection variables and fiducial cell structures is necessary for which we refer to several papers in the literature, see e.g. \cite{AshtekarQuantumGeometryand,CorichiLoopquantizationof,AshtekarQuantumExtensionof,AshtekarQuantumTransfigurarationof} and references within.
A detailed discussion of the fiducial cell structures shows that under a change of the fiducial length $L_o \mapsto \alpha L_o$ the variables transform as

\begin{equation}\label{eq:sclaingcb}
b \longmapsto b \quad , \quad c \longmapsto \alpha c \quad , \quad p_b \longmapsto \alpha p_b \quad , \quad p_c \longmapsto p_c \;.
\end{equation}
\noindent
Obviously, physical quantities can not depend on this fiducial structures and must be independent of this rescaling.

Let us now discuss the classical solution and how the integration constants can be fixed by physical input. By solving the equations of motion, the metric \eqref{eq:metricansatzconn} is determined. Hereby, as already mentioned, the Hamiltonian analysis shows that $N_T$ is a Lagrange multiplier and hence is purely gauge. The remaining system has four kinematic degrees of freedom $(c,p_c)$, $(b, p_b)$, where the Hamiltonian constraint determines two of them, leading to two remaining physical degrees of freedom on the constraint surface. Therefore, solving the equations of motion, we expect two integration constants which should be determined by two initial conditions, or in the language of constrained systems, two Dirac observables. We can make this explicit by solving the equations of motion for the lapse given by

$$
N_T = \frac{\gamma \;\text{sign}(p_c) \sqrt{|p_c|}}{b} \;,
$$
\noindent
i.e.

\begin{align}
\dot{b} = -\frac{1}{2 b} \left(b^2 + \gamma^2\right) \quad , & \quad \dot{c} = -2 c \;,\\
\dot{p_b} = \frac{p_b}{2 b^2}  \left(b^2 - \gamma^2\right) \quad , & \quad \dot{p_c} = 2 p_c \;,
\end{align}

\noindent
where the dot denotes derivatives w.r.t. $T$. We can now integrate the equations for $c$, $p_c$ and $b$ and solve the equation for $p_b$ by using the Hamiltonian constraint, thus yielding the solutions

\begin{align}
b(T) = \pm \gamma\sqrt{A e^{-T}-1} \quad ,& \quad c(T) = c_o e^{-2T} \\
p_b(T) = -\frac{2 c p_c}{b+ \frac{\gamma^2}{b}} = \mp \frac{2c_o p_c^o}{\gamma} \sqrt{\frac{e^T}{A} \left(1- \frac{e^T}{A}\right)} \quad ,&\quad p_c(T) = p_c^o e^{2T} \;.
\end{align}

\noindent
From the solution we can read off that the integration constant $A = e^{-T_o}$ simply produces a shift of $T_o$ in the $T$ coordinate, i.e. it is non-physical. This agrees with the above discussion of only two physical degrees of freedom and can be made manifest by writing the solutions in a coordinate free way, e.g. parametrised by $p_c$. Without loss of generality we can then set $A = 1$. Due to the scaling behaviour \eqref{eq:sclaingcb}, the integration constants have to scale as 

$$
c_o \longmapsto \alpha c_o \quad, \quad p_c^o \longmapsto p_c^o \;,
$$
\noindent
under a rescaling of the fiducial length $L_o$. This indicates that $c_o$ cannot be physical. We can thus construct the metric \eqref{eq:metricansatzconn} out of our solutions and 

$$
N_T = \frac{\gamma \;\text{sign}(p_c) \sqrt{|p_c|}}{b}  = \pm \frac{\text{sign}(p_c^o) \sqrt{p_c^o} e^T}{\sqrt{e^{-T}-1}} \;,
$$
\noindent
as

\begin{equation}
\dd s^2 = - \frac{p_c^o e^{2T}}{e^{-T}-1} \dd T^2 +  \frac{4 c_o^2 |p_c^o|}{\gamma^2 L_o^2} \left(e^{-T}-1\right) \dd x^2 + |p_c^o|e^{2T} \dd \Omega_2^2 \;. 
\end{equation}
\noindent
Redefining now the coordinates as 

\begin{equation}\label{eq:classcoordinatechange}
\tau = \sqrt{|p_c^o|} e^T \quad, \quad y = \frac{2 c_o \sqrt{|p_c^o|}}{\gamma L_o}x \;,
\end{equation}
\noindent
leads to

\begin{equation}\label{eq:metricclfinal}
\dd s^2 = -\frac{1}{\frac{\sqrt{|p_c^o|}}{\tau}-1} \dd \tau^2 + \left(\frac{\sqrt{|p_c^o|}}{\tau} -1\right) \dd y^2 + \tau^2 \dd \Omega_2^2 \;,
\end{equation}
\noindent
from which we see that by identifying $\sqrt{|p_c^o|} = 2M = R_{hor}$, where $M$ is the ADM mass of the black hole and $R_{hor}$ the horizon radius, this metric is indeed the classical Schwarzschild interior solution. Note that we need only one physical input parameter, namely $M$ (or equivalently $R_{hor}$) to fix uniquely the metric, i.e. the physical spacetime. The other integration constant $c_o$ does not appear in the metric \eqref{eq:metricclfinal} after a coordinate transformation, showing that it is purely gauge and has no physical relevance. This is in agreement with the fact that $c_o$ scales under a fiducial cell rescaling.

We can view this also in terms of Dirac observables. The phase space funtions

\begin{equation}\label{eq:DOclass}
\mathcal{R}_{hor} = \sqrt{|p_c|} \left(\frac{b^2}{\gamma^2} + 1\right) \quad , \quad \mathcal{D} = c p_c 
\end{equation}
\noindent
are both Dirac observables as they (weakly) commute with the Hamiltonian constraint. Giving these two Dirac observables together with the Hamiltonian constraint, the system is completely determined from a Hamiltonian point of view. Nonetheless, under a rescaling of the fiducial cell $L_o \mapsto \alpha L_o$ we find according to \eqref{eq:sclaingcb}

\begin{equation}
\mathcal{R}_{hor} \longmapsto \mathcal{R}_{hor} \quad , \quad \mathcal{D} \longmapsto \alpha \mathcal{D} \;.
\end{equation}
\noindent
Therefore, $\mathcal{D}$ can not be physical and hence not fixed by physical input, as it depends on the non-physical fiducial cell. Due to this, $\mathcal{D}$ cannot appear in the final form of the metric, which is verified by \eqref{eq:metricclfinal}. As the rescaling $L_o \mapsto \alpha L_o$ is not a gauge transformation in the canonical sense, i.e. in the canonical analysis of physical degrees of freedom, this transformation is not taken into account. Moreover, as $R_{hor}$, $\mathcal{D}$ and $\mathcal{H}$ span the space of Dirac observables, it is easy to see that there can not exist another Dirac observable which is fiducial cell independent as this new one needs to be a combination of $R_{hor}$, $\mathcal{D}$ and $\mathcal{H}$, which always scales. 

Also $\mathcal{D}/L_o$ does not solve the problem. Indeed, although this combination is invariant under rescaling, it depends on the coordinate choice $x$ as $L_o$ is the coordinate size of the fiducial cell. As proposed in \cite{BodendorferEffectiveQuantumExtended}, one could use instead the physical size of the fiducial cell $\mathscr L_o = \int_0^{L_o} \left.\sqrt{g_{xx}}\right|_{T = T_{ref}} dx$ at a reference point $T_{ref}$. The combination $\mathcal{D}/\mathscr L_o$ is then independent of fiducial cell rescaling and also independent of the coordinate $x$, but then the problem is shifted into a $T_{ref}$-dependence.

To sum up, the integration constants can be fixed in a gauge (i.e. coordinate) independent way by specifying values of Dirac observables. In the classical Schwarzschild black hole setting, there exists only one physical Dirac observable which represents the size of the horizon or equivalently the mass of the black hole. The other Dirac observable depends on fiducial structures and furthermore can be removed from the final metric by using a residual diffeomorphism. Hence, it cannot be determined by physical input. As the metric is independent of it, the specific value of this Dirac observable does not affect the physics, as it should. 
Let us stress here that these features are not visible at the Hamiltonian level. There only one constraint, the Hamiltonian constraint generating time evolution is left. As the quantities occurring in the Hamiltonian picture are all integrated over the fiducial cell and hence independent of the $x$-coordinate, the canonical transformation corresponding to a rescaling $x \mapsto y = \frac{2 c_o \sqrt{|p_c^o|}}{\gamma}x$ (cfr. Eq. \eqref{eq:classcoordinatechange}) corresponds to the identity transformation on the phase space level and therefore there exists no non-trivial first class constraint generating it. Consistently from the Hamiltonian perspective, in fact, we find two Dirac observables as we have one first class constraint for four degrees of freedom thus yielding two physical d.o.f. on the reduced phase space. We can remove one of these Dirac observables only by going back to the non-canonical components of the metric, which are $x$-coordinate dependent. 
In turn, this is possible as the spacetime metric and the metric entering the Hamiltonian differ by an arbitrary compactification of the $t$-direction.
Therefore, at the Hamiltonian level, all quantities differing only by a fiducial cell rescaling have to be viewed as equivalent, which adds another ``constraint'' (not in the sense of Dirac) to the system.
The true solution space is therefore the space spanned by the values of $\mathcal{R}_{hor}$ and $\mathcal{D}$, modulo the equivalence classes of fiducial cell rescaling.
This space is again one-dimensional and fits the observation that the spacetime metric has only one free parameter, the size of the horizon.
Due to this identification, fixing a value of $\mathcal{D}$, there exists always a fiducial cell rescaling such that $\mathcal{D} = 1$.
The second observable is therefore not removed by a canonical transformation/diffeomorphism from the Hamiltonian framework, but rather by modding out equivalence classes of fiducial cell rescaling.

The fact that this is possible highly depends on the Hamiltonian and the solutions. As we will see, for many effective polymer Hamiltonians the second Dirac observable cannot be removed from the final metric.
In the following, we will discuss in detail how the situation looks in recent effective polymer models of black holes. 

\section{Integration constants in effective polymer models}\label{sec:effectivequantumtheory}

In this section we discuss different polymer models of black holes and how the integration constants can be fixed gauge independently by defining Dirac observables and assigning physical input to them.

Hereby, we refer to effective polymer models as models where part of the phase space variables are replaced by their complex exponentials (point holonomies) in the Hamiltonian, allowing a polymer quantisation inspired by full LQG\footnote{Let us recall that this effective prescription is motivated by the quantum theory where weak discontinuity of the polymer representation implies that only the exponentiated version rather than bare momenta do exist as well-defined operators on the polymer Hilbert space \cite{AshtekarMathematicalStructureOf,AshtekarQuantumGravityShadow,CorichiPolymerQuantumMechanics}. At the semi-classical level, this translates into expressing the dependence on the momenta in any phase space function as a linear combination of their point holonomies of which the $\sin$ function is a simple choice commonly adopted in the literature \cite{AshtekarMathematicalStructureOf, AshtekarLoopQuantumCosmology}.}. For black hole models, the replacement

\begin{equation}
c \longmapsto \frac{\sin\left(\delta_c c\right)}{\delta_c} \quad, \quad b \longrightarrow \frac{\sin\left(\delta_b b\right)}{\delta_b} \;,
\end{equation}
\noindent
is usually done, where $\delta_c$, $\delta_b$ are the polymerisation scales controlling the onset of quantum effects\footnote{Note that there are many proposals of polymerisation which include choosing different functions or polymerising only parts of the phase space or different choices for the polymerisation scales \cite{BenAchourPolymerSchwarzschildBlack,ModestoSemiclassicalLoopQuantum,AssanioussiPerspectivesonthe,DaporCosmologicaleffectiveHamiltonian,BojowaldCovarianceinmodels}. Such different models can be motivated by physical inputs or full theory based results and arguments like general covariance and anomaly-free realisations of the constraint algebra at the quantum level. However, here we do not consider such alternative choices for simplicity.}. These scales should be thought as generic phase space functions remaining of order Planck scale in a suitable classical limit. In a regime where $\delta_c c$, $\delta_b b$ is small, we get back the classical equations due to $\sin(\delta_c c)/\delta_c \simeq c$, $\sin(\delta_b b)/\delta_b \simeq b$. The choice of the polymerisation scales classifies the corresponding scheme. The commonly adapted schemes available in the literature are classified as follows:

\begin{enumerate}
	\item The simplest one is the so-called $\mu_o$-scheme where the polymerisation scales are chosen to be constant 
	so they are not phase space dependent at all. A selection of $\mu_o$-models is \cite{AshtekarQuantumGeometryand,ModestoLoopquantumblack,CampigliaLoopquantizationof,ModestoSemiclassicalLoopQuantum,ModestoBlackHoleinterior}.
	
	\item More generic, one can allow $\delta_c$, $\delta_b$ to be any phase space function of $p_b$ and $p_c$ which is then called $\bar{\mu}$-scheme as in \cite{BoehmerLoopquantumdynamics,ChiouPhenomenologicalloopquantum} (see also \cite{ChiouPhenomenologicaldynamicsof,JoeKantowski-Sachsspacetimein} for the cosmological Kantowski-Sachs setting).
		
	\item A more recent development is provided by the so-called generalised $\mu_o$ schemes where $\delta_c$, $\delta_b$ are phase space dependent only through Dirac observables \cite{CorichiLoopquantizationof,OlmedoFromblackholesto,AshtekarQuantumExtensionof,AshtekarQuantumTransfigurarationof}.
\end{enumerate}

How to precisely fix the polymerisation scales is a delicate procedure based on different arguments in different works. These arguments usually depend on the dynamical trajectories and how the integration constants are fixed. In the following, we show that in contrast to the classical case, two physically relevant Dirac observables can exist. We carefully fix the integration constants by means of horizon or mass Dirac observables and discuss how physical requirements on the polymerisation scales due to curvature or plaquette arguments induce relations between the Dirac observables.

\subsection{In $(v_1,P_1)$, $(v_2,P_2)$ variables}\label{sec:BMM}

Let us begin with a model recently proposed by the authors \cite{BodendorferEffectiveQuantumExtended}, where the strategy of fixing the integration constants by using Dirac observables was first introduced. The authors introduced new variables $(v_1,P_1)$, $(v_2,P_2)$, which are (in the interior of the black hole) related to connection variables via

 \begin{align}
 \left(p_b\right)^2 = -8 v_2 \quad &, \quad |p_c| = \left(24 v_1\right)^{\frac{2}{3}}\;, 
 \label{connvaraibles1}
 \\
 b = \text{sign}(p_b)\; \frac{\gamma}{4}\; \sqrt{-8 v_2} \;P_2 \quad &, \quad c = -\text{sign}(p_c)\; \frac{\gamma}{8}\; \left(24 v_1\right)^{\frac{1}{3}}\; P_1  \;.
 \label{connvaraibles2}
 \end{align}
 
 \noindent
 with the Poisson brackets 
 
 $$
 \left\{v_i, v_j\right\} = 0\quad,\quad \left\{P_i, P_j\right\} = 0 \quad ,\quad  \left\{v_i, P_j\right\} = \delta_{ij} \;.
 $$
  \noindent
 In these variables, the classical Hamiltonian Eq. \eqref{eq:classHamconn} becomes
 
 \begin{equation}\label{eq:HamclassvP}
 H_{cl}=\sqrt{n}\mathcal{H}_{cl} \quad,\quad \mathcal{H}_{cl} = 3v_1P_1P_2+v_2P_2^2-2 \approx 0\;,
 \end{equation}
 
 \noindent
 where $n = N a + B^2$ is a Lagrange multiplier, as defined before.
 
The metric components can be reconstructed by the relations\footnote{Note that this relation matches with \eqref{connvaraibles1},\eqref{connvaraibles2} and \eqref{eq:metricmap} only up to a constant factor. This is due to a factor $1/4$ in front of the action which was neglected in \cite{BodendorferEffectiveQuantumExtended} and could be re-translated into $G = \frac{1}{4}$ instead of $G=1$. In the following we keep notation and results of \cite{BodendorferEffectiveQuantumExtended} and do not translate them according to this factor.} 
 
 \begin{equation}
 a = \frac{v_2}{2} \left(\frac{2}{3 v_1}\right)^{\frac{2}{3}} \quad ,\quad \mathscr b = \left(\frac{3 v_1}{2}\right)^{\frac{1}{3}} \;.
 \end{equation}
 \noindent
 Quantum effects are taken care of by means of the following  polymerisation scheme
 
 \begin{equation}\label{eq:polyvP}
 P_1 \longmapsto \frac{\sin\left(\lambda_1 P_1\right)}{\lambda_1} \quad, \quad P_2 \longrightarrow \frac{\sin\left(\lambda_2 P_2\right)}{\lambda_2} \;,
 \end{equation}
 \noindent
 where $\lambda_1$, $\lambda_2$ are the polymerisation scales and should be thought of being of Planck size and constant. Translating the variables back to $(c,b)$ and requiring 
 
 \begin{align}
 \lambda_1 P_1 = \delta_c c \quad , \quad \lambda_2 P_2 = \delta_b b\;,
 \end{align}
 
 \noindent
 leads to a relation of the polymerisation scales 
 
 \begin{align}\label{eq:polyvP1}
 \delta_c &= \pm \frac{8}{\gamma} \frac{\lambda_1}{\sqrt{|p_c|}}\;,\\
 \delta_b &= \pm \frac{4 \lambda_2}{\gamma |p_b|}\;, \label{eq:polyvP12}
 \end{align}
 
 \noindent
 according to which the polymerisation scheme \eqref{eq:polyvP} with constant $\lambda_1$, $\lambda_2$ corresponds to a specific $\bar{\mu}$-scheme in connection variables, given by the above phase space dependent polymerisation scales.
 
 The effective Hamiltonian for this setting reads 
 
 \begin{equation}\label{Heff1}
 H_{\text{eff}} = \sqrt{n} \mathcal{H}_{\text{eff}}\quad , \quad \mathcal{H}_{\text{eff}} = 3v_1 \frac{\sin\left(\lambda_1 P_1\right)}{\lambda_1} \frac{\sin\left(\lambda_2 P_2\right)}{\lambda_2} + v_2 \frac{\sin\left(\lambda_2 P_2\right)^2}{\lambda_2^2} - 2 \approx 0 \;.
 \end{equation}
 
 \noindent
 The resulting equations of motion can be solved analytically for $n = \mathscr L_o^2$ (see \cite{BodendorferEffectiveQuantumExtended} for details), leading to the following solutions for the effective dynamics
 
 \begin{align}
 v_1(r) &= \frac{2 C^2 \lambda_1^2 \sqrt{n}^3}{\lambda_2^3} D \frac{\frac{\lambda_2^6}{16 C^2 \lambda_1^2 n^3} \left( \frac{\sqrt{n} r}{\lambda_2} + \sqrt{1+\frac{n r^2}{\lambda_2^2}} \right)^6 +1 }{\left( \frac{\sqrt{n} r}{\lambda_2} + \sqrt{1+\frac{n r^2}{\lambda_2^2}} \right)^3} \; ,\label{solutionv1quant}
 \\
 v_2(r) &= 2 n \left(\frac{\lambda_2}{\sqrt{n}}\right)^2 \left(1+\frac{n r^2}{\lambda_2^2}\right) \left( 1 - \frac{3 C D}{2 \lambda_2} \frac{1}{\sqrt{1+\frac{n r^2}{\lambda_2^2}}} \right) \; , \label{solutionv2quant}
 \end{align}
 \begin{align}
 P_1(r) &= \frac{2}{\lambda_1} \cot^{-1}\left( \frac{\lambda_2^3}{4 C \lambda_1 \sqrt{n}^3} \left( \frac{\sqrt{n} r}{\lambda_2} + \sqrt{1+\frac{n r^2}{\lambda_2^2}} \right)^3\right) \; , \label{solutionP1quant}
 \\
 P_2(r) &= \frac{1}{\lambda_2} \cot^{-1}\left(\frac{\sqrt{n} r}{\lambda_2}\right) + \frac{\pi}{\lambda_2} \theta\left(-\frac{\sqrt{n}\; r}{\lambda_2}\right) \;, \label{solutionP2quant}
 \end{align}
 
 \noindent
 which translates to the metric components as
 
 \begin{align}
 \mathscr b &= \left(\frac{3 v_1}{2}\right)^{\frac{1}{3}} = \frac{\mathscr L_o}{\lambda_2}\left( 3 D C^2 \lambda_1^2\right)^\frac{1}{3} \frac{\left(\frac{\lambda_2^6}{16 C^2 \lambda_1^2 \mathscr L_o^6} \left( \frac{\mathscr L_o r}{\lambda_2} + \sqrt{1+\frac{\mathscr L_o^2 r^2}{\lambda_2^2}} \right)^6 +1 \right)^\frac{1}{3}}{\left( \frac{\mathscr L_o r}{\lambda_2} + \sqrt{1+\frac{\mathscr L_o^2 r^2}{\lambda_2^2}} \right)}\; , \label{bquant}\\
 \bar{a} &= \frac{v_2}{2 L_o^2} \left(\frac{2}{3 v_1}\right)^{\frac{2}{3}} \notag \\ 
 &= \frac{\mathscr L_o^2}{L_o^2}\left(\frac{\lambda_2}{\mathscr L_o}\right)^4 \left(1+\frac{\mathscr L_o^2 r^2}{\lambda_2^2}\right) \left( 1 - \frac{3 C D}{2 \lambda_2} \frac{1}{\sqrt{1+\frac{\mathscr L_o^2 r^2}{\lambda_2^2}}} \right) \frac{ \left(\frac{1}{3 D C^2 \lambda_1^2}\right)^{\frac{2}{3}} \left( \frac{\mathscr L_o r}{\lambda_2} + \sqrt{1+\frac{\mathscr L_o^2 r^2}{\lambda_2^2}} \right)^2}{\left(\frac{\lambda_2^6}{16 C^2 \lambda_1^2 \mathscr L_o^6} \left( \frac{\mathscr L_o r}{\lambda_2} + \sqrt{1+\frac{\mathscr L_o^2 r^2}{\lambda_2^2}} \right)^6 +1 \right)^{\frac{2}{3}}} \; , \label{aquant} 
 \end{align}
 \noindent
 where $C,D$ are integration constants. The gauge and hence the $r$-coordinate is chosen such that 

$$
N = \frac{\mathscr L_o^2}{L_o^2 \bar{a}} \quad , \quad \bar{B} = 0\;.
$$

\noindent
As discussed in the previous section, there are two integration constants $C$, $D$, which we fix by means of physical input, i.e. Dirac observables. As was done in \cite{BodendorferEffectiveQuantumExtended}, we can take the limits $r\rightarrow \pm \infty$ and re-express the physically meaningless coordinate $r$ in terms of the areal radius $\mathscr b_\pm := \mathscr b(r\rightarrow\pm \infty)$ so that by suitably rescaling the time coordinate $t \rightarrow \tau$ by a constant factor we get the asymptotic metrics

\begin{align}
\dd s_+^2 \simeq& -\left(1- \left(\frac{3}{2} D\right)^{\frac{4}{3}} \frac{C}{\mathscr L_o}\frac{1}{\mathscr b}\right) \dd \tau^2 + \frac{1}{1- \left(\frac{3}{2} D\right)^{\frac{4}{3}} \frac{C}{\mathscr L_o}\frac{1}{ \mathscr b}} \dd \mathscr b^2 + \mathscr b^2 \dd\Omega_2^2 \;, \\
\dd s_-^2 \simeq& -\left(1-\frac{3 C D \mathscr L_o}{\lambda_2^2} \frac{\left(3 D C^2 \lambda_1^2\right)^{\frac{1}{3}}}{\mathscr b}\right) \dd\tau^2 + \frac{1}{1-\frac{3 C D \mathscr L_o}{\lambda_2^2} \frac{\left(3 D C^2 \lambda_1^2\right)^{\frac{1}{3}}}{\mathscr b}} \dd \mathscr b^2 + \mathscr b^2 \dd\Omega_2^2 \;.
\end{align}
\noindent
Obviously, the asymptotic regions are described by Schwarzschild spacetimes with masses

\be
2 M_{BH} = \left(\frac{3}{2} D\right)^{\frac{4}{3}} \frac{C}{\mathscr L_o} \quad , \quad
2 M_{WH} = \frac{3 C D \mathscr L_o}{\lambda_2^2} \left(3 D C^2 \lambda_1^2\right)^{\frac{1}{3}}  \; , \label{Fonshell}
\ee

\noindent
where we call the masses in the $r \rightarrow +\infty$ and $r \rightarrow - \infty$ regions respectively black hole and white hole mass and refer to black hole and white hole regions correspondingly. Note that these names have no deeper meaning as they can be exchanged arbitrarily without affecting the physics, and are hence just for convenience. Eq. \eqref{Fonshell} now relates the integration constants $C,D$ to the physical quantities of the black hole and white hole mass. Furthermore, as in the classical theory, we can write down off-shell expressions for Dirac observables corresponding on-shell to the two masses given by

\begin{align}
2 \mathcal{M}_{BH} =&\; 3 v_1 \frac{\sin\left(\lambda_1 P_1\right)}{\lambda_1} \frac{\left( \frac{3}{2} v_1 \cos^2\left(\frac{\lambda_1 P_1}{2}\right) \right)^\frac{1}{3}}{\lambda_2 \cot\left(\frac{\lambda_2 P_2}{2}\right)} \label{F} \;,
\\
2 \mathcal{M}_{WH} =&\; 3 v_1 \frac{\sin\left(\lambda_1 P_1\right)}{\lambda_1} \left( \frac{3}{2} v_1 \sin^2\left(\frac{\lambda_1 P_1}{2}\right) \right)^\frac{1}{3} \frac{\cot\left(\frac{\lambda_2 P_2}{2}\right)}{\lambda_2} \label{Fbar} \;.
\end{align} 

\noindent
Computing the classical limit (i.e. $\lambda_1, \lambda_2 \rightarrow 0$) gives $2 \mathcal M_{BH} \rightarrow (3v_1/2)^{(4/3)} P_1 P_2 $, which is (up to a factor as discussed above) exactly the classical horizon Dirac observable of \eqref{eq:DOclass}. For $\mathcal{M}_{WH}$ this limit does not exist and depends on how exactly the double limit  $\lambda_1, \lambda_2 \rightarrow 0$ is performed. This reflects the fact that this observable does not exist classically. Of course, we could simply multiply by suitable powers of $\lambda_1$ and $\lambda_2$ to reach a well defined limit. This introduces then fiducial cell dependencies and, combining with $\mathcal{M}_{BH}$, the classical Dirac observable for $\mathcal{D}$ (cfr. \eqref{eq:DOclass}) can be reproduced.
 
 \begin{figure}[t!]
 	\centering\includegraphics[scale=0.75]{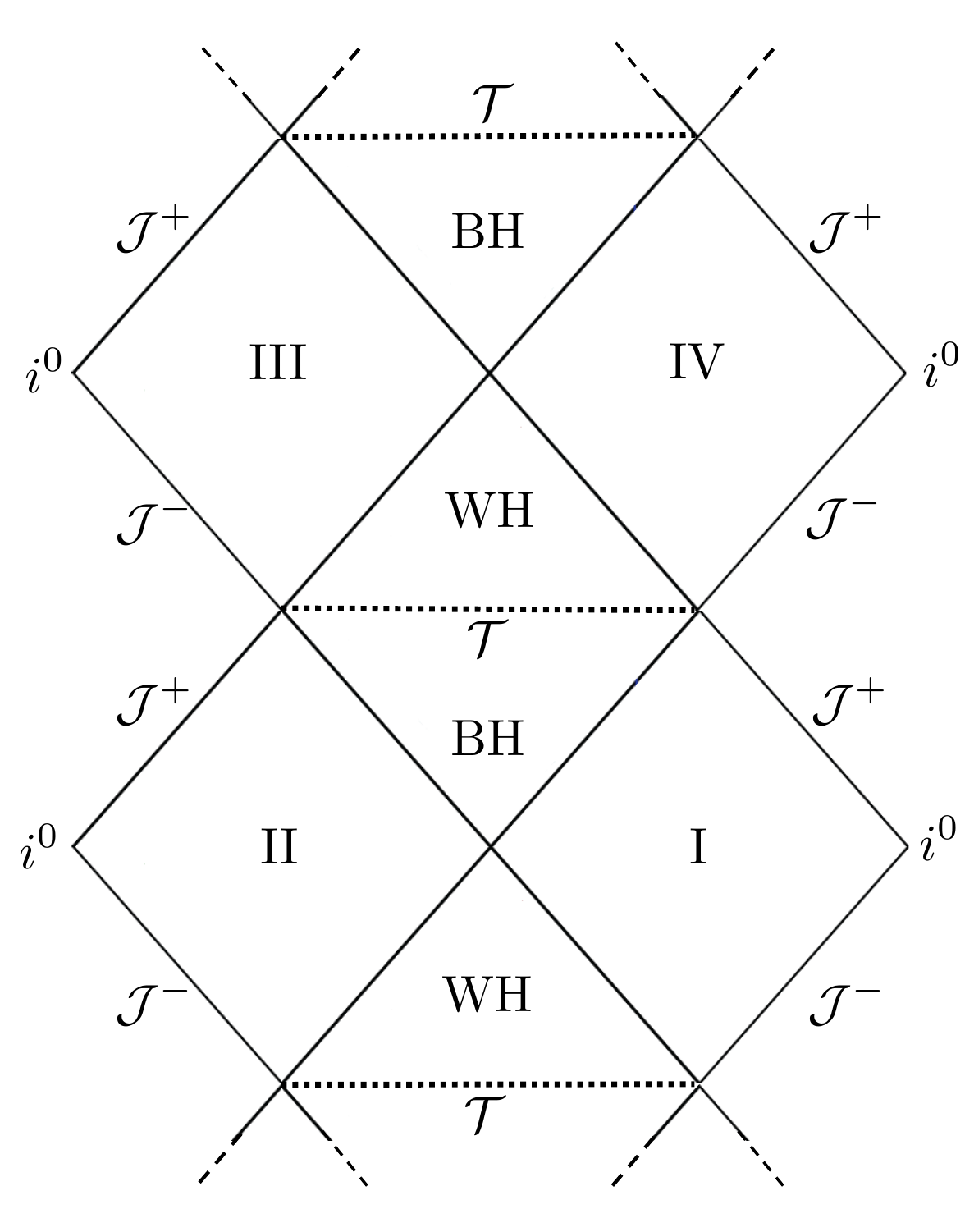}
 	\caption{Penrose diagram for the Kruskal extension of the full quantum corrected polymer Schwarzschild spacetime.}
 	\label{Penrosediag2}
 \end{figure}
 
Let us at this point recall some of the main features of the quantum corrected spacetime described by the metric coefficients \eqref{bquant}, \eqref{aquant}. The Penrose diagram is given in Fig. \ref{Penrosediag2}. It is an infinite tower of asymptotically Schwarzschild spacetimes of (alternating) masses $M_{BH}$ and $M_{WH}$. The would be Schwarzschild singularity is replaced by the spacelike transition surface of topology $\mathbb{R} \times \mathbb{S}^2$, where the areal radius reaches its minimal value given by
$$
\mathscr b_{\mathcal{T}} = 2^{1/12}(\lambda_1\lambda_2)^{1/4}(M_{BH}M_{WH})^{1/8} \;.
$$

\noindent
This surface represents the transition between trapped and anti-trapped regions and hence the transition from black to white hole interior regions and vice versa.
There are two horizons characterised by $a(r_{\pm}) = 0$ whose area is given by $A_H^{\pm} = 4 \pi \mathscr b(r_{\pm})^2$.
It is important to notice that the two masses are not fixed up to this point. The model allows in principle to choose both masses independently from each other. The masses alternate going though the Penrose diagram as the roles of $M_{BH}$ and $M_{WH}$ become exchanged going from one asymptotic region to another one.

\begin{figure}[t!]
	\centering
	\includegraphics[width=7.75cm,height=5.5cm]{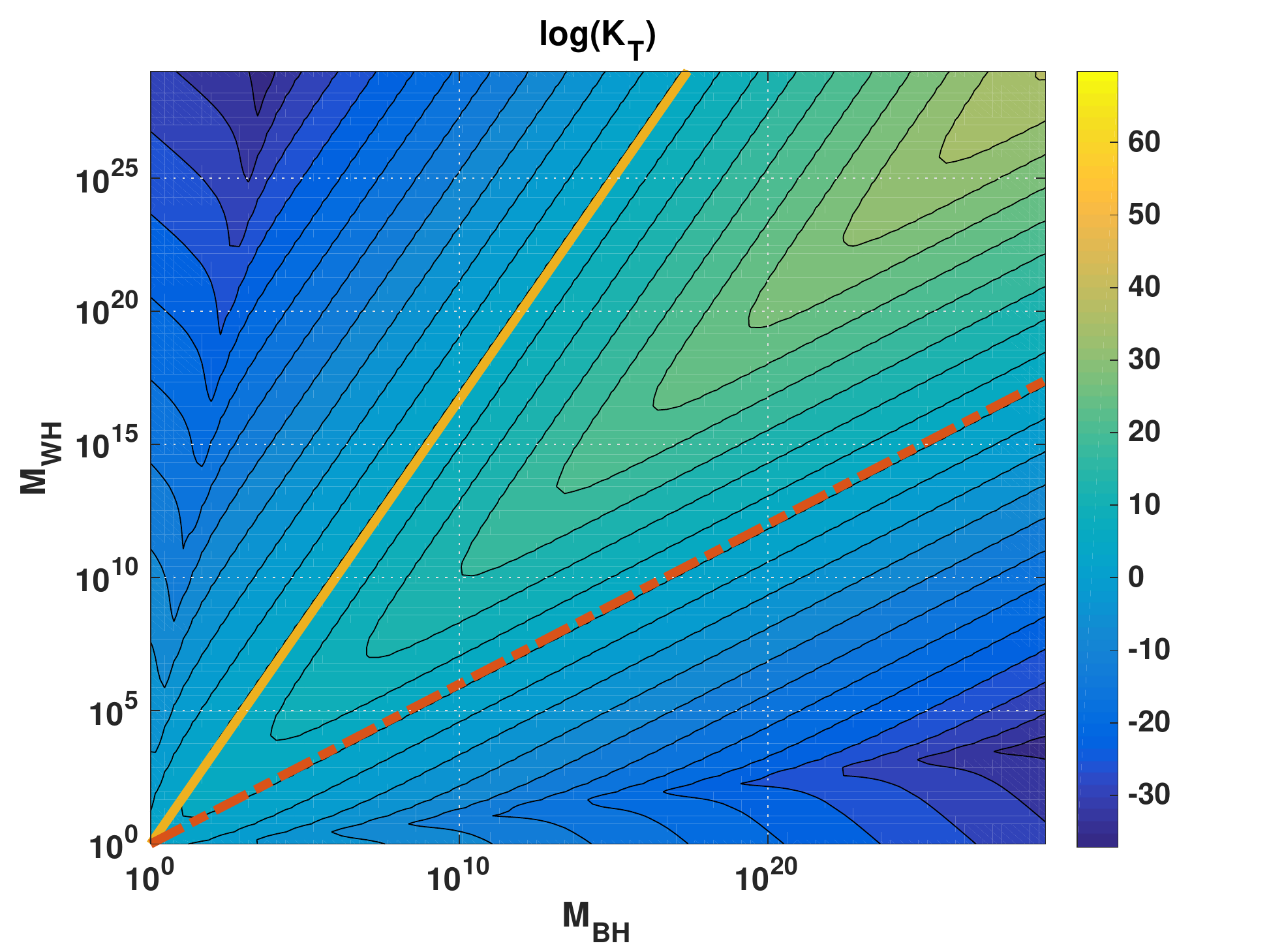}
	\caption{The colour scale encodes the value of the logarithm of the Kretschmann scalar at the transition surface as a function of the black hole $M_{BH}$ and white hole mass $M_{WH}$ for $\mathscr L_o\lambda_1= \lambda_2/\mathscr L_o =1$. Both axis are logarithmically. Finite non-zero curvatures for large masses can only be achieved by following a level line asymptotically given by Eq. \eqref{eq:BHWH} for $\beta = \frac{5}{3}$ and $\beta = \frac{3}{5}$. Different values of $\bar{m}$ correspond to different choices of the level line. The yellow line corresponds to $\beta = \frac{5}{3}$ and the red dashed line to $\beta = \frac{3}{5}$.}
	\label{fig:Colorkretschmann}
\end{figure}

A relation between the two masses can be found by adding a quantum condition. These arguments are usually heuristic and refer to conditions for plaquettes or curvature. For the presented model the authors chose the requirement of a mass independent unique upper curvature bound. The Kretschmann scalar reaches its maximal value close to the transition surface. A plot of the Kretschmann scalar at the transition surface as a function of the two masses is given in Fig. \ref{fig:Colorkretschmann}.
To achieve a unique mass independent curvature scale at which quantum effects become relevant, we need to fix a relation between the two masses, which for large masses approximately is a level line of $\mathcal{K}(r_\mathcal{T})$. This fixes the relation\footnote{Note that also $\beta = -1$ is a solution to this problem as analytic computations confirm and also can be seen in Fig. \ref{fig:Colorkretschmann} close to the axes. As this requires Planck size black hole and white hole masses and we expect effective polymer models not to be valid for such small masses, we excluded this solutions in \cite{BodendorferEffectiveQuantumExtended}.} to be

\begin{equation}\label{eq:BHWH}
M_{WH} \sim M_{BH}^\beta  \quad , \quad \beta = \frac{5}{3} \;, \frac{3}{5} \;.
\end{equation}

\noindent
This additional condition then fixes a relation between the masses, i.e. selects a relation among the initial conditions. 

In principle, we can rewrite the set of Dirac observables $(\mathcal M_{BH}, \mathcal M_{WH})$ in terms of the horizons $(\mathcal R_{BH}$, $\mathcal R_{WH})$. Abstractly, this simply amounts to 

$$
R_{BH} = b(r_+; M_{BH}, M_{WH}) \quad, \quad R_{WH} = b(r_-; M_{BH}, M_{WH}) \;,
$$

\noindent
on-shell, where the additional arguments denote that all $C$ and $D$ are replaced by $M_{BH}$ and $M_{WH}$ according to \eqref{Fonshell}. As $b(r_\pm)$ is an involved expression (cfr. \cite{BodendorferEffectiveQuantumExtended}), we do not report it here explicitly. Re-expressing then $(M_{BH}, M_{WH})$ by their off-shell Dirac observables gives the off-shell expressions for $(\mathcal R_{BH}, \mathcal R_{WH})$.

\subsection{Connection variables based models}\label{sec:intconst}

As discussed also in the classical setting, in the loop quantum gravity community  connection variables are mostly used. Different models were worked out by using different polymerisation schemes, as discussed above.
In what follows, we focus on a selection of previous models taken as representatives of the different polymerisation schemes. We discuss how horizon and mass Dirac observables can be constructed in these models as well as the subsequent restrictions on the initial conditions.

\subsubsection{(Generalised) $\mu_o$-schemes}
In this section we focus on the work of \cite{CorichiLoopquantizationof, AshtekarQuantumTransfigurarationof} and the notation therein. In contrast to them, we treat the polymerisation scales as purely constant at the beginning. This is the starting point of the original $\mu_o$-schemes of \cite{AshtekarQuantumGeometryand,ModestoLoopquantumblack,CampigliaLoopquantizationof,ModestoBlackHoleinterior}.
The effective quantisations is achieved by replacing the connection variables $(c, b)$ by the $\sin$-function, i.e.

\begin{equation}
c \longmapsto \frac{\sin\left(\delta_c c\right)}{\delta_c} \quad, \quad b \longrightarrow \frac{\sin\left(\delta_b b\right)}{\delta_b} \;,
\end{equation}
\noindent
where $\delta_c$, $\delta_b$ are the polymerisation scales and should be thought of Planck size. Note that as $c$ scales with the fiducial cell (cfr. \eqref{eq:sclaingcb}), the $\mathscr L_o$-independent physical polymerisation scales are $\delta_b$ and $\mathscr L_o \delta_c$, where $\mathscr L_o$ is the physical size of the fiducial cell (cfr. Sec. \ref{sec:Classical}).

The effective model is then described by the effective Hamiltonian ($G=1$)

\begin{equation}\label{eq:AOSeffH}
H = N_T \mathcal{H} \quad , \quad \mathcal{H} = - \frac{\sin{(\delta_b b)}}{2 \gamma^2 \text{sign}(p_c) \sqrt{|p_c|} \delta_b} \left( 2 \frac{\sin{(\delta_c c)}}{\delta_c} p_c + \left(\frac{\sin{(\delta_b b)}}{\delta_b}+ \frac{\gamma^2 \delta_b}{\sin{(\delta_b b)}} \right) p_b \right) \approx 0 \;.
\end{equation}
\noindent
With the choice 

$$
N_T = \frac{\gamma \;\text{sign}(p_c) \sqrt{|p_c|} \delta_b}{\sin{(\delta_b b)}} \;,
$$
\noindent
the equations of motions are\footnote{The works of \cite{CorichiLoopquantizationof,OlmedoFromblackholesto,AshtekarQuantumExtensionof,AshtekarQuantumTransfigurarationof} are not a $\mu_o$-scheme as the polymerisation schemes depend on the phase space through Dirac observables. However, as discussed in \cite{BodendorferAOSnote}, the above equations of motion follow only for purely constant $\delta_b$, $\delta_c$ from the Hamiltonian \eqref{eq:AOSeffH}.}

\begin{align}
\dot{b} = -\frac{1}{2} \left(\frac{\sin{(\delta_b b)}}{\delta_b} + \frac{\gamma^2 \delta_b}{\sin{(\delta_b b)}}\right) \quad,&\quad \dot{c} =  -2 \frac{\sin(\delta_c c)}{\delta_c} \;, \label{eq:EoMmuo1}
\\
\dot{p}_b = \frac{p_b}{2} \cos{(\delta_b b)} \left( 1- \frac{\gamma^2 \delta_b^2}{\sin(\delta_b b)^2} \right) \quad,& \quad \dot{p}_c = 2 p_c \cos(\delta_c c) \;. \label{eq:EoMmuo2}
\end{align}
\noindent
These equations are analytically solved by 

\begin{align}
c &= \frac{2}{\delta_c} \tan^{-1}\left(C e^{-2 T}\right) \;, \label{eq:AOScsol}\\
p_c &= D \left(C^2 e^{-2 T} + e^{2 T} \right) \;, \\
b &= \frac{1}{\delta_b} \cos^{-1}\left( b_o \tanh\left(\frac{1}{2} b_o T + A\right)\right) \;, \label{sol:AOSb}
\end{align}
\noindent
with $b_o = \sqrt{1+\gamma^2 \delta_b^2}$ and 

\begin{equation}\label{eq:AOSpbsol}
p_b = - 2 \frac{\sin(\delta_c c)}{\delta_c} \frac{\sin(\delta_b b)}{\delta_b} \frac{p_c}{\left(\frac{\sin(\delta_b b)^2}{\delta_b^2} + \gamma^2\right)} \;,
\end{equation}
\noindent
where we used the Hamiltonian constraint \eqref{eq:AOSeffH}\footnote{At this point the results are again the same for the $\mu_o$ and the generalised $\mu_o$-scheme of \cite{CorichiLoopquantizationof,OlmedoFromblackholesto,AshtekarQuantumExtensionof,AshtekarQuantumTransfigurarationof}. The subtle point is that \cite{CorichiLoopquantizationof,OlmedoFromblackholesto,AshtekarQuantumExtensionof,AshtekarQuantumTransfigurarationof} assume the equation of motions \eqref{eq:EoMmuo1}, \eqref{eq:EoMmuo2} but they do not follow from the Hamiltonian \eqref{eq:AOSeffH} (cfr. \cite{BodendorferAOSnote}). Nonetheless, in \cite{AshtekarQuantumExtensionof} it is shown that there exists a constrained Hamiltonian system leading to  \eqref{eq:EoMmuo1}, \eqref{eq:EoMmuo2} and $\mathcal{H} \approx 0$.
}. As in the classical part, there are three integration constants $A,C,D$, where $A$ simply induces a shift in the $T$ coordinate\footnote{Redefining $T' = T - 2A/b_o$ absorbs $A$ in a coordinate transformation which does not affect the physics.}. Without loss of generality, we can set $A = 0$. Note that, $T \in \left[-T_{\text{max}}, T_{\text{max}}\right]$ with $T_{\text{max}} = 2 \tanh^{-1}(1/b_o)/b_o$ describes the interior region.

We can relate the integration constants $C$, $D$ to the physical quantities of the black and white hole horizon. The horizons are located at the point $T_{BH/WH}$ where $g_{xx}$ vanishes. As neither $p_c$ nor $\sin(\delta_c c)/\delta_c$ become zero, the horizon condition is

\begin{equation}
g_{xx}(T_{BH/WH}) = 0 \quad \Leftrightarrow \quad p_b(T_{BH/WH}) = 0 \quad \Leftrightarrow \quad \sin(\delta_b b(T_{BH/WH})) = 0 \;.
\end{equation}
\noindent
This is the case at the boundaries of the $T$ interval, namely at $T_{BH/WH} = \pm T_{\text{max}} = \pm2 \tanh^{-1}(1/b_o)/b_o$, where we call $T_{BH} = +T_{\text{max}}$ the black hole horizon and $T_{WH}= -T_{\text{max}}$ the white hole horizon. The areal radii of the horizons are then given by

\begin{equation}\label{eq:onshellDOAOS}
R_{BH} = \left.\sqrt{|p_c|}\right|_{T=T_{BH}} = \sqrt{D\; \left(\frac{C^2}{B_o^2} + B_o^2\right)} \quad , \quad R_{WH} = \left.\sqrt{|p_c|}\right|_{T=T_{WH}} = \sqrt{D\; \left(C^2 B_o^2 + \frac{1}{B_o^2}\right)} \;,
\end{equation}
\noindent
with $B_o = \exp(2\tanh^{-1}(1/b_o)/b_o)$. Inverting these equations gives the integration constants in terms of the horizon radii

\begin{equation}\label{eq:CDAOS}
C^2 = \frac{R_{WH}^2 B_o^2-\frac{R_{BH}^2}{B_o^2}}{R_{BH}^2 B_o^2-\frac{R_{WH}^2}{B_o^2}} \quad , \quad D = \frac{R_{BH}^2 B_o^2 - \frac{R_{WH}^2}{B_o^2}}{B_o^4 - \frac{1}{B_o^4}} \;.
\end{equation}

\noindent
Note that $C$ becomes imaginary for

$$
\frac{R_{WH}^2 B_o^2-\frac{R_{BH}^2}{B_o^2}}{R_{BH}^2 B_o^2-\frac{R_{WH}^2}{B_o^2}} < 0 \;.
$$
\noindent
As long as $R_{BH/WH}$ remains real-valued and furthermore the metric is real-valued, this causes no problems.
It is also possible to prove that the metric, i.e. $p_c$ and $p_b$ are symmetric under the exchange of $R_{BH} \longleftrightarrow R_{WH}$ (i.e $C^2 \mapsto 1/C^2$, $D \mapsto DC^2$) and $T \longmapsto -T$.

Solving the solutions for the integration constants $C$, $D$ and expressing $T$ as a function of $c$ from Eq. \eqref{eq:AOScsol} gives phase-space expressions for them.
Specifically, we have

\begin{equation}
p_c \sin\left(\delta_c c\right) = 2 C D \quad , \quad e^{-2T} = \frac{\tan\left(\frac{\delta_c c}{2}\right)}{C} \;,
\end{equation}
\noindent
and inserting this into Eq. \eqref{sol:AOSb} yields

\begin{equation}
C = \tan\left(\frac{\delta_c c}{2}\right) \left(\frac{b_o + \cos\left(\delta_b b\right)}{b_o - \cos\left(\delta_b b\right)}\right)^{\frac{2}{b_o}}
\end{equation}
\noindent
Rearranging these according to \eqref{eq:onshellDOAOS} gives the Dirac observables for the horizons

\begin{align}
\mathcal{R}_{BH} &= \left[ \frac{p_c \sin(\delta_c c)}{2} \left(\frac{\tan\left(\frac{\delta_c c}{2}\right)}{B_o^2} \left(\frac{b_o+\cos(\delta_b b)}{b_o-\cos(\delta_b b)}\right)^{\frac{2}{b_o}} + \frac{B_o^2}{\tan\left(\frac{\delta_c c}{2}\right)}\left(\frac{b_o-\cos(\delta_b b)}{b_o+\cos(\delta_b b)}\right)^{\frac{2}{b_o}}\right)  \right]^{\frac{1}{2}} \label{eq:RBHDOAOS}\;,\\
\mathcal{R}_{WH} &= \left[ \frac{p_c \sin(\delta_c c)}{2} \left(B_o^2 \tan\left(\frac{\delta_c c}{2}\right) \left(\frac{b_o+\cos(\delta_b b)}{b_o-\cos(\delta_b b)}\right)^{\frac{2}{b_o}} + \frac{1}{B_o^2 \tan\left(\frac{\delta_c c}{2}\right)}\left(\frac{b_o-\cos(\delta_b b)}{b_o+\cos(\delta_b b)}\right)^{\frac{2}{b_o}} \right) \right]^{\frac{1}{2}} \label{eq:RWHDOAOS}\;.
\end{align}
\noindent
Note that we can not simply take the limit $\delta_b,\delta_c \rightarrow 0$. This will not give back the classical solutions. The reason for this is that the equations of motion \eqref{eq:EoMmuo1}, \eqref{eq:EoMmuo2} converge only point-wise and not uniformly to the classical equations. As such integrating the equations and taking the limit does not commute. Nonetheless, there is no problem as the solutions need to be well approximated by the classical solutions at the horizons (or in general in the classical regime), which is a different limit. 
The advantage of writing down these observables is the full control over the integration constants and their physical content. Being Dirac observables, these quantities are fully gauge independent and free of any coordinate choice. Furthermore, in contrast to the classical situation, both observables are independent of fiducial cell structures, i.e. both of them have physical meaning. Giving the Dirac observables specific values fixes all integration constants and makes the solutions unique.

Furthermore, note that $R_{BH/WH} \neq 2M_{BH/WH}$\footnote{To be precise $R_{BH/WH} \neq 2M_{BH/WH}$ for $M_{BH/WH}$ the ADM mass. This is usually meant by talking about ``the mass of a black hole'' as the ADM mass is the gravitational mass experienced by a distant observer. What still hold true by definition is $R_{BH/WH} = 2M_{\text{Misner-Sharp}}(R_{BH/WH})$, where $M_{\text{Misner-Sharp}}(b)$ is the Misner-Sharp mass (cfr. p. 40 in \cite{SzabadosQuasi-LocalEnergy} and references therein or Sec. \ref{sec:newvariables}), which is a quasi-local measure of energy enclosed in a sphere of areal radius $b$. Depending on the asymptotic behaviour of the spacetime at spatial infinity the Misner-Sharp mass (at infinity) and the AMD mass might be a priory completely different objects.} opposite to what is stated in \cite{CorichiLoopquantizationof,AshtekarQuantumExtensionof,AshtekarQuantumTransfigurarationof} (and also other references as e.g. \cite{BoehmerLoopquantumdynamics}). To really speak about the mass of the black hole, it is necessary to construct the exterior spacetime and check the asymptotic behaviour. Only if the metric is asymptotically a Schwarzschild spacetime, the black hole mass can be read off. As we saw in Sec. \ref{sec:BMM} (and \cite{BodendorferEffectiveQuantumExtended}), the relation between black hole horizon radius and black hole mass can be non-trivial. Of course, even if the asymptotic spacetime is not Schwarzschild, but asymptotically flat, in principle the ADM mass can be constructed. But also here a detailed analysis of the exterior spacetime is necessary and the interpretation as black hole is a priori not guaranteed. In \cite{CorichiLoopquantizationof} this analysis of the exterior spacetime has not been performed. In \cite{AshtekarQuantumExtensionof,AshtekarQuantumTransfigurarationof}, the exterior spacetime was analysed, but it is not asymptotically Schwarzschild as noted in \cite{BrahmaCommentonAOS}. Indeed, in \cite{ModestoSemiclassicalLoopQuantum} it has been studied which polymerisation and $\mu_o$-scheme yields asymptotically Schwarzschild spacetime as will be discussed in Sec. \ref{sec:modesto}. 

In contrast to the classical case, also in these models, we have two physical quantities to fix the two integration constants. Fixing the integration constants simply in coordinates to match the classical metric at the black hole horizon transfers the classical non-physical ambiguity in the effective quantum theory where it has physical effect. Concluding, in effective quantum models the fixing of integration constants needs a careful analysis.

The most recent polymer black hole models based on the $\mu_o$-solutions discussed in this section are the generalised $\mu_o$-models of \cite{AshtekarQuantumExtensionof,AshtekarQuantumTransfigurarationof}. 
In these works the  polymerisation scales are related to full LQG parameters by means of quantum geometry arguments based on rewriting the curvature in terms of the holonomies of the gravitational connection along suitably chosen plaquettes enclosing the minimal area at the transition surface. This introduces a mass dependence of the polymerisation scales.
As in previous papers, although a detailed analysis of the integration constants in terms of Dirac observables as it is presented here was not carried out, a relation between black hole and white hole horizon was implicitly fixed (see below). In the light of the above discussion, we see that actually there are two free input parameters, which can be specified at will. This gives an additional degree of freedom and we can study if the plaquette argument can be satisfied even with constant polymerisation scales. After a detailed discussion which was given in \cite{AshtekarQuantumTransfigurarationof}, the mathematical requirement is

\begin{align}\label{eq:plaquette}
2 \pi \delta_c \delta_b |p_b|_{\mathcal{T}} = \Delta \quad , \quad 
4 \pi \delta_b^2 |p_c|_{\mathcal{T}} = \Delta \;,
\end{align} 

\noindent
where subscript $\mathcal{T}$ means evaluation at the transition surface and $\Delta$ is the area gap predicted by the full theory of LQG. The spacelike transition surface is identified by the time $\mathcal{T} = \ln(C)/2$ as $\dot{p_c}(\mathcal{T}) = 0$. Evaluating the solutions \eqref{eq:AOScsol}-\eqref{eq:AOSpbsol} at the transition surface, the conditions \eqref{eq:plaquette} then give

\begin{align}
\frac{8 \pi C D \delta_b^2}{b_o^2} \sqrt{1-b_o^2\tanh\left(\frac{1}{4}b_o \ln(C)\right)^2}\cosh\left(\frac{1}{4}b_o \ln(C)\right)^2 &= \Delta \;,\label{eq:plaquette1}\\
8 \pi CD \delta_b^2 &= \Delta \;. \label{eq:plaquette2}
\end{align}

\noindent
These equations should be seen as conditions for $\delta_b$ and $\delta_c$ as $C$ may also contain $\delta_c$. We can simplify the equations by dividing the first one by the second yielding

\begin{equation}\label{eq:plaquettesimplifed}
\frac{1}{b_o^2} \sqrt{1-b_o^2\tanh\left(\frac{1}{4}b_o \ln(C)\right)^2}\cosh\left(\frac{1}{4}b_o \ln(C)\right)^2 = 1\;,
\end{equation}

\noindent
which now is an equation for $\delta_b$ in terms of $C$ only (recall that $b_o = \sqrt{1+\gamma^2 \delta_b^2}$). The solution is complicated and not necessarily unique. In any case, a solution gives $\delta_b$ as a function of $C$ only. This means that we have the possibility to obtain a horizon (mass) independent $\delta_b$ (as it was initially assumed for a proper $\mu_o$-scheme) only if $C$ is independent of the horizons. For 

\begin{equation}\label{eq:AOSRadiusrelation}
R_{WH} = \alpha R_{BH} \;,
\end{equation}

\noindent
with $\alpha$ a dimensionless constant, Eq. \eqref{eq:CDAOS} yields

\begin{equation}
C^2 = \frac{\alpha^2 B_o^2-\frac{1}{B_o^2}}{B_o^2 -\frac{\alpha^2}{B_o^2}} \quad , \quad D = R_{BH}^2 \frac{B_o^2-\frac{\alpha^2}{B_o^2}}{B_o^4-\frac{1}{B_o^4}} \;,
\end{equation}

\noindent
i.e. $C$ is horizon independent\footnote{Note that the condition $\frac{\partial}{\partial R_{BH}} C^2(R_{WH}(R_{BH}),R_{BH})$ has only solutions of the form \eqref{eq:AOSRadiusrelation}.}, while $D$ goes as $R_{BH}^2$. Although this yields a constant $\delta_b$ according to \eqref{eq:plaquettesimplifed}, it is in conflict with \eqref{eq:plaquette2} as this gives

$$
\delta_b^2 = \frac{\Delta}{8 \pi C D}\;,
$$

\noindent
which can only be satisfied if $\delta_b \propto R_{BH}^{-1}$ as $D$ is horizon dependent. This is a contradiction and shows that no solutions to the plaquette equations \eqref{eq:plaquette1} and \eqref{eq:plaquette2} with horizon independent $\delta_b$ and $\delta_c$ exist. This further shows that if \eqref{eq:plaquette1} and \eqref{eq:plaquette2} are imposed, the polymerisation can not be a pure $\mu_o$-scheme and hence a construction as in \cite{AshtekarQuantumTransfigurarationof} is necessary, which on the other hand loses the connection to the initial Hamiltonian \eqref{eq:AOSeffH} (cfr. \cite{BodendorferAOSnote}).

\begin{figure}[t!]
	\centering
	\includegraphics[width=7.75cm,height=5.5cm]{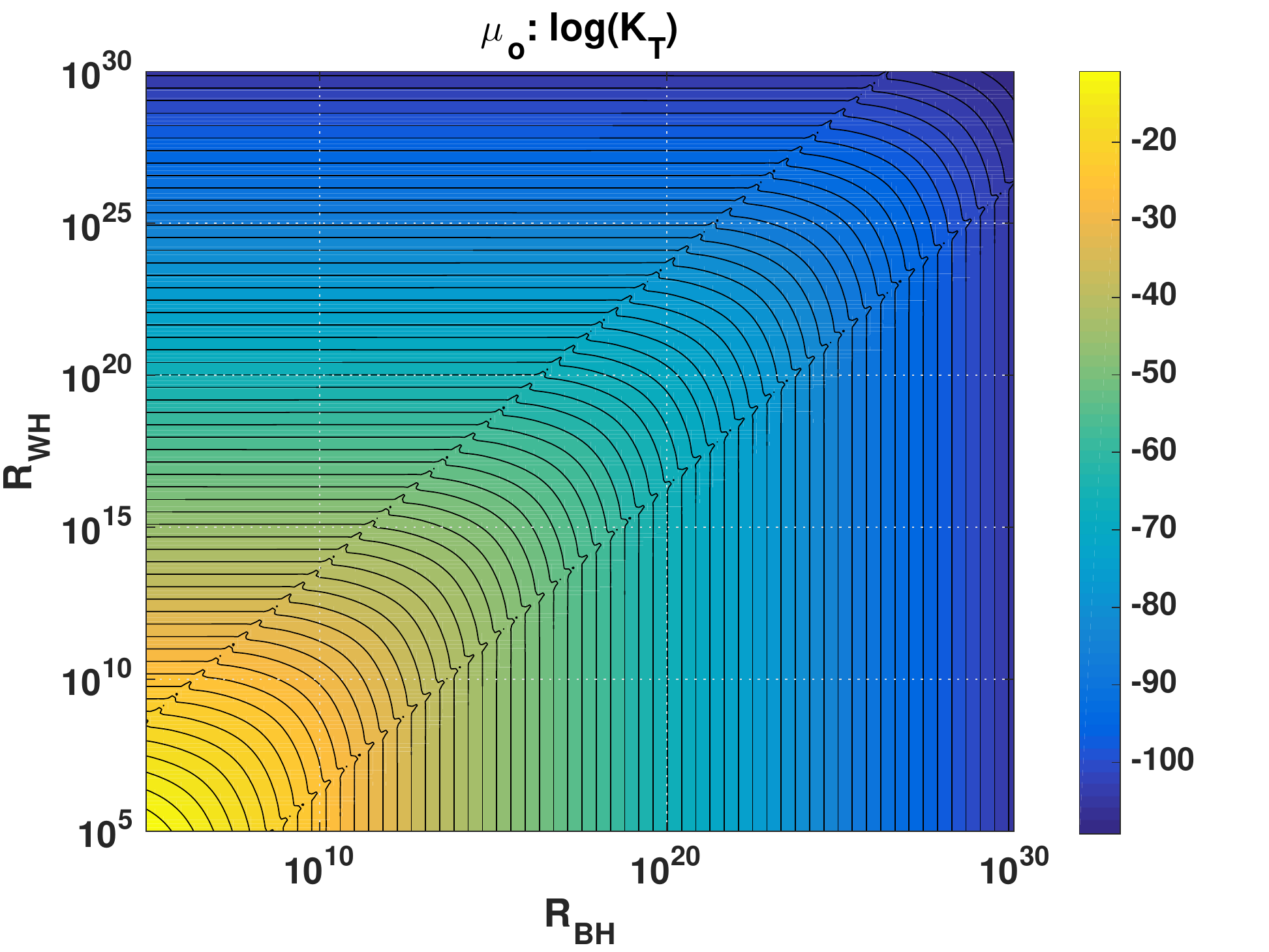}
	\caption{The colour scale encodes the value of the logarithm of the Kretschmann scalar at the transition surface as a function of the black hole $R_{BH}$ and white hole horizon $R_{WH}$ for $L_o=1$, $\delta_b=1$, $\delta_c =1$ and $\gamma = 0.2375$ (cfr. \cite{DomagalaBlack-holeentropy,MeissnerBlack-holeentropy}). Both axis are logarithmically.}
	\label{fig:ColorkretschmannAOS}
\end{figure}

As an alternative strategy, we could follow the argument of \cite{BodendorferEffectiveQuantumExtended} and fix the polymerisation scales to be constant, but fix a relation between the horizons in such a way that the curvature is bounded by a unique mass independent scale. For this, we report the Kretschmann scalar at the transition surface as a function of $R_{BH}$ and $R_{WH}$ in Fig. \ref{fig:ColorkretschmannAOS}.
To achieve a mass independent upper bound for the curvature, the relation between the black hole and white hole horizon needs to be a level line of Fig. \ref{fig:ColorkretschmannAOS}. As can be seen from the plot this, induces a maximal value for the black hole and white hole horizon depending on which level line is picked.

\begin{figure}[t!]
	\centering
	\includegraphics[width=7.75cm,height=5.5cm]{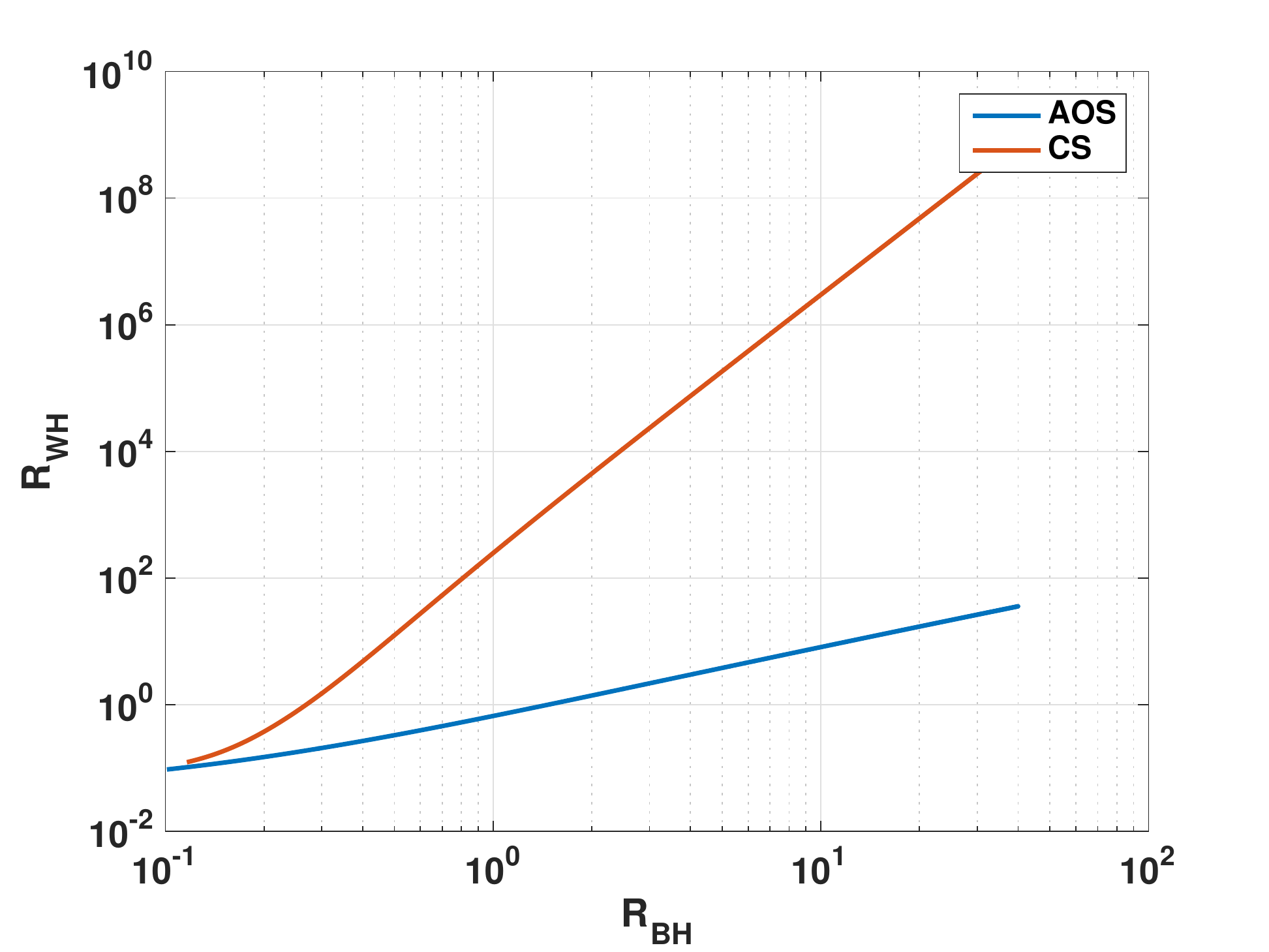}
	\caption{Relation between $R_{BH}$ and $R_{WH}$ in \cite{CorichiLoopquantizationof} (CS) and \cite{AshtekarQuantumExtensionof,AshtekarQuantumTransfigurarationof} (AOS). The plots show Eq. \eqref{eq:RBH-RWHmuo} for $\delta_b = \sqrt{\Delta}/2m$, $\delta_c = \sqrt{\Delta}/L_o$ for \cite{CorichiLoopquantizationof}  and $\delta_b = \left(\sqrt{\Delta}/(\sqrt{2 \pi} \gamma^2 m)\right)^{1/3}$, $\delta_c = \left(\gamma \Delta^2/(4 \pi^2 m)\right)^{1/3}/(2 L_o)$ for \cite{AshtekarQuantumExtensionof,AshtekarQuantumTransfigurarationof}. The parameters are $L_o=1$, $\delta_b=1$, $\delta_c =1$, $\Delta = 1$ and $\gamma = 0.2375$ (cfr. \cite{DomagalaBlack-holeentropy,MeissnerBlack-holeentropy}).}
	\label{fig:HorizonRelationCSAOS}
\end{figure}

As a last point, let us recall how the integration constants in the generalised $\mu_o$-schemes \cite{CorichiLoopquantizationof,AshtekarQuantumExtensionof,AshtekarQuantumTransfigurarationof} are actually chosen. There, the integration constants are fixed directly by expressing them in terms of only one free parameter $m$ using an argument coming from the classical solution. After rescaling $T \mapsto T' = T +2 \tanh^{-1}(1/b_o)/b_o$ of the solutions of \cite{CorichiLoopquantizationof,AshtekarQuantumExtensionof,AshtekarQuantumTransfigurarationof} the integration constants are in the language used in the present paper given by

\begin{equation}
C = \mp \frac{\gamma L_o \delta_c}{8 m} B_o^2 \quad , \quad D = \frac{4 m^2}{B_o} \;.
\end{equation}
\noindent
Note that in these papers, it is claimed that $2m$ is the black hole horizon or in fact that $m$ is the black hole mass. As discussed above, this can only be justified if there is an exterior metric which is asymptotically Schwarzschild, which is not true in all three papers (cfr. \cite{BrahmaCommentonAOS}). Also, the horizons are fixed in both cases at the moment where $C$ and $D$ are fixed. They are given by

\begin{equation}\label{eq:RBH-RWHmuo}
R_{BH} = 2 m \sqrt{1+ \frac{\delta_c^2 \gamma^2 L_o^2}{64 m^2}} \quad , \quad R_{WH} = 2m \sqrt{\frac{1}{B_o^4} + \frac{\delta_c^2 \gamma^2 L_o^2 B_o^4}{64 m^2}} \;,
\end{equation}
\noindent
which shows that $2m$ is the black hole horizon only up to quantum corrections. Note that $R_{BH}$ and $R_{WH}$ are related to each other as there is only one free parameter $m$ left. The only difference between the approach of \cite{CorichiLoopquantizationof} and \cite{AshtekarQuantumExtensionof,AshtekarQuantumTransfigurarationof} is how the polymerisation scales $\delta_c$ and $\delta_b$ are chosen and how they depend on the parameter $m$. This gives different behaviours $R_{WH} (R_{BH})$ (see Fig. \ref{fig:HorizonRelationCSAOS}).

\subsubsection{Modesto approach}\label{sec:modesto}

Another similar but sightly different approach was presented by Modesto \cite{ModestoSemiclassicalLoopQuantum,ModestoSelf-dualBlackHoles}. After doing a holonomy argument Modesto arrives at the effective Hamiltonian constraint

\begin{equation}\label{eq:MeffH}
H = N_T \mathcal{H} \quad , \quad \mathcal{H} = - \frac{\sin{(\sigma(\delta) \delta b)}}{2 G \gamma^2 \text{sign}(p_c) \sqrt{|p_c|} \delta} \left( 2 \frac{\sin{(\delta c)}}{\delta} p_c + \left(\frac{\sin{(\sigma(\delta) \delta b)}}{\delta}+ \frac{\gamma^2 \delta}{\sin{( \sigma(\delta) \delta b)}} \right) p_b \right) \approx 0 \;,
\end{equation}

\noindent
with only one polymerisation scale $\delta$ and the function $\sigma(\delta)$, which is initially not specified.
The effective Hamiltonian is very similar to the one discussed in the previous section (cfr. \eqref{eq:AOSeffH}), only the constant polymerisation scales have different forms. This leads to formally similar equations.
Due to this, we can follow the previous construction and find for $N_T = (\gamma \text{sign}(p_c) \sqrt{|p_c|} \delta)/\sin(\sigma(\delta) \delta b)$

\begin{align}
\dot{b} = -\frac{1}{2} \left(\frac{\sin{(\sigma(\delta) \delta b)}}{\delta} + \frac{\gamma^2 \delta}{\sin{(\sigma(\delta) \delta b)}}\right) \quad,&\quad \dot{c} =  -2 \frac{\sin(\delta c)}{\delta} \;,
\\
\dot{p}_b = \frac{p_b}{2} \sigma(\delta) \cos{(\delta b)} \left( 1- \frac{\gamma^2 \delta^2}{\sin(\sigma(\delta) \delta b)^2} \right) \quad,& \quad \dot{p}_c = 2 p_c \cos(\delta c) \;,
\end{align}
\noindent
as equations of motion, which are solved by

\begin{align}
c &= \frac{2}{\delta} \tan^{-1}\left(C e^{-2 T}\right) \;,
\label{eq:csolMod}\\
p_c &= D \;\left(C^2 e^{-2 T} + e^{2 T} \right) \;, \label{eq:pcsolMod}\\
b &= \frac{1}{\sigma(\delta) \delta} \cos^{-1}\left( b_o \tanh\left(\frac{1}{2} b_o \sigma(\delta) T + A\right)\right) \;, \label{sol:Mb}
\\
p_b& = - 2 \frac{\sin(\delta c)}{\delta} \frac{\sin(\sigma(\delta) \delta b)}{\delta} \frac{p_c}{\left(\frac{\sin(\sigma(\delta) \delta b)^2}{\delta^2} + \gamma^2\right)} \;,
\label{eq:pbsolMod}
\end{align}
\noindent
with $b_o = \sqrt{1+\gamma^2 \delta^2}$.
Obviously, as in the previous setting, there are three integration constants $A$, $C$, $D$, where $A$ is simply a shift in the $T$-coordinate which we can set to zero without loss of generality. The two remaining integration constants $C$, $D$ can now be related to the black and white hole horizon radius by 

\begin{equation}\label{eq:onshellDOM}
R_{BH} = \left.\sqrt{|p_c|}\right|_{T=T_{BH}} = \sqrt{D\; \left(\frac{C^2}{B_o^2} + B_o^2\right)} \quad , \quad R_{WH} = \left.\sqrt{|p_c|}\right|_{T=T_{WH}} = \sqrt{D\; \left(C^2 B_o^2 + \frac{1}{B_o^2}\right)} \;,
\end{equation}
\noindent
where 
$$
p_b(T_{BH/WH}) = 0\;, \;T_{BH/WH} = \pm T_{max} = \pm \frac{2}{\sigma(\delta) b_o} \tanh^{-1}\left(\frac{1}{b_o}\right) \;,\; B_o = \exp\left(\frac{2}{\sigma(\delta) b_o} \tanh^{-1}\left(\frac{1}{b_o}\right)\right) \;.
$$ Note that the definitions of $T_{max}$, $b_o$, $B_o$ are different with respect to the previous section, but the construction works exactly along the same steps.

As discussed in \cite{ModestoSemiclassicalLoopQuantum}, we can reconstruct the metric coefficients $\bar{a}$ and $\mathscr{b}$ via \eqref{eq:metricmap}, leading to

\begin{align}
\bar{a} &= -\frac{16 C^2 D \left(1-b_o^2 \tanh ^2\left(\frac{1}{2} b_o T \sigma(\delta )\right)\right)}{L_o^2 \left(C^2 e^{-2 T}+e^{2 T}\right) \left(b_o^2 \tanh ^2\left(\frac{1}{2} b_o T \sigma(\delta )\right)-\delta^2 \gamma ^2-1\right)^2} \;,
\\
\mathscr b &= \sqrt{D\; \left(C^2 e^{-2 T} + e^{2T}\right)} \;,
\\
N_T^2 &= \frac{\delta^2 \gamma^2  \left( D  \left(C^2e^{-2 T}+e^{2 T}\right)\right)}{1-b_o^2 \tanh^2\left(\frac{1}{2} b_o T \sigma (\delta )\right)} \;.
\end{align}

\noindent
These expressions are extendible beyond $T_{BH/WH}$, thus providing us with the exterior solution as the analytic extension of the interior metric. As such, we can study the asymptotic regions $T \rightarrow \pm \infty$, $\mathscr{b} \rightarrow \infty$.
In the $T \rightarrow \infty$ limit, we find 

$$
\mathscr{b}_+ := \mathscr{b}(T \rightarrow + \infty) \simeq \sqrt{D} \;e^{T} \;,
$$

\noindent
and

$$
\tanh^2\left(\frac{b_o \sigma(\delta)}{2} T \right) = \left(\frac{e^{b_o \sigma(\delta) T}-1}{e^{b_o \sigma(\delta) T}+1}\right)^2 \simeq \left(1-2 e^{-b_o \sigma(\delta) T}\right)^2 \simeq 1-4 \left(\frac{\mathscr{b}_+}{\sqrt{D}}\right)^{-b_o \sigma(\delta)} \;.
$$

\noindent
Asymptotically, the metric is then described by

\begin{equation}
\dd s^2_+ \simeq -N_T^2 \dd T^2 - \bar{a} \dd x^2 + \mathscr{b}_+^2 \dd \Omega_2^2
\end{equation}
\noindent
with 

\begin{align}
N_T^2 &\simeq -\frac{\delta^2 \gamma^2 \mathscr{b}_+^2}{\delta^2 \gamma^2 - 4 b_o^2 \left(\frac{\mathscr{b}_+}{\sqrt{D}}\right)^{-b_o \sigma(\delta)}} \;,
\\
\bar{a} &\simeq \frac{C^2 D^{2-b_o \sigma(\delta)} \gamma^2 \delta^2}{L_o^2 b_o^4} \left( \mathscr{b}_+^{2b_o \sigma(\delta)-2} - \frac{4 b_o^2 D^{\frac{b_o \sigma(\delta)}{2}}}{\delta^2 \gamma^2} \frac{1}{\mathscr{b}_+^{2-b_o \sigma(\delta)}} \right) \;.
\end{align}

\noindent
From this expression, we conclude that the only possibility for asymptotically Minkowski spacetime is 

\begin{equation}\label{eq:sigma}
\sigma(\delta) = \frac{1}{b_o} = \frac{1}{\sqrt{1+\gamma^2 \delta^2}} \;,
\end{equation}

\noindent
which is exactly the result of \cite{ModestoSemiclassicalLoopQuantum}. Changing the coordinates according to $T \mapsto \mathscr b_+ = \sqrt{D} e^T$ and $x \mapsto \tau := 2 C \sqrt{D} \gamma \delta/(L_o b_o^2) x$ gives the asymptotic line element

\begin{equation}\label{eq:approx+}
\dd s_+^2 \simeq -\left(1 - \frac{4 b_o^2 \sqrt{D}}{\delta^2 \gamma^2 \mathscr{b}_+}\right) \dd \tau^2 + \frac{1}{1- \frac{4 b_o^2 \sqrt{D}}{\gamma^2 \delta^2 \mathscr{b}_+}} \dd \mathscr{b}_+^2 + \mathscr{b}_+^2 \dd \Omega_2^2\;,
\end{equation}

\noindent
which is a Schwarzschild metric with mass

\begin{equation}\label{eq:MBHmodesto}
M_{BH} = \frac{2 b_o^2 \sqrt{D}}{\gamma^2 \delta^2} \;.
\end{equation}

\noindent
Note that the choice \eqref{eq:sigma} was crucial here to identify the mass. The (ADM) mass is only asymptotically defined and hence it is crucial to have the right asymptotic behaviour. Again, as discussed in the previous section, having only the interior metric as in \cite{CorichiLoopquantizationof} or not asymptotic Schwarzschild spacetime as in \cite{AshtekarQuantumExtensionof,AshtekarQuantumTransfigurarationof} different possibly inequivalent notions of mass do appear and what is meant by ``black hole mass'' needs to be specified. In this approach, on the other hand, the exterior metric exists and the corresponding spacetime is asymptotically isometric to the Schwarzschild solution, hence $M_{BH}$ is a well-defined quantity.

We can repeat the last steps for the other asymptotic region, i.e. $T \rightarrow -\infty$, yielding

$$
\mathscr{b}_- := \mathscr{b}(T \rightarrow - \infty) \simeq \sqrt{D}C \;e^{-T} \;,
$$

\noindent
and 

$$
\tanh^2\left(\frac{b_o \sigma(\delta)}{2} T \right) = \left(\frac{e^{b_o \sigma(\delta) T}-1}{e^{b_o \sigma(\delta) T}+1}\right)^2 \simeq \left(-1+2 e^{b_o \sigma(\delta) T}\right)^2 \simeq 1-4 \left(\frac{\mathscr{b}_-}{\sqrt{D}C}\right)^{-b_o \sigma(\delta)} \;.
$$
\noindent
Asymptotically we find then 

\begin{align}
N_T^2 &\simeq -\frac{\delta^2 \gamma^2 \mathscr{b}_-^2}{\delta^2 \gamma^2 - 4 b_o^2 \left(\frac{\mathscr{b}_-}{\sqrt{D}C}\right)^{-b_o \sigma(\delta)}} \;,
\\
\bar{a} &\simeq \frac{C^{2-2 b_o \sigma(\delta)} D^{2-b_o \sigma(\delta)} \gamma^2 \delta^2}{L_o^2 b_o^4} \left( \mathscr{b}_-^{2b_o \sigma(\delta)-2} - \frac{4  b_o^2(DC^2)^{\frac{b_o \sigma(\delta)}{2}}}{\delta^2 \gamma^2} \frac{1}{\mathscr{b}_-^{2-b_o \sigma(\delta)}} \right) \;.
\end{align}
\noindent
Consistently with the other asymptotic region, for Eq. \eqref{eq:sigma} being satisfied, the spacetime becomes asymptotically Schwarzschild spacetime. Changing the coordinates to $T \mapsto \mathscr b_- = \sqrt{D}C e^{-T}$ and $x \mapsto \tau := \sqrt{D}\gamma \delta/(L_o b_o^2) x$ gives the asymptotic metric

\begin{equation}\label{eq:approx-}
\dd s_-^2 \simeq -\left(1 - \frac{4 b_o^2 \sqrt{D}C}{\delta^2 \gamma^2 \mathscr{b}_-}\right) \dd \tau^2 + \frac{1}{1- \frac{4 b_o^2 \sqrt{D} C}{\gamma^2 \delta^2 \mathscr{b}_-}} \dd \mathscr{b}_-^2 + \mathscr{b}_-^2 \dd \Omega_2^2\;,
\end{equation}
\noindent
which is a Schwarzschild spacetime of mass

\begin{equation}\label{eq:MWHmodesto}
M_{WH} = \frac{2 b_o^2 \sqrt{D} C}{\delta^2 \gamma^2} \;.
\end{equation}
\noindent
Note that, as in the previous section, in the limit $\delta\to0$ the solutions \eqref{eq:csolMod}-\eqref{eq:pbsolMod} do not reduce to the classical result. As discussed before, this is on the one hand due to possible hidden $\delta$ in the integration constants $C$ and $D$ as well as the non-uniform convergence of the equations of motion to the classical ones. Nevertheless, the relevant requirement is quantum effects to be negligible in the ``classical regime''. This is the case in the region far away from the transition surface, where the spacetime geometry is well approximated by the classical solution asymptotically. As the exterior metric exists in this case, we can check this explicitly and find the consistency condition \eqref{eq:sigma}. Although $\delta$ still appears in the approximate spacetime \eqref{eq:approx+}, \eqref{eq:approx-}, it is classical as it should.

Mass Dirac observables can be constructed by inverting the solutions \eqref{eq:csolMod}-\eqref{eq:pbsolMod} for the integration constants and replacing the resulting expressions in \eqref{eq:MBHmodesto} and \eqref{eq:MWHmodesto}. To this aim, let us first invert Eq. \eqref{eq:csolMod} to get

\begin{equation}\label{eq:TincMod}
e^{-2T} = \frac{\tan\left(\frac{\delta c}{2}\right)}{C} \;.
\end{equation}

\noindent
Inserting then Eq. \eqref{eq:TincMod} into Eq. \eqref{sol:Mb} gives (with \eqref{eq:sigma})

$$
\cos\left(\frac{\delta b}{b_o}\right) = b_o \frac{1-e^{-T}}{1+e^{-T}} = b_o \frac{1- \sqrt{\frac{\tan\left(\delta c\right)}{C}}}{1+ \sqrt{\frac{\tan\left(\delta c\right)}{C}}} \;,
$$
\noindent
which solving for $C$ yields

\begin{equation}
C = \tan\left(\frac{\delta c}{2}\right) \left( \frac{b_o + \cos\left(\frac{\delta b}{b_o}\right)}{b_o - \cos\left(\frac{\delta b}{b_o}\right)} \right)^2 \;.
\end{equation}

\noindent
By noticing that

\begin{equation}
p_c \sin\left(\delta c\right) = 2 C D \;,
\end{equation}

\noindent
the phase space expression for $D$ can be easily constructed from the above result for $C$. Now we have phase space expressions for the integration constants $C$, $D$, which we can use to construct the phase space functions for the masses

\begin{align}
\mathcal{M}_{BH} &= \frac{2 b_o^2}{\gamma^2 \delta^2} \sqrt{p_c} \cos\left(\frac{\delta c}{2}\right)\;  \left( \frac{b_o - \cos\left(\frac{\delta b}{b_o}\right)}{b_o + \cos\left(\frac{\delta b}{b_o}\right)} \right) \;,
\\
\mathcal{M}_{WH} &= \frac{2 b_o^2}{\gamma^2 \delta^2} \sqrt{p_c} \sin\left(\frac{\delta c}{2}\right)\;  \left( \frac{b_o + \cos\left(\frac{\delta b}{b_o}\right)}{b_o - \cos\left(\frac{\delta b}{b_o}\right)} \right) \;.
\end{align}
Again, we see that the masses are in principle independent from each other and the integration constants can be fixed by fixing the physically relevant black hole and white hole masses instead.
In a similar way we could now also compute the Dirac observables corresponding to the horizon. This leads to similar expressions as those of Eqs. \eqref{eq:RBHDOAOS}-\eqref{eq:RWHDOAOS}.

Let us compare with the original paper \cite{ModestoSemiclassicalLoopQuantum}. There, the polymerised equations are solved and the integration constants are fixed (see Eq. (20) in \cite{ModestoSemiclassicalLoopQuantum}\footnote{Eq. (17) in arXiv:0811.2196 [gr-qc].}) by

$$
C = \mp \frac{\gamma \delta p_b^0}{8 m} \mathcal{P}(\delta)^{-\frac{2}{\sigma(\delta) \delta}}\quad, \quad D = \pm 4 m^2 \mathcal{P}(\delta)^{\frac{2}{\sigma(\delta) \delta}}\;, 
$$
\noindent
where $m$ is claimed to be the Schwarzschild mass, $p_b^0$ is not fixed, $\mathcal{P}(\delta) = \frac{b_o-1}{b_o+1}$ and we performed a coordinate transformation $T' = T -\ln(2m)-\ln(\mathcal{P}(\delta))/\sigma(\delta)b_o$ as $A$ is not chosen to be zero in \cite{ModestoSemiclassicalLoopQuantum}. We further notice that $B_o = \mathcal{P}(\delta)^{-\frac{1}{\sigma(\delta) b_o}}$.
In \cite{ModestoSemiclassicalLoopQuantum} the condition $\sigma(\delta) b_o = 1$ is chosen to ensure asymptotic flatness, which also simplifies the computations.
We can now compute the black and white hole horizons as

\begin{equation}
R_{BH} = 2 m \sqrt{1+\left(\frac{\gamma \delta p_b^0}{8m}\right)^2} \quad , \quad R_{WH} = 2 m \sqrt{\mathcal{P}(\delta)^4 + \left(\frac{\gamma \delta p_b^0}{8m \mathcal{P}(\delta)^2}\right)^2} \;,
\end{equation}
\noindent
and similarly the masses 

\begin{equation}
M_{BH} = 4 m \frac{b_o^2 \mathcal{P}(\delta)}{\gamma^2 \delta^2}\quad , \quad M_{WH} = M_{BH} \frac{\gamma \delta p_b^0}{8 m \mathcal{P}(\delta)^2} = \frac{b_o^2 p_b^0}{2 \gamma \delta \mathcal{P}(\delta)}\;,
\end{equation}
\noindent
from which we see that $m$ is not the Schwarzschild mass of the black hole and only proportional to it. Furthermore, the not yet fixed integration constant $p_b^0$ controls the white hole mass $M_{WH}$.
In \cite{ModestoSemiclassicalLoopQuantum}, a minimal area argument motivated by full LQG is used to fix

$$
p_b^0 = \frac{\Delta}{4 \pi \delta \gamma m} \;,
$$
\noindent
where $\Delta$ is the area gap in LQG. This relates the two masses as

\begin{equation}\label{eq:MWHofMBHMod}
M_{WH} = \frac{\Delta b_o^4}{2 \pi \gamma^4 \delta^4} \frac{1}{M_{BH}} \;.
\end{equation}
\noindent
This strategy is similar to what was done in \cite{BodendorferEffectiveQuantumExtended} as the quantum argument selects the initial data.
We could further ask if there is a relation between the two masses, which satisfies the transition surface plaquette argument of \cite{AshtekarQuantumExtensionof,AshtekarQuantumTransfigurarationof} or the maximal curvature argument of \cite{BodendorferEffectiveQuantumExtended}. 

\begin{figure}[t!]
	\centering
	\includegraphics[width=7.75cm,height=5.5cm]{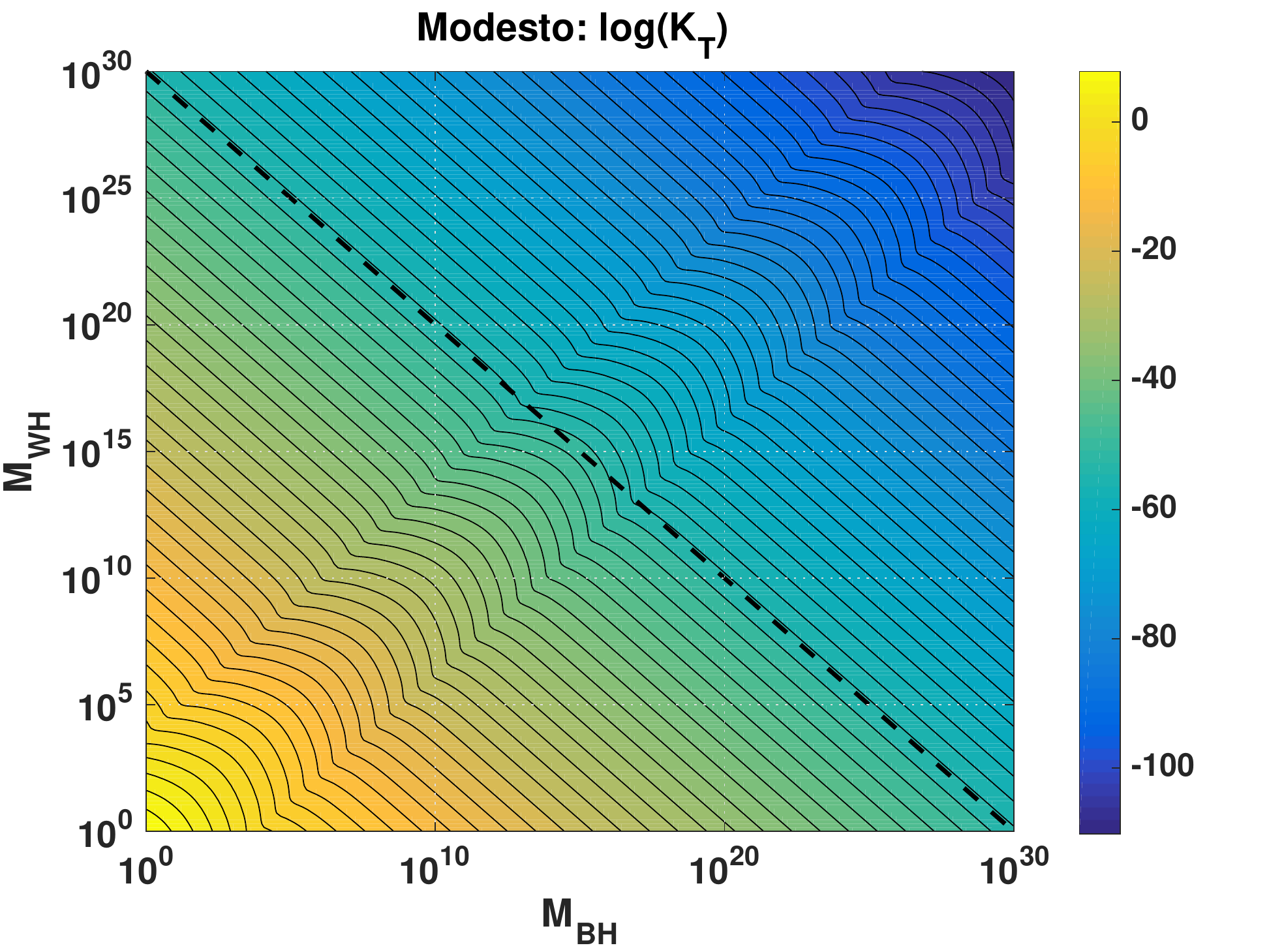}
	\caption{The colour scale encodes the value of the logarithm of the Kretschmann scalar at the transition surface as a function of the black hole $M_{BH}$ and white hole mass $M_{WH}$ for $\delta=1$ and $\gamma = 0.2375$ (cfr. \cite{DomagalaBlack-holeentropy,MeissnerBlack-holeentropy}). The black dashed line corresponds to $M_{WH} \propto 1/M_{BH}$. Both axis are logarithmically.}
	\label{fig:ColorkretschmannMod}
\end{figure}

Studying the Kretschmann scalar at the transition surface in terms of the black hole and white hole mass leads to Fig. \ref{fig:ColorkretschmannMod}. If we want to impose the condition of a unique upper curvature scale we have to fix a relation between $M_{BH}$ and $M_{WH}$. As the plot shows, this is exactly true for the relation \eqref{eq:MWHofMBHMod}.

\subsection{Other approaches}

At this point, it shall be noted that there are numerous other approaches, which have not been discussed here. For instance, we did not discuss any $\bar{\mu}$-schemes  \cite{BoehmerLoopquantumdynamics,ChiouPhenomenologicalloopquantum} (see also \cite{ChiouPhenomenologicaldynamicsof,JoeKantowski-Sachsspacetimein} for the cosmological Kantowski-Sachs setting). As the results in these approaches are mainly of numerical nature, it is much harder to analyse the Dirac observables of the system. Interesting to point out is the work \cite{BoehmerLoopquantumdynamics}, where the authors mention a dependence of the white hole horizon from the initial value $p_b^0$. A detailed analysis of this dependence was not performed, but it was already noticed that in principle $R_{WH}$ can be tuned by changing $p_b^0$ even if $R_{BH}$ is fixed.

Other approaches as \cite{BenAchourPolymerSchwarzschildBlack} include considerations about the anomaly-freedom of the hypersurface deformation algebra, i.e. polymerisation functions which are not necessarily the $\sin$-function. In the work \cite{BenAchourPolymerSchwarzschildBlack}, Dirac observables are not discussed, although their fixing of integration constants is consistent with the above discussions and furthermore one of them is redundant similar to the classical setting.

A discussion of integration constants and Dirac observables in further approaches to non-singular black holes as e.g. limiting curvature mimetic gravity \cite{ChamseddineNonsingularBlackHole} or other approaches not mentioned here is shifted elsewhere. It is important to stress that fixing the integration constants is a subtle point. In many approaches, they are fixed in a chart w.r.t. certain coordinates by means of asymptotic behaviour. This works in the classical setting where only one Dirac observable contains physical information and the other one is redundant. In the effective quantum theory, this is a priori not guaranteed and the situation changes, as discussed above. Fixing the integration constants in a given chart and demanding classicality might lead to the missing of the second, now not redundant, Dirac observable. Hence, we want to stress the necessity of a detailed analysis of the integration constants, which is an important issue in many polymer models and also other approaches as e.g. \cite{ChamseddineNonsingularBlackHole} and continuously upcoming models as e.g. \cite{AssanioussiPerspectivesonthe}.

In the next section we present a polymer model which satisfies the condition on a unique upper bound of the Kretschmann curvature without superselecting certain integration constants and a further class of models where only one physical Dirac observable exits.

\section{New variables for polymer black holes: Curvature variables}\label{sec:newvariables}

Let us now come back to the model previously proposed by the authors in \cite{BodendorferEffectiveQuantumExtended}. As discussed in Sec. \ref{sec:BMM}, the selection of specific relations between black hole and white hole masses was necessary to ensure a unique mass independent curvature upper bound in the effective quantum theory. The heart of the problem is rooted in the fact that $P_1$ is not exactly proportional to (the square root of) the Kretschmann scalar unless the integration constant entering the proportionality factor is selected to be independent of the mass. Thus, the canonical momentum $P_1$ comes to be proportional to the Kretschmann scalar only after restricting to a certain subset of initial conditions.

A possible way out might be to introduce new canonical variables in which one of the momenta is exactly the square root of the Kretschmann scalar. To this aim, let us look at the expression of the Kretschmann scalar in $(v_1,P_1,v_2,P_2)$-variables\footnote{This can be obtained as follows. Starting from metric variables $(a,p_a,\mathscr b,p_{\mathscr b})$, the Kretschmann scalar $\mathcal K=R_{\mu\nu\alpha\beta}R^{\mu\nu\alpha\beta}$ can be explicitly computed as a function of the metric coefficients $a,\mathscr b$ and their first and second $r$-derivatives, namely $\mathcal K=\mathcal K(a,a',a'',\mathscr b,\mathscr b',\mathscr b'')$. Using then the expressions of $a', \mathscr b'$ and $a'',\mathscr b''$ as functions respectively of $P_1,P_2$ and $P_1',P_2'$ given by the equations of motion together with the definitions of $v_1$ and $v_2$ in terms of $a,b$ (i.e., $v_1=\frac{2}{3} \mathscr b^3$ and $v_2=2a\mathscr b^2$) and the Hamiltonian constraint, the Kretschmann scalar can be expressed in terms of the variables $(v_1,P_1,v_2,P_2)$ as in \eqref{KretschmannvP}.}:

\be\label{KretschmannvP}
\mathcal K(v_1,P_1,v_2,P_2)=12\left(\frac{3}{2}v_1\right)^{\frac{2}{3}}P_1^2P_2^2\;.
\ee
\noindent
Let us then introduce the following new variables

\be\label{vPtokj}
v_k=\left(\frac{3}{2}v_1\right)^{\frac{2}{3}}\frac{1}{P_2}  \quad,\quad  v_j=v_2-\frac{3v_1P_1}{2P_2} \quad,\quad k=\left(\frac{3}{2}v_1\right)^{\frac{1}{3}}P_1P_2 \quad,\quad j=P_2 \;.
\ee

\noindent
As can be easily checked by direct computation, the map $(v_1,v_2,P_1,P_2)\mapsto(v_k,v_j,k,j)$ defined by \eqref{vPtokj} is a canonical transformation, i.e., the variables \eqref{vPtokj} satisfy the following canonical Poisson brackets

\be
\begin{aligned}
	&\{v_k,k\}=1\quad,\quad\{v_j,j\}=1\\
	&\{k,j\}=\{v_k,v_j\}=\{k,v_j\}=\{j,v_k\}=0\;.
\end{aligned}
\ee

\noindent
Note that the canonical momentum $k$ conjugate to $v_k$ is in fact the square root of the Kretschmann scalar \eqref{KretschmannvP} up to a numerical factor\footnote{This was actually our starting point for the introduction of the new variables. Requiring that one of the momenta ($k$) is directly proportional to the Kretschmann scalar and keeping the other momentum unchanged ($j=P_2$), the corresponding canonical configuration variables can be determined via the generating function approach. In principle, we could have considered a transformation affecting also the canonical momentum $P_2$. Let us remark that $P_2$ is well suited for the model as its polymerisation is sensitive to small volume corrections ($\sim 1/b$). No better choice for the second momentum with a clear on-shell interpretation is known so far, so we focused on the simplest choice $j=P_2$.  Moreover, this choice keeps the simple form of the Hamiltonian unchanged and hence the corresponding quantum theory can still be analytically solved along the same steps of \cite{BodendorferEffectiveQuantumExtended}.
}. From a off-shell point of view, let us notice that

\begin{equation}\label{eq:kMMS}
k \stackrel{\mathcal{H}\approx 0}{\approx} R_{\mu \nu \alpha \beta} \epsilon^{\mu \nu} \epsilon^{\alpha \beta} = \mathscr b \left(1- \frac{\mathscr b'^2}{N}\right) = \frac{2 M_{\text{Misner-Sharp}}(\mathscr b)}{\mathscr b^3}\;,
\end{equation}
\noindent
where $\epsilon^{\mu \nu} = g^{\mu \alpha} g^{\nu \beta} \epsilon_{\alpha \beta}$ with $\epsilon_{\alpha \beta} \dd x^\alpha \wedge \dd x^{\beta} = \mathscr b^2 \sin\theta \dd \theta \wedge \dd \phi$ is the volume two-form of the $r,t=const.$ two-sphere and $M_{\text{Misner-Sharp}}$ is the Misner-Sharp mass (see e.g. p. 40 in \cite{SzabadosQuasi-LocalEnergy} and references therein). $M_{\text{Misner-Sharp}}(\mathscr b)$ measures the gravitational mass enclosed in the constant $t$-sphere of areal radius $\mathscr b$. This provides us with a off-shell interpretation for the variable $k$ which is then related to the Riemann curvature tensor via Eq. \eqref{eq:kMMS}. Consistently, the above interpretation of $k$ as proportional to the square root of the Kretschmann scalar is recovered on-shell from Eq. \eqref{eq:kMMS}.
As the momentum $j=P_2$ is not modified by the canonical transformation \eqref{vPtokj}, its on-shell interpretation in terms of the the angular components of the extrinsic curvature still holds \cite{BodendorferEffectiveQuantumExtended}. Thus, we have now a new set of canonical variables whose canonical momenta are directly related to the Kretschmann scalar and the extrinsic curvature, respectively. As we will discuss in the following, a polymerisation scheme based on these variables turns out to be well suited for achieving a unique curvature upper bound at which quantum effects become dominant without any further restriction on the initial conditions for the effective dynamics of the model.  

\subsection{Classical theory}

Let us then rewrite the Hamiltonian constraint \eqref{eq:HamclassvP} in the new variables. Inverting the relations \eqref{vPtokj} to express $(v_1,P_1,v_2,P_2)$ in terms of $(v_k,k,v_j,j)$, we have

\be\label{newvarham}
H_{\text{cl}}=\sqrt{n}\,\mathcal H_{\text{cl}}\qquad,\qquad\mathcal H_{\text{cl}}=3v_kkj+v_jj^2-2\approx0\;.
\ee

\noindent
Note that the remarkably simple structure (functional dependence) of the Hamiltonian constraint remains exactly the same in the new canonical variables (compare Eqs.  \eqref{newvarham} and \eqref{eq:HamclassvP}). The corresponding equations of motion are given by

\be\label{eomjk}
\begin{cases}
	v_k'=3\sqrt{n}\,v_k\,j\\
	v_j'=3\sqrt{n}\,v_k\,k+2\sqrt{n}\,v_j\,j\\
	j'=-\sqrt{n}\,j^2\\
	k'=-3\sqrt{n}\,k\,j
\end{cases}
\;.
\ee

\noindent
According to the transformation properties of the variables $(v_1,P_1,v_2,P_2)$ under fiducial cell rescaling $L_o\mapsto\alpha L_o$ (cfr. \cite{BodendorferEffectiveQuantumExtended}) 

\begin{equation}
v_1\longmapsto v_1\quad,\quad P_1\longmapsto \alpha P_1\quad,\quad v_2\longmapsto\alpha^2v_2\quad,\quad P_2\longmapsto\alpha^{-1}P_2\;,
\end{equation}
\noindent
the variables \eqref{vPtokj} transform as

\be\label{fidcellrescaling}
v_k\longmapsto\alpha v_k\quad,\quad k\longmapsto k\quad,\quad v_j\longmapsto\alpha^2v_j\quad,\quad j\longmapsto\alpha^{-1}j\;,
\ee

\noindent
i.e., as expected, the product of the configuration variables and their canonically conjugate momenta (and hence their Poisson bracket) is a density weight 1 object in $t$-direction, and the equations of motion are invariant under rescaling of the fiducial cell. Physical quantities can thus only depend on the combinations $v_k/L_o, k, v_j/L_o^2, L_o j$ in $t$-chart or the coordinate independent quantities $v_k/\mathscr L_o, k, v_j/\mathscr L_o^2, \mathscr L_o j$. Note that $k$ does not depend on any fiducial structure compatible with its interpretation as a spacetime curvature scalar.

As in the new variables the Hamiltonian and hence the corresponding equations of motion have the same form as in the previous variables, the solution strategy is the same as in \cite{BodendorferEffectiveQuantumExtended} (see also Sec. \ref{sec:BMM}) thus yielding the solutions

\begin{align}
j(r)&=\frac{1}{\sqrt{n}\,r}\;,\\
k(r)&=\frac{C}{r^3}\;,\\
v_k(r)&=D\,r^3\;,\\
v_j(r)&=n\,r^2\left(2-\frac{3CD}{\sqrt{n}\,r}\right)\;,
\end{align}

\noindent
where $\sqrt{n}=const.=\mathscr L_o$, and only two of the four integration constants are left as the one encoding a shift in the $r$-coordinate has been set to zero and we get rid off the other one by using the Hamiltonian constraint.
The two remaining integration constants $C$ and $D$ can be fixed in a gauge invariant way by means of Dirac observables. As already discussed in Sec. \ref{sec:Classical}, in the classical case, there is only one fiducial cell independent Dirac observable which on-shell can be identified with the horizon radius and hence it is uniquely specified by the black hole mass. In the new variables it reads (cfr. Eq. \eqref{eq:DOclass})

\be\label{classicaldo1}
2 \mathcal M_{BH} = \mathcal R_{BH}= k\,(v_k\,j)^{\frac{3}{2}}\;,
\ee

\noindent
whose on-shell expression yields

\be\label{massDObs}
2M_{BH}=R_{BH}=C\left(\frac{D}{\sqrt{n}}\right)^{\frac{3}{2}}\;.
\ee

\noindent
Therefore, specifying the mass of the black hole provides us with one condition for a combination of both the integration constants $C$ and $D$. The metric coefficients can then be written as

\begin{align}
\mathscr b(r)&=\sqrt{v_k\,j}=\sqrt{\frac{D}{\sqrt{n}}}\,r\;,\label{clmetriccoeffjkvariables1}\\
a(r)&=\frac{j\,v_j+k\,v_k}{2v_k\,j^2}=\frac{n\sqrt{n}}{D}\left(1-\frac{CD}{\sqrt{n}\,r}\right)\label{clmetriccoeffjkvariables2}\;,
\end{align}

\noindent
which can be recast into a coordinate independent form by expressing $a$ in terms of $\mathscr b$ as

\be
a(\mathscr b)=\frac{\mathscr L_o^3}{D}\left(1-\left(\frac{D}{\mathscr L_o}\right)^{\frac{3}{2}}\frac{C}{ \mathscr b}\right)=\frac{\mathscr L_o^3}{D}\left(1-\frac{2M_{BH}}{\mathscr b}\right)\;,
\ee 

\noindent
where we used $\sqrt{n}=\mathscr L_o$. For $N(\mathscr b)=\left(1-\frac{2M_{BH}}{\mathscr b}\right)^{-1}$ , i.e. $B^2=Na-n=0$, the line element then reads 

\be
\dd s^2=-\frac{\mathscr L_o^3}{D\,L_o^2}\left(1-\frac{2M_{BH}}{\mathscr b}\right)\dd t^2+\frac{D}{\mathscr L_o\left(1-\frac{2M_{BH}}{\mathscr b}\right)}\dd r^2+\frac{D}{\mathscr L_o}\,r^2\,\dd\Omega_2^2\;,
\ee

\noindent
so that, by means of the coordinate redefinition\footnote{As expected from the discussion of Sec. \ref{sec:Classical}, coherently with having only one fiducial cell independent Dirac observable, in the classical theory we can get rid off one integration constant by absorbing it into a coordinate redefinition.} $t\mapsto\tau=\sqrt{\frac{\mathscr L_o^3}{D\,L_o^2}}\,t$ and $r\mapsto \mathscr b=\sqrt{\frac{D}{\mathscr L_o}}\,r$, the classical Schwarzschild solution

\be
\dd s^2=-\left(1-\frac{2M_{BH}}{\mathscr b}\right)\dd\tau^2+\left(1-\frac{2M_{BH}}{\mathscr b}\right)^{-1}\dd \mathscr b^2+\mathscr b^2\,\dd\Omega_2^2
\ee

\noindent
is recovered. This also provides us with an \textit{on-shell} interpretation for the canonical momenta. Indeed, substituting the above expressions for the metric coefficients into the definitions of $k$ and $j$, we get

\be\label{classicalonshell}
k(\mathscr b)=\left(\frac{D}{\mathscr L_o}\right)^{\frac{3}{2}}\frac{C}{\mathscr b^3}\overset{\eqref{massDObs}}{=}\frac{2M_{BH}}{\mathscr b^3}\quad,\quad \mathscr L_o\,j(\mathscr b)=\left(\frac{D}{\mathscr L_o}\right)^{\frac{1}{2}}\frac{1}{\mathscr b}
\ee

\noindent
from which we see that the on-shell value of $k$ is related to (the square root of) the Kretschmann scalar by $\mathcal K=12\,k^2$, while $\mathscr L_o\,j$ is related to the angular components of the extrinsic curvature $\frac{1}{\mathscr b}=\sqrt{N(\mathscr b)}\,K_\theta^\theta=\sqrt{N(\mathscr b)}\,K_\phi^\phi$. Therefore, as discussed in the next section, the polymerisation of the model would involve two scales controlling the onset of quantum effects which can be distinguished into Planck curvature quantum effects ($k$-sector) and small area quantum effects ($j$-sector), respectively.

\subsection{Effective polymer model}\label{sec:CVeffectivemodel}

As in the previous models, effective quantum effects obtained by classical polymerisation with the $\sin$-function, i.e. 

\begin{align}
&k\longmapsto \frac{\sin(\lambda_k\,k)}{\lambda_k}\;,\label{kpolymerisation}\\ 
&j\longmapsto \frac{\sin(\lambda_j\,j)}{\lambda_j}\;,\label{jpolymerisation}
\end{align}
\noindent
where we keep $\lambda_j$ and $\lambda_k$ constant. As we will discuss later in Sec. \ref{sec:comparison}, this polymerisation choice does not correspond to a $\mu_o$-scheme in connection variables as the polymerisation scales turn out to be phase space dependent in those variables. The classical behaviour is recovered in the $\lambda_k k\ll1$, $\lambda_j j\ll1$ regime for which we have $\sin(\lambda_kk)/\lambda_k\simeq k$ and $\sin(\lambda_jj)/\lambda_j\simeq j$. On the other hand, as $k$ is related to the square root of the Kretschmann scalar, the polymerisation \eqref{kpolymerisation} leads to corrections in the Planck curvature regime. Moreover, as we will discuss in the next section, the fact that $k$ is directly proportional to the Kretschmann scalar with no pre-factors involving the integration constants (cfr. Eq. \eqref{classicalonshell}) allows us to achieve a universal mass-independent curvature upper bound with purely constant polymerisation scales for all initial conditions. In turn, according to the on-shell interpretation of $j$ (cfr. second equation in \eqref{classicalonshell}), the polymerisation \eqref{jpolymerisation} will give corrections in the regime in which the angular components of the extrinsic curvature become large. As expected from the $1/\mathscr b$ factor in \eqref{classicalonshell}, this is the case for small radii of the $r,t=const.$ $2$-sphere which allows us to interpret the polymerisation of the $j$-sector as giving small length quantum effects.

The above interpretation is compatible with dimensional considerations. Indeed, according to the behaviour \eqref{fidcellrescaling} of $j$ and $k$ under fiducial cell rescaling $\mathscr L_o\mapsto\alpha\mathscr L_o$, the polymerisation scales $\lambda_k$ and $\lambda_j$ have to transform accordingly as

\be\label{fidcellpolscale}
\lambda_k\longmapsto\lambda_k \qquad,\qquad \lambda_j\longmapsto \alpha\,\lambda_j
\ee

\noindent
so that the scale invariant physical quantities are respectively given by $\lambda_k$ and $\lambda_j/\mathscr L_o$. Recalling then the definitions \eqref{vPtokj}, $k$ and $j$ have dimensions

\be
[k]=[\mathscr bP_1P_2]=L^{-2}\qquad,\qquad [j]=[P_2]=L^{-2}\;,
\ee 

\noindent
where $L$ denotes the dimension of length. Therefore, due to the products $\lambda_kk$ and $\lambda_jj$ being dimensionless, the physical scales have the following dimensions

\be
[\lambda_k]=\left[\frac{1}{k}\right]=L^2\qquad,\qquad \left[\frac{\lambda_j}{\mathscr L_o}\right]=\left[\frac{1}{\mathscr L_o\,j}\right]=L
\ee

\noindent
which are compatible with them controlling Planck curvature and Planck length quantum corrections, respectively.

The polymerised effective Hamiltonian then reads as

\be\label{eq:hampolyjk}
H_{\text{eff}}=\sqrt{n}\,\mathcal H_{\text{eff}}\quad,\quad\mathcal H_{\text{eff}}=3v_k\frac{\sin(\lambda_k\,k)}{\lambda_k}\frac{\sin(\lambda_j\,j)}{\lambda_j}+v_j\frac{\sin^2(\lambda_j\,j)}{\lambda_j^2}-2\approx0\;,
\ee

\noindent
and the corresponding equations of motion are given by

\be
\begin{cases}
	v_k'=3\sqrt{n}\,v_k\,\cos(\lambda_kk)\,\frac{\sin(\lambda_jj)}{\lambda_j}\\
	v_j'=3\sqrt{n}\,v_k\,\frac{\sin(\lambda_kk)}{\lambda_k}\,\cos(\lambda_jj)+2v_j\,\sqrt{n}\,\frac{\sin(\lambda_jj)}{\lambda_j}\,\cos(\lambda_jj)\\
	k'=-3\sqrt{n}\,\frac{\sin(\lambda_kk)}{\lambda_k}\,\frac{\sin(\lambda_jj)}{\lambda_j}\\
	j'=-\sqrt{n}\,\frac{\sin^2(\lambda_jj)}{\lambda_j^2}\\
\end{cases}\;.
\ee

\noindent
Note that the equations of motion for the effective dynamics in the new variables have the same form of the ones in $(v_1,P_1,v_2,P_2)$-variables with the replacements $v_1\leftrightarrow v_k$, $v_2\leftrightarrow v_j$, $P_1\leftrightarrow k$, and $P_2\leftrightarrow j$ (cfr. Eqs. (3.5)-(3.8) in \cite{BodendorferEffectiveQuantumExtended}). Therefore, the solutions will have the same form given by (cfr. Eqs. (3.26)-(3.29) in \cite{BodendorferEffectiveQuantumExtended})

\begin{align}
v_k(r)&=\frac{2DC^2\lambda_k^2\sqrt{n}^3}{\lambda_j^3}\frac{\frac{\lambda_j^6}{16C^2\lambda_k^2n^3}\left(\frac{\sqrt{n}\,r}{\lambda_j}+\sqrt{1+\frac{nr^2}{\lambda_j^2}}\right)^6+1}{\left(\frac{\sqrt{n}\,r}{\lambda_j}+\sqrt{1+\frac{nr^2}{\lambda_j^2}}\right)^3}\;,\label{effdynsol1}\\
v_j(r)&=2n\left(\frac{\lambda_j}{\sqrt{n}}\right)^2\left(1+\frac{nr^2}{\lambda_j^2}\right)\left(1-\frac{3CD}{2\lambda_j}\frac{1}{\sqrt{1+\frac{nr^2}{\lambda_j^2}}}\right)\;,\label{effdynsol2}\\
k(r)&=\frac{2}{\lambda_k}\cot^{-1}\left(\frac{\lambda_j^3}{4C\lambda_k\sqrt{n}^3}\left(\frac{\sqrt{n}\,r}{\lambda_j}+\sqrt{1+\frac{nr^2}{\lambda_j^2}}\right)^3\right)\;,\label{effdynsol3}\\
j(r)&=\frac{1}{\lambda_j}\cot^{-1}\left(\frac{\sqrt{n}r}{\lambda_j}\right)+\frac{\pi}{\lambda_j}\theta\left(-\frac{\sqrt{n}r}{\lambda_j}\right)\;,\label{effdynsol4}
\end{align}

\noindent
where $C$, $D$ are the integration constants which, according to the scaling behaviours \eqref{fidcellpolscale}, transform as $C\mapsto C$ and $D\mapsto\alpha D$ under a fiducial cell rescaling, and we use the same gauge $\sqrt{n}=const.=\mathscr L_o$ as in the classical case.

Given the solutions of the effective dynamics, we can now reconstruct the metric components $a$ and $\mathscr b$ as phase space functions by means of analogous relations to Eqs. \eqref{clmetriccoeffjkvariables1}, \eqref{clmetriccoeffjkvariables2} with polymerised momenta\footnote{For this, we use the same polymerisation as we used in the Hamiltonian \eqref{eq:hampolyjk} as this is the most natural and a consistent choice. Nevertheless, there might be room for arguments to choose different polymerisations at this point. One consequence of this choice is the fact that $\left\{a,b\right\} = \mathcal{O}(\lambda_j^2) + \mathcal{O}(\lambda_k^2)$, in contrast to $\left\{a,b\right\} = 0$, classically. Although not reported here, there exist other possible polymerisation choices, which preserve the classical Poisson-commutativity.}. Specifically, we get 

\begin{align}
\mathscr b^2(r)&=v_k(r)\frac{\sin(\lambda_jj(r))}{\lambda_j}=\frac{2DC^2\lambda_k^2\sqrt{n}^3}{\lambda_j^4}\frac{1}{\sqrt{1+\frac{nr^2}{\lambda_j^2}}}\frac{\frac{\lambda_j^6}{16C^2\lambda_k^2n^3}\left(\frac{\sqrt{n}\,r}{\lambda_j}+\sqrt{1+\frac{nr^2}{\lambda_j^2}}\right)^6+1}{\left(\frac{\sqrt{n}\,r}{\lambda_j}+\sqrt{1+\frac{nr^2}{\lambda_j^2}}\right)^3}\;,\label{qmetriccoeff1}\\
a(r)&=\frac{1}{2v_k(r)}\frac{\lambda_j^2}{\sin^2(\lambda_jj(r))}\left(v_j(r)\frac{\sin(\lambda_jj(r))}{\lambda_j}+v_k(r)\frac{\sin(\lambda_kk(r))}{\lambda_k}\right)\nonumber\\
&=\frac{\lambda_j^6}{2DC^2\lambda_k^2\sqrt{n}^3}\left(1+\frac{nr^2}{\lambda_j^2}\right)^{\frac{3}{2}}\left(1-\frac{CD}{\lambda_j\sqrt{1+\frac{nr^2}{\lambda_j^2}}}\right)\frac{\left(\frac{\sqrt{n}\,r}{\lambda_j}+\sqrt{1+\frac{nr^2}{\lambda_j^2}}\right)^3}{\frac{\lambda_j^6}{16C^2\lambda_k^2n^3}\left(\frac{\sqrt{n}\,r}{\lambda_j}+\sqrt{1+\frac{nr^2}{\lambda_j^2}}\right)^6+1}\,,\label{qmetriccoeff2}
\end{align}

\noindent
and the line element reads

\be\label{effmetricjk}
\dd s^2=-\frac{a(r)}{L_o^2}\dd t^2+\frac{\mathscr L_o^2}{a(r)}\dd r^2+\mathscr b^2(r)\left(\dd\theta^2+\sin^2\theta\dd\phi^2\right)\;,
\ee

\noindent
where, as stated before, $\sqrt{n}=\mathscr L_o$ and we used the expression of the metric coefficient $\bar a=a/L_o^2$. Note that all solutions \eqref{effdynsol1}-\eqref{effdynsol4} as well as the metric coefficients \eqref{qmetriccoeff1} and \eqref{qmetriccoeff2} are smoothly well-defined in the whole $r$ domain $r\in(-\infty,+\infty)$, which describes both the interior and exterior regions.

As already discussed throughout the paper, the remaining integration constants ($C$ and $D$) in the solutions of the effective dynamics can be fixed in a gauge independent way by means of Dirac observables. The latter can be determined as follows. First, we consider the effective quantum corrected metric in the two asymptotic regions $r\to\pm\infty$, express the metric coefficient $a$ in terms of the areal radius $\mathscr b$, and rescale the coordinates so that Schwarzschild solution is recovered asymptotically. This allows us to read off the corresponding on-shell expression for the fiducial cell independent mass Dirac observables by looking at the metric coefficients in the two asymptotic regions. These on-shell quantities will of course depend only on the two integration constants and on the polymerisation scales. Finally, the off-shell expressions of the Dirac observables can be determined by solving the solutions of the effective dynamics in terms of the integration constants.

In the $r\to+\infty$ limit, the metric coefficients in Eqs. \eqref{qmetriccoeff1} and \eqref{qmetriccoeff2} then yield

\be\label{bplus}
\mathscr b_+^2:=\mathscr b^2(r\to+\infty)=\frac{D}{\sqrt{n}}\,r^2\qquad,\qquad a_+:=a(r\to+\infty)=\frac{n\sqrt{n}}{D}\left(1-\frac{CD}{\sqrt{n}\,r}\right)
\ee

\noindent
from which it follows that

\be
a(\mathscr b(r\to+\infty))=\frac{n\sqrt{n}}{D}\left(1-\left(\frac{D}{\sqrt{n}}\right)^{\frac{3}{2}}\frac{C}{\mathscr b}\right)\;.
\ee

\noindent
Thus, similarly to the classical case, by means of the coordinates rescaling $r\mapsto \mathscr b=\sqrt{\frac{D}{\mathscr L_o}}r$ and $t\mapsto\tau=\sqrt{\frac{\mathscr L_o^3}{DL_o^2}}\,t$ for $\sqrt{n}=\mathscr L_o$ the metric \eqref{effmetricjk} reduces to the classical Schwarzschild solution in the $\mathscr b(r\to+\infty)$ asymptotic region. Hence, the on-shell expression for the black hole mass Dirac observables is given by

\be\label{onshelldo1}
2 M_{BH}=C\left(\frac{D}{\sqrt{n}}\right)^{\frac{3}{2}}\;.
\ee

\noindent
On the other hand, in the limit $r\to-\infty$, we have

\begin{align}\label{bminus}
\mathscr b_-^2&:=\mathscr b^2(r\to-\infty)=\frac{16DC^2\lambda_k^2}{\sqrt{n}}\left(\frac{\sqrt{n}}{\lambda_j}\right)^6|r|^2\;,\\
a_-&:=a(r\to-\infty)=\frac{n\sqrt{n}}{16DC^2\lambda_k^2}\left(\frac{\lambda_j}{\sqrt{n}}\right)^6\left(1-\frac{CD}{\sqrt{n}|r|}\right)\;,
\end{align}

\noindent
from which it follows that

\be
a(\mathscr b(r\to-\infty))=\frac{n\sqrt{n}}{16DC^2\lambda_k^2}\left(\frac{\lambda_j}{\sqrt{n}}\right)^6\left(1-\frac{4nDC^2\lambda_k}{\lambda_j^3}\sqrt{\frac{D}{\sqrt{n}}}\,\frac{1}{\mathscr b}\right)\;.
\ee

\noindent
By means of the coordinate rescaling $r\mapsto \mathscr b=4C\lambda_k\left(\frac{\mathscr L_o}{\lambda_j}\right)^3\sqrt{\frac{D}{\mathscr L_o}}(-r)$ and $t\mapsto\tau=\sqrt{\frac{\mathscr L_o^3}{DL_o^2}}\,\frac{\lambda_j^3}{4C\lambda_k\mathscr L_o^3}\,t$ for $\sqrt{n}=\mathscr L_o$, the metric \eqref{effmetricjk} reduces to the classical Schwarzschild solution in the $\mathscr b(r\to-\infty)$ asymptotic region. The on-shell expression for the white hole mass Dirac observable is thus given by

\be\label{onshelldo2}
2 M_{WH}=\frac{4\lambda_kC^2\sqrt{n}^3}{\lambda_j^3}\left(\frac{D}{\sqrt{n}}\right)^{\frac{3}{2}}=8C\lambda_k\left(\frac{\sqrt{n}}{\lambda_j}\right)^3 M_{BH}\;.
\ee

Therefore, the two asymptotic regions $\mathscr b(r\to\pm\infty)$ are described by Schwarzschild spacetimes with asymptotic masses $M_{BH}$ and $M_{WH}$, respectively. Specifying these two quantities completely determines the two integration constants as can be seen by inverting the relations \eqref{onshelldo1} and \eqref{onshelldo2}, namely

\be\label{eq:CDjk}
C=\frac{\lambda_j^3}{4\lambda_k\sqrt{n}^3}\frac{M_{WH}}{M_{BH}}\qquad,\qquad D=\sqrt{n}\left(\frac{8 \lambda_k\sqrt{n}^3}{\lambda_j^3}\frac{M_{BH}^2}{M_{WH}}\right)^{\frac{2}{3}}\;.
\ee

\noindent
Using then the solutions \eqref{effdynsol1}-\eqref{effdynsol4} of the effective dynamics to determine the expressions of $C$ and $D$ in terms of the phase space variables, and substituting them into Eqs. \eqref{onshelldo1} and \eqref{onshelldo2} yields the following off-shell expressions for the Dirac observables

\begin{align}
2\mathcal{M}_{BH}&=\frac{\sin(\lambda_kk)}{\lambda_k}\cos\left(\frac{\lambda_kk}{2}\right)\left(\frac{2v_k}{\lambda_j\cot\left(\frac{\lambda_jj}{2}\right)}\right)^{\frac{3}{2}}\;,\\
2\mathcal{M}_{WH}&=\frac{\sin(\lambda_kk)}{\lambda_k}\sin\left(\frac{\lambda_kk}{2}\right)\left(\frac{2v_k}{\lambda_j}\cot\left(\frac{\lambda_jj}{2}\right)\right)^{\frac{3}{2}}\;,
\end{align} 

\noindent
which are fiducial cell independent as can be easily checked by means of the transformation behaviours \eqref{fidcellrescaling} and \eqref{fidcellpolscale} under fiducial cell rescalings. Moreover, in the limit $\lambda_j,\lambda_k\to0$, $\mathcal M_{BH}$ reduces to the classical Dirac observable \eqref{classicaldo1} while $\mathcal M_{WH}$ is not well-defined in this limit coherently with it not being present at the classical level where there is only one fiducial cell independent Dirac observable identified on-shell with the black hole mass.

As before, the physical phase space is two-dimensional.
The kinematical phase space has dimension four, and the first class Hamiltonian constraint removes two degrees of freedom.
The solution space, i.e. the physical phase space can then be parametrised by the observables $\mathcal{M}_{BH}$ and $\mathcal{M}_{WH}$ (or equivalently $C$ and $D$), which provide a global set of coordinates.
The same holds true for all previously discussed models.
In this respect, we notice that the above mass observables have non-trivial Poisson brackets, i.e.
\begin{equation}
\left\{\mathcal{M}_{BH}\,,\,\mathcal{M}_{WH}\right\} = \frac{3}{2 \lambda_j} \left(\lambda_k \mathcal{M}_{BH}^2 \mathcal{M}_{WH}^2\right)^\frac{1}{3} \;.
\end{equation}

\noindent
Therefore, $\mathcal{M}_{BH}^{1/3}$ and $\mathcal{M}_{WH}^{1/3}$ are canonically conjugate (up to a constant factor that can be reabsorbed) as 
\begin{equation}
\left\{\mathcal{M}_{BH}^\frac{1}{3}\,,\,\mathcal{M}_{WH}^\frac{1}{3}\right\} = \frac{\lambda_k^\frac{1}{3}}{6 \lambda_j} \;.
\end{equation}

Having the relation of the integration constants $C$, $D$ to the two masses we can rewrite the metric as 

\begin{equation}
\dd s^2 = - \frac{a(x)}{\lambda_j^2} d\tau^2 + \frac{\lambda_j^2}{a(x)} \dd x^2 + \mathscr b(x)^2 \dd \Omega_2^2 \;,
\end{equation}
\noindent
where we rescaled the coordinates $x = \mathscr L_o r /\lambda_j$, $\tau = \lambda_j t /L_o $ and 

\begin{align}
\mathscr b^2(x)&=\frac{1}{2} \left(\frac{\lambda_k}{M_{BH} M_{WH}}\right)^{\frac{2}{3}}\frac{1}{\sqrt{1+x^2}}\frac{M_{BH}^2\left(x+\sqrt{1+x^2}\right)^6+M_{WH}^2}{\left(x+\sqrt{1+x^2}\right)^3}\;,\\
\frac{a(x)}{\lambda_j^2}&=2 \left(\frac{M_{BH} M_{WH}}{\lambda_k}\right)^{\frac{2}{3}}\left(1-\left(\frac{M_{BH} M_{WH}}{\lambda_k}\right)^{\frac{1}{3}}\frac{1}{\sqrt{1+x^2}}\right)\frac{\left(1+x^2\right)^{\frac{3}{2}}\left(x+\sqrt{1+x^2}\right)^3}{M_{BH}^2\left(x+\sqrt{1+x^2}\right)^6+M_{WH}^2}\,.
\end{align}
\noindent
Note that in the final line element $\lambda_j$ does not appear any more and hence its precise value can not have any physical meaning. Consistently, it will not appear in later computation in any physical expressions.

As in \cite{BodendorferEffectiveQuantumExtended}, we can check what happens with initial conditions given in the black hole asymptotic region evolved to the white hole asymptotic region. At a given value of $\mathscr b$ on the black hole side, which is considered large and in the classical regime, initial conditions can be specified by 

\begin{equation}
v_j(\mathscr b)\quad ,\quad v_k(\mathscr b) \quad, \quad k(\mathscr b) \simeq 0 \quad ,\quad  j(\mathscr b) \simeq 0 \;,
\end{equation}

\noindent
and a specific value of $M_{BH}$ and $M_{WH}$. Following the spacetime evolution towards the white hole classical regime up to the same value of $\mathscr b$ gives

\begin{equation}
v_j\mapsto v_j \quad ,\quad v_k \mapsto v_k\quad, \quad k \mapsto \frac{\pi}{\lambda_k}- k \quad ,\quad  j \mapsto \frac{\pi}{\lambda_j} - j \;.
\end{equation}

\noindent
This furthermore transforms the Dirac observables for the masses according to 

\begin{equation}
2 \mathcal{M}_{BH} \longmapsto 2 \mathcal{M}_{WH} \quad , \quad 2 \mathcal{M}_{WH} \longmapsto 2 \mathcal{M}_{BH} \;.
\end{equation}

\noindent
An observer starting at the black hole side who specified a value for $M_{BH}$ and $M_{WH}$ travelling on to the white hole side would observe that his $M_{BH}$ coincides with the value of $M_{WH}$ of an observer living on the white hole side and vice versa.

\subsection{Onset of quantum effects}

In the classical regime, the polymerisation functions (sin functions) can be approximated by their arguments. By looking at the solution \eqref{effdynsol3} for $k(r)$ for positive and large $r$, we see that the approximation $\sin(\lambda_kk)\simeq\lambda_kk$ and $\sin(\lambda_j j)\simeq\lambda_j j$  holds true for

\be
\frac{\mathscr L_o\,r}{\lambda_j}\gg1\qquad,\qquad\frac{2r^3}{C\lambda_k}\gg1
\ee

\noindent
or equivalently, using then Eq. \eqref{bplus} for the areal radius $\mathscr b_+$ in the $r\to+\infty$ limit, the classical regime for positive and large $r$ is given by the coordinate-free conditions

\be
\frac{\mathscr L_o}{\lambda_j}\sqrt{\frac{\mathscr L_o}{D}}\,\mathscr b_+\gg1\qquad,\qquad\frac{2}{C\lambda_k}\left(\frac{\mathscr L_o}{D}\right)^{\frac{3}{2}}\,\mathscr b_+^3\gg1\;.
\ee

\noindent
In particular, recalling the on-shell expression for the black hole mass Dirac observable \eqref{onshelldo1}, we find that the classical regime corresponds to

\be\label{classicalregime1}
\mathscr b_+ \gg \left( 8 \lambda_k \frac{\left(M_{BH}\right)^2}{M_{WH}} \right)^{\frac{1}{3}} \quad,\quad \frac{M_{BH}}{\mathscr b_+^3} \ll \frac{1}{\lambda_k}\;.
\ee

\noindent
Of special interest is the second condition, which rewritten in terms of the classical Kretschmann scalar of the black hole side gives

\begin{equation}\label{eq:curvscaleBH}
\mathcal{K}_{cl}^{BH} = \frac{48 M_{BH}^2}{\mathscr b_+^6} \ll \frac{48}{\lambda_k^2} \;,
\end{equation}

\noindent
thus providing us with a unique mass independent scale of onset of curvature effects without restricting any integration constants (as it was needed in \cite{BodendorferEffectiveQuantumExtended} or \cite{ModestoSemiclassicalLoopQuantum}). 

\noindent
Similarly, for large and negative $r$, the asymptotic classical  Schwarzschild spacetime is reached for

\be
\frac{\mathscr L_o|r|}{\lambda_j}\gg1\qquad,\qquad\frac{32C\lambda_k\mathscr L_o^6|r|^3}{\lambda_j^6}\gg1\;.
\ee

\noindent
Using then the expressions \eqref{bminus} for $\mathscr b_-$ and \eqref{onshelldo2} for the on-shell white hole mass Dirac observable, we get that the classical regime in the negative $r$ branch is given by

\be\label{classicalregime2}
\mathscr b_- \gg \left( 8 \lambda_k \frac{M_{BH}^2}{M_{WH}} \right)^{\frac{1}{3}} \frac{M_{WH}}{M_{BH}} = \left( 8 \lambda_k \frac{M_{WH}^2}{M_{BH}} \right)^{\frac{1}{3}} \quad, \quad \frac{M_{WH}}{\mathscr b_-^3} \ll \frac{1}{\lambda_k} \;.
\ee

\noindent
Again, the second equation re-expressed in terms of the classical Kretschmann scalar of the white hole side gives

\begin{equation}\label{eq:curvscaleWH}
\mathcal{K}_{cl}^{WH} = \frac{48 M_{WH}^2}{\mathscr b_-^6} \ll \frac{48}{\lambda_k^2} \;,
\end{equation}

\noindent
which also on the white hole side defines a unique curvature scale at which quantum effects become relevant.
Therefore, according to the second expressions in Eqs. \eqref{classicalregime1} and \eqref{classicalregime2}, the polymerisation scale $\lambda_k$ is related to the inverse Planck curvature and quantum effects become negligible in the low curvature regime. On the other hand, we interpret the first conditions in Eqs. \eqref{classicalregime1} and \eqref{classicalregime2} as small volume effects.

We can now check whether there is a possibility that the quantum effects reach the horizons. As discussed below, for large masses the horizons are approximately located at $\mathscr b_+ \simeq 2 M_{BH}$ and $\mathscr b_- \simeq 2 M_{WH}$, respectively. For the black hole side we conclude from Eq. \eqref{classicalregime1}

\begin{equation}
2 M_{BH} \gg \left( 8 \lambda_k \frac{M_{BH}^2}{M_{WH}} \right)^{\frac{1}{3}} \quad , \quad 2 M_{BH} \lambda_k \ll 8 \left(M_{BH}\right)^3 \;,
\end{equation}

\noindent
which is always satisfied for large black hole and white hole masses. Similar results can be found for the white hole side. Hence, for astrophysical black holes the horizon is always classical and quantum effects are suppressed.

An important question remaining is: Are there choices of $M_{BH}$ and $M_{WH}$ for which the small volume effects become relevant earlier than the high curvature effects? On the black hole side, we can deduce from Eq. \eqref{classicalregime1} that quantum effects become relevant at the length scales

\begin{equation}
\mathscr b_+ \gg \left( 8 \lambda_k \frac{M_{BH}^2}{M_{WH}} \right)^{\frac{1}{3}} \quad,\quad \mathscr b_+ \gg \left(M_{BH} \lambda_k\right)^{\frac{1}{3}}\;.
\end{equation}
\noindent
We can ask when the left length scale is actually larger than the second one, i.e. 

$$
\left( 8 \lambda_k \frac{M_{BH}^2}{M_{WH}} \right)^{\frac{1}{3}} > \left(M_{BH} \lambda_k\right)^{\frac{1}{3}} \;,
$$

\noindent
which leads to the condition

\begin{equation}\label{eq:QEffbound1}
\frac{M_{WH}}{M_{BH}} < 8 \;.
\end{equation}
\noindent
Similar considerations taking into account Eq. \eqref{classicalregime2} leads to

 \begin{equation}\label{eq:QEffbound2}
 \frac{M_{WH}}{M_{BH}} > \frac{1}{8} \;.
 \end{equation}
 
\noindent
Therefore, in the regime $1/8 <  \frac{M_{WH}}{M_{BH}} < 8$ the finite 2-sphere area effects become relevant earlier than the high curvature effects.

\begin{figure}[t!]
	\centering
	\subfigure[]
	{\includegraphics[width=7.75cm,height=5.5cm]{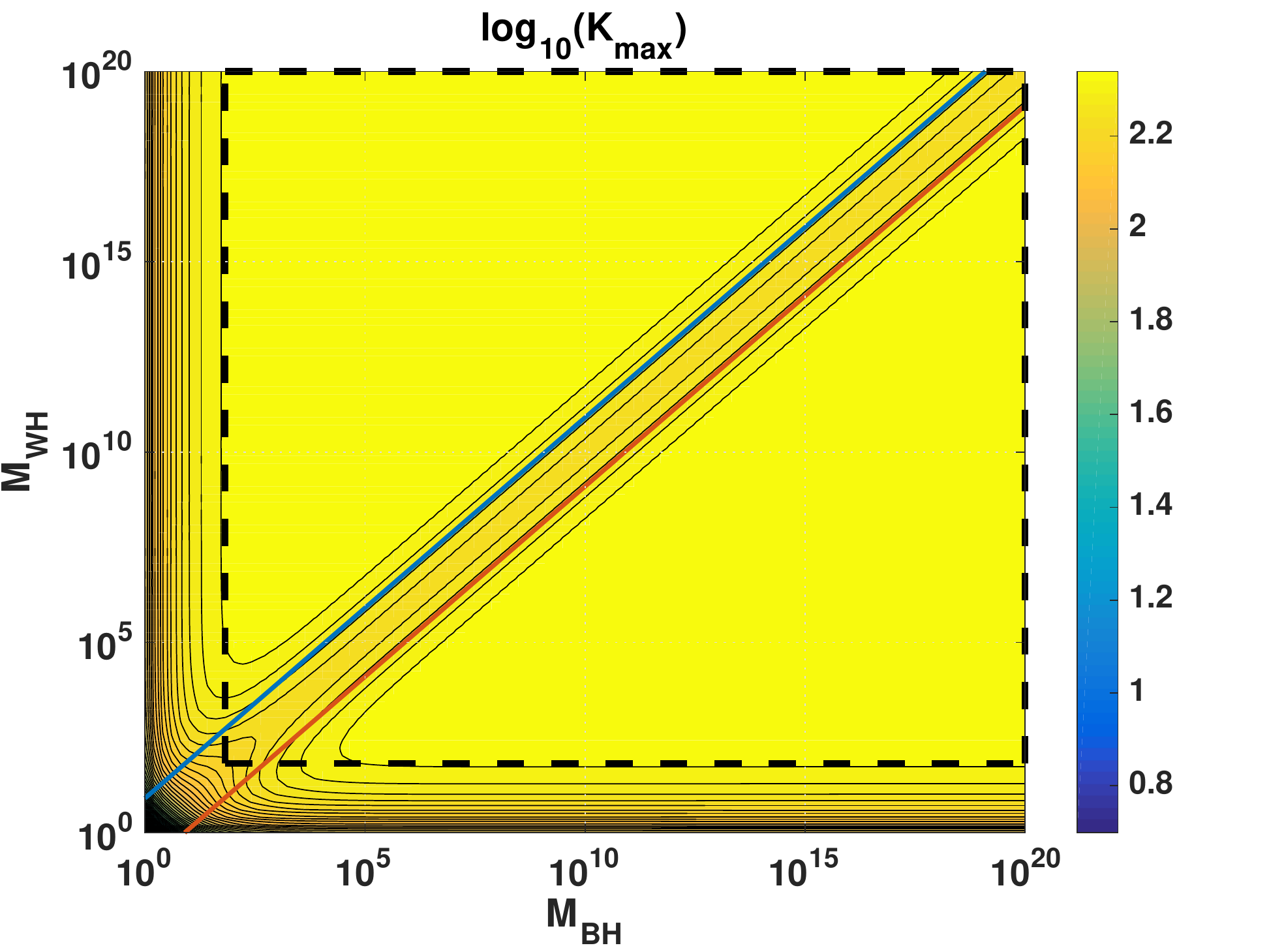}}
	\hspace{2mm}
	\subfigure[]
	{\includegraphics[width=7.75cm,height=5.5cm]{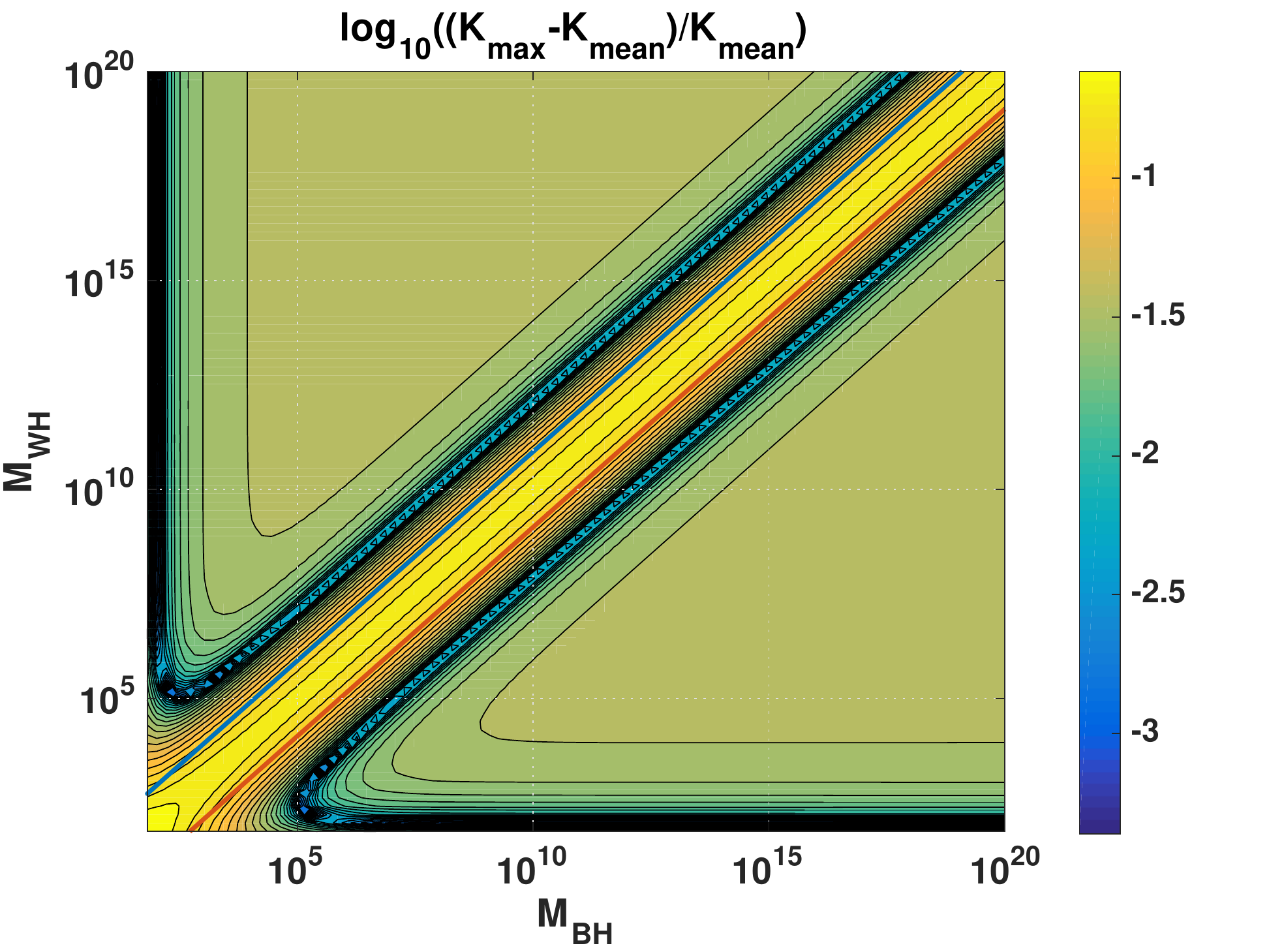}}
	\caption{Logarithm of the maximal value of the Kretschmann scalar (a) and the deviation of the Kretschmann scalar from its mean value (mean over all masses in the black dashed box) (b) as a function of $M_{BH}$ and $M_{WH}$ in logarithmic axis. The maximal value of the Kretschmann scalar remains largely independent of the masses. The two colour lines represent the boundaries of Eqs. \eqref{eq:QEffbound1} and \eqref{eq:QEffbound2}. For the plot the maximal value of the Kretschmann scalar is computed numerically. The parameters are settled to $\lambda_j = \lambda_k = \mathscr L_o = 1$.}
	\label{fig:Kmax}
\end{figure}

\begin{figure}[t!]
	\centering
	\subfigure[]
	{\includegraphics[width=7.75cm,height=5.5cm]{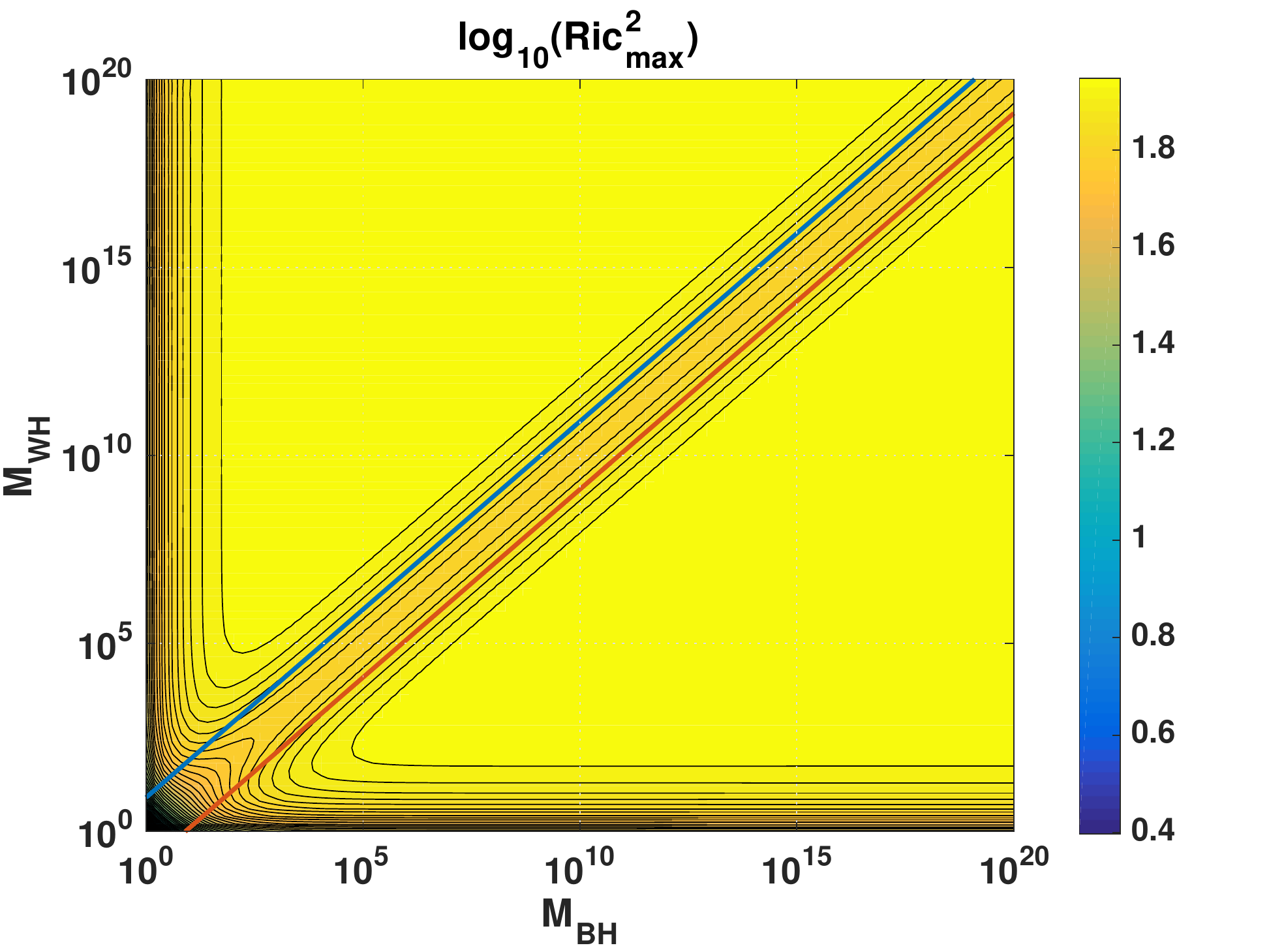}}
	\hspace{2mm}
	\subfigure[]
	{\includegraphics[width=7.75cm,height=5.5cm]{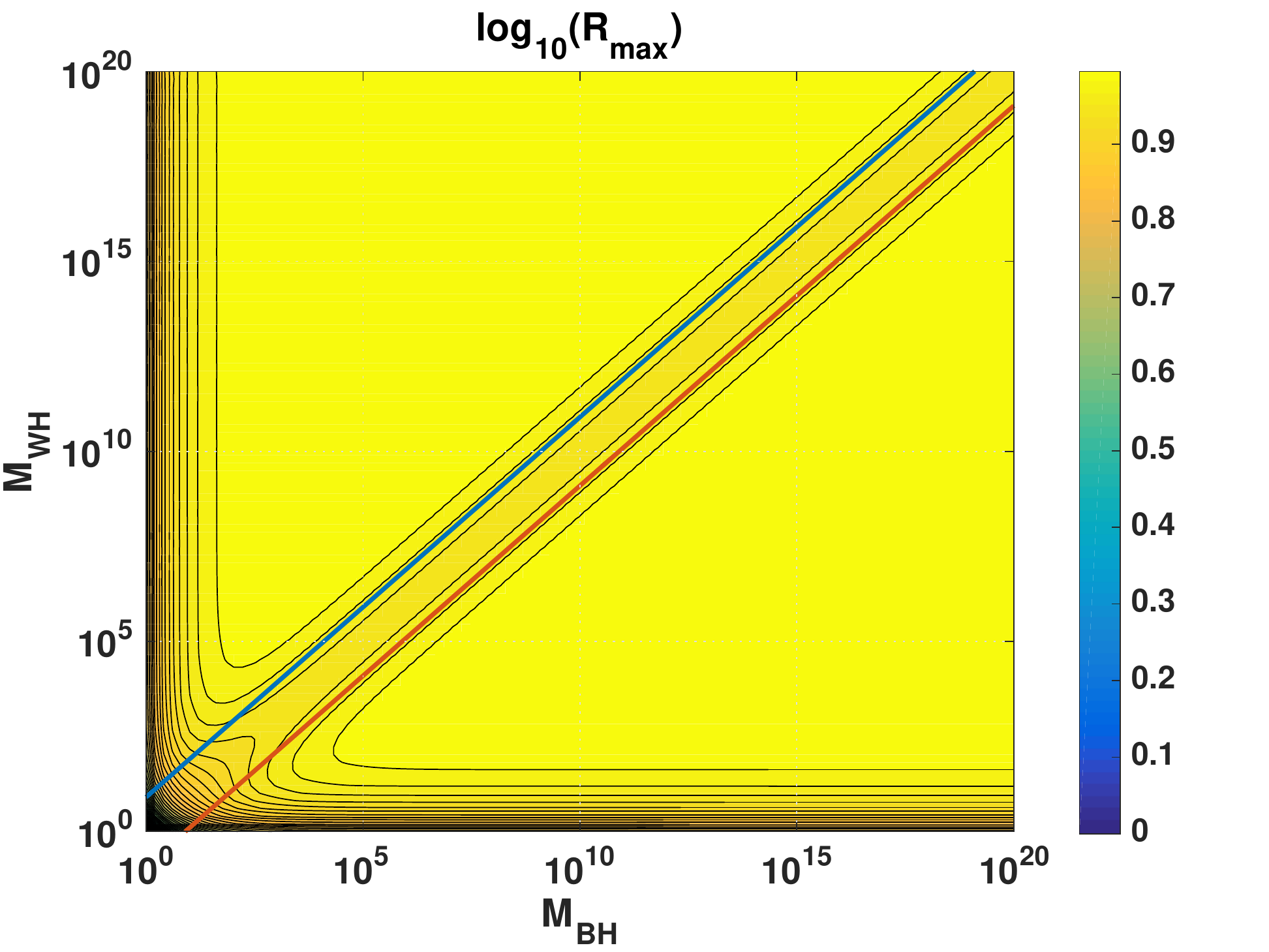}}
	\caption{Other curvature invariants as $Ric^2 = R_{\mu \nu} R^{\mu \nu}$ in (a) and the Ricci-scalar $R$ show the same behaviour as $\mathcal K$ at the transition surface and remain bounded. Within the large mass expansion it is $\mathcal K = \mathcal K_{cl} + \mathcal{O}(\lambda_k^2)$, $R_{\mu \nu} R^{\mu \nu} = \mathcal{O}(\lambda_k^4)$, $R^2 = \mathcal{O}(\lambda_k^4)$, and the Weyl scalar becomes $C_{\mu \nu \alpha \beta} C^{\mu \nu \alpha \beta} = \mathcal K - 2 R_{\mu \nu } R^{\mu \nu} + \frac{1}{3} R^2 = \mathcal K_{cl} + \mathcal{O}(\lambda_k^2)$. Therefore all curvature scalars admit the same behaviour at the transition surface and reduce to the classical expression in the classical regime. In the limit $\lambda_k \rightarrow 0$ the $R^2$ and $R_{\mu \nu} R^{\mu \nu}$ vanish in agreement with the Schwarzschild case. The parameters are set to $\lambda_j = \lambda_k = \mathscr L_o = 1$.}
	\label{fig:curvinv}
\end{figure}

The discussion so far focused on the the classical regime and when it fails to hold. As in \cite{BodendorferEffectiveQuantumExtended}, we can check what happens to the Kretschmann scalar in the deep quantum regime, i.e. at the transition surface. In Fig. \ref{fig:Kmax} the maximal value of the Kretschmann scalar is shown. In accordance with the second equation of Eqs. \eqref{classicalregime1} and \eqref{classicalregime2} for a wide choice of masses, the value of the Kretschmann scalar at the transition surface remains unchanged.
The same argument can be made for other curvature invariants as $R^2$, $R_{\mu \nu} R^{\mu \nu}$ or $C_{\mu \nu \alpha \beta} C^{\mu \nu \alpha \beta}$ (Weyl scalar), which leads to the same conclusion (see Fig.~\ref{fig:curvinv}).
For large masses, we find the expressions

\begin{align*}
&\mathcal K = \mathcal K_{cl} + \mathcal{O}(\lambda_k^2)\;,\\
&R_{\mu \nu} R^{\mu \nu} = \mathcal{O}(\lambda_k^4)\;,\\
&R^2 = \mathcal{O}(\lambda_k^4)\;,\\
&C_{\mu \nu \alpha \beta} C^{\mu \nu \alpha \beta} = \mathcal K - 2 R_{\mu \nu } R^{\mu \nu} + \frac{1}{3} R^2 = \mathcal K_{cl} + \mathcal{O}(\lambda_k^2)\;,
\end{align*}
\noindent
and therefore the result remains analogous.
The discussion also extends to the effective stress energy tensor ($T_{\mu \nu}^{\text{(eff.)}} := G_{\mu \nu}/8\pi$), which is just composed of Ricci scalar and Ricci tensor terms.


\noindent

Both arguments lead to the conclusion that the relation between the two masses can be left unspecified still leading to an unique upper curvature bound. Nevertheless, there are interesting specific choices. 

\begin{figure}[t!]
	\centering
	\subfigure[]
	{\includegraphics[width=7.75cm,height=5.5cm]{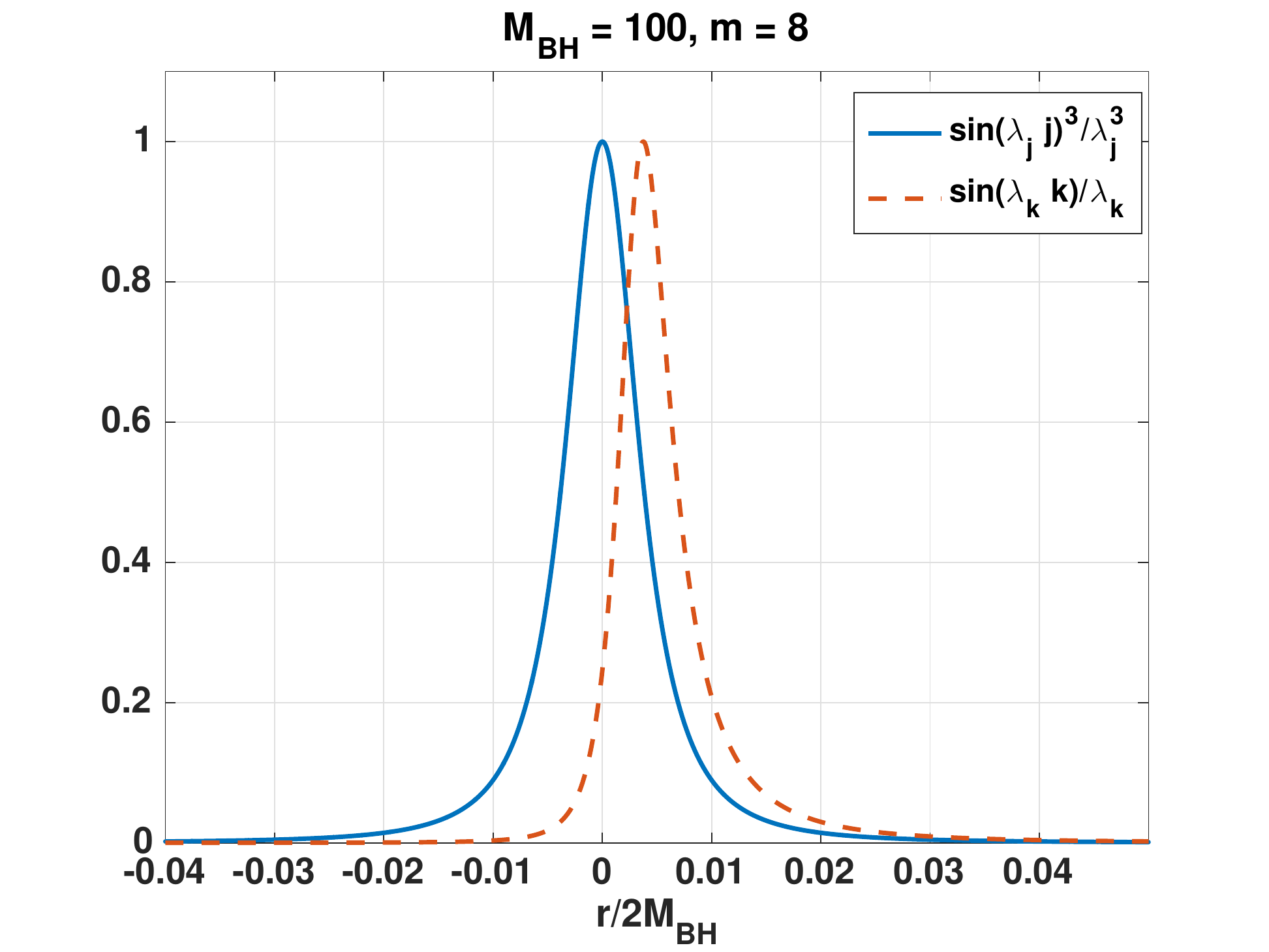}}
	\hspace{2mm}
	\subfigure[]
	{\includegraphics[width=7.75cm,height=5.5cm]{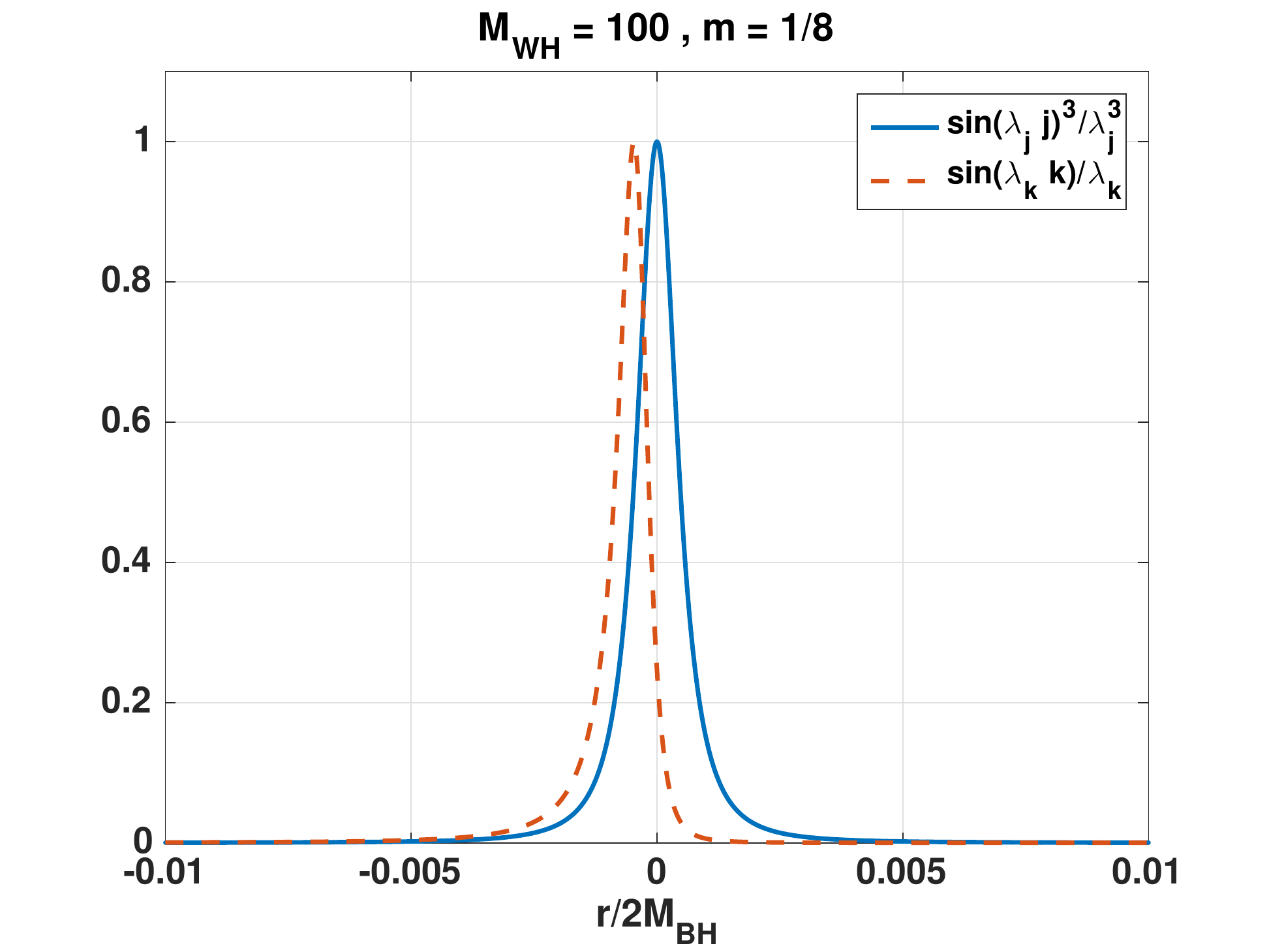}}
	\caption{$\sin(\lambda_j j)^3/\lambda_j^3$ compared to $\sin(\lambda_k k)/\lambda_k$ for $m = 8$ (a) and $m = 1/8$ (b). The parameters are $\lambda_j = \lambda_k = \mathscr L_o = 1$.}
	\label{fig:onset}
\end{figure}

A particular interesting class of relations between the masses is

\begin{equation}\label{eq:symbounce}
M_{WH} = m M_{BH} \;,
\end{equation} 
\noindent
for a dimensionless number $m$. For this relation, we find that the first equation in Eq. \eqref{classicalregime1} becomes a curvature scale as

\begin{equation}\label{eq:curvscaleBH2}
\mathscr b_+ \gg  \left(\frac{8 \lambda_k}{m} M_{BH}\right)^{\frac{1}{3}}  \quad\Longleftrightarrow\quad \frac{48 M_{BH}^2}{\mathscr b_+^6} \ll \frac{3 m^2}{4 \lambda_k^2} \;,
\end{equation}
\noindent
The same hold true for the white hole side and Eq. \eqref{classicalregime2} for which we find

\begin{equation}\label{eq:curvscaleWH2}
\mathscr b_- \gg \left(8 \lambda_k m M_{WH}\right)^{\frac{1}{3}} \quad\Longleftrightarrow\quad \frac{48 M_{WH}^2}{\mathscr b_+^6} \ll \frac{3}{4 \lambda_k^2  m^2} \;,
\end{equation}
\noindent
Checking furthermore the classical limit for 

$$
\mathscr L_o^3 \frac{\sin(\lambda_j j)^3}{\lambda_j^3} \simeq \frac{4 \lambda_k \mathscr L_o^3}{\lambda_j^3} \begin{cases}
\frac{1}{m} \frac{2 M_{BH}}{b_+^3} &\;,\; r \rightarrow + \infty\\
m \frac{2 M_{WH}}{b_-^3} &\;,\; r \rightarrow - \infty
\end{cases}\;,
$$
\noindent
we find that it is actually up to a $m$-dependent numerical factor proportional to $\sin(\lambda_k k)/\lambda_k$, i.e. the square root of the Kretschmann scalar.
In agreement with the above computation for $m = 8$ the new curvature scale at the black hole side \eqref{eq:curvscaleBH2} agrees with the curvature scale of the $k$-sector \eqref{eq:curvscaleBH}. While for this value the curvature scale \eqref{eq:curvscaleWH2} is smaller than \eqref{eq:curvscaleWH}, i.e. coming from the white hole side, quantum effects of the $j$-sector are relevant first. 
The same result can be found for $m = 1/8$ where the quantum effects match on the white hole side. Fig. \ref{fig:onset} shows this graphically.
Of particular interest in then the case $m = 1$, which means the value of the masses is the same. In this case, there are coming from both sides quantum effects of the $j$-sector become first relevant at the Kretschmann curvature scale $3/4 \lambda_k^2$, while effects of the $k$-sector become relevant at higher curvatures ($48/\lambda_k^2$). 
\begin{figure}[t!]
	\centering\includegraphics[scale=0.4]{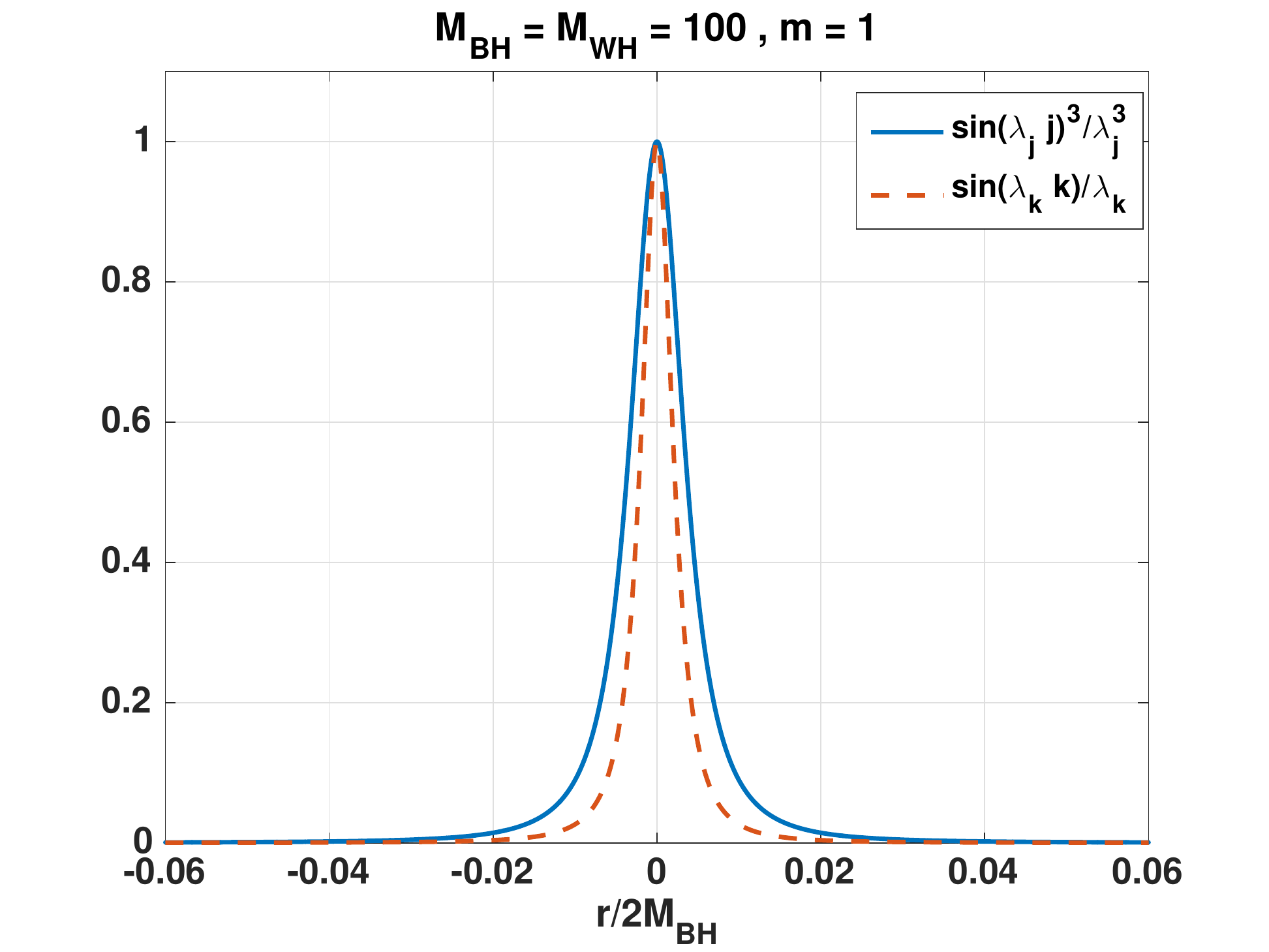}
	\caption{$\sin(\lambda_j j)^3/\lambda_j^3$ compared to $\sin(\lambda_k k)/\lambda_k$ for $m = 1$. The curve of $j$ encloses completely $k$, i.e. the dominant contribution for quantum effects comes from $j$. Coming from both sides the onset of quantum effects is at the Kretschmann curvature scale $3/4 \lambda_k^2$. Parameters are $\lambda_j = \lambda_k = \mathscr L_o = 1$}
	\label{fig:symmetric}
\end{figure}
Fig. \ref{fig:symmetric} shows that from both sides quantum effects become relevant due to the $j$-sector at the same curvature and $k$ always plays a sub dominant role.

Note that we can generically interpret the first Eqs. in \eqref{classicalregime1},\eqref{classicalregime2} as curvature scales, depending on the asymmetry of the two sides. This can be seen by rewriting the first Eqs. in \eqref{classicalregime1},\eqref{classicalregime2} as 

\begin{equation}
\mathcal{K}^{BH}_{cl} = \frac{48 M_{BH}^2}{\mathscr b_+^6} \ll \frac{3}{4 \lambda_k^2} \left(\frac{M_{BH}}{M_{WH}}\right)^2 \quad ,\quad \mathcal{K}^{WH}_{cl} = \frac{48 M_{WH}^2}{\mathscr b_-^6} \ll \frac{3}{4 \lambda_k^2} \left(\frac{M_{WH}}{M_{BH}}\right)^2\;. 
\end{equation}
\noindent
For equal masses this gives a unique scale.

A second possibility is 

\begin{equation}\label{eq:lengtheffBH}
M_{WH} = \frac{M_{BH}^2}{m} \;,
\end{equation}
\noindent
where $m$ is a constant of dimension mass, and the corresponding inverse relation

\begin{equation}\label{eq:lengtheffWH}
M_{BH} = \frac{M_{WH}^2}{m} \;.
\end{equation}
\noindent
For the first case \eqref{eq:lengtheffBH}, we see that the first condition of \eqref{classicalregime1} becomes

\begin{equation}
\mathscr b_+ \gg \left(8 \lambda_k m\right)^\frac{1}{3} \;,
\end{equation}
\noindent
and hence is a proper length scale. Hence, for large black hole masses, coming from the black hole side, one would observe first quantum effects coming from the $k$-polymerisation at the Kretschmann curvature scale $48/\lambda_k^2$ and afterwards effects coming from small 2-sphere area effects ($j$-polymerisation) at the length scale $(8 m \lambda_k)^\frac{1}{3}$.
For Eq. \eqref{eq:lengtheffWH} the same is true for \eqref{classicalregime2} as coming from the white hole side.

Both possibilities Eq. \eqref{eq:symbounce} for $m = 1$ and Eqs. \eqref{eq:lengtheffBH}-\eqref{eq:lengtheffWH} seem to be physically special. The first option produces a symmetric bounce with a unique onset of quantum effects on both sides, while the second options lead to sensible finite 2-sphere area effects. In principle the presented model allows to not relate the two masses at all and still leading to sensible curvature effects, but if one wants to specify a relation these options seems to be physically special.

\section{Effective quantum corrected spacetime structure}\label{sec:spacetimestructure}

The presented model provides the usual qualitative features. The classical singularity is replaced by a transition surface which connects a trapped and anti-trapped region whose past and future boundaries identify two horizons corresponding to black and white hole horizons, respectively. In this section, we discuss these features more precisely and construct the Penrose diagram of the effective quantum corrected spacetime.

Let us start by recalling the asymptotic behaviour. As the solutions for the metric coefficients \eqref{qmetriccoeff1} and \eqref{qmetriccoeff2} and hence the metric itself are analytic for all $r \in \left(-\infty, \infty\right)$ they provide us also with a solution for the exterior of the black hole as the analytic continuation of the interior metric. As such we can study the asymptotic behaviour, which was done in Sec. \ref{sec:CVeffectivemodel}. The asymptotic spacetime geometries for $r\rightarrow \pm \infty$ are described by

\begin{align}
ds_+^2 &\simeq -\left(1-\frac{2M_{BH}}{\mathscr b}\right) \dd\tau^2 + \frac{1}{1-\frac{2M_{BH}}{\mathscr b}} \dd \mathscr b^2 +\mathscr b^2 \dd \Omega_2^2 \\
ds_-^2 &\simeq -\left(1-\frac{2M_{WH}}{\mathscr b}\right) \dd\tau^2 + \frac{1}{1-\frac{2M_{WH}}{\mathscr b}} \dd \mathscr b^2 +\mathscr b^2 \dd \Omega_2^2 \;,
\end{align}  

\noindent
which correspond to two Schwarzschild spacetimes of masses $M_{BH}$ and $M_{WH}$, respectively.

Next, we can determine the horizons. The Killing horizons are given by the condition 

\begin{align}
&a(r_s^{(\pm)}) \stackrel{!}{=} 0 \quad \Leftrightarrow \quad 
\notag
\\
 &\left.\frac{\lambda_j^6}{2DC^2\lambda_k^2\sqrt{n}^3}\left(1+\frac{nr^2}{\lambda_j^2}\right)^{\frac{3}{2}}\left(1-\frac{CD}{\lambda_j\sqrt{1+\frac{nr^2}{\lambda_j^2}}}\right)\frac{\left(\frac{\sqrt{n}\,r}{\lambda_j}+\sqrt{1+\frac{nr^2}{\lambda_j^2}}\right)^3}{\frac{\lambda_j^6}{16C^2\lambda_k^2n^3}\left(\frac{\sqrt{n}\,r}{\lambda_j}+\sqrt{1+\frac{nr^2}{\lambda_j^2}}\right)^6+1} \right|_{r = r_s^{(\pm)}}\stackrel{!}{=} 0 \;.
\end{align}

\noindent
The only term that can vanish is

$$
1-\frac{CD}{\lambda_j\sqrt{1+\frac{nr^2}{\lambda_j^2}}} = 0 \;,
$$
\noindent
which leads to 

\begin{equation}
r_s^{(\pm)} = \pm \sqrt{\frac{C^2 D^2}{\mathscr L_o^2} - \frac{\lambda_j^2}{\mathscr L_o^2}} = \frac{\lambda_j}{\mathscr L_o} \sqrt{\left(\frac{M_{BH} M_{WH}}{\lambda_k}\right)^\frac{1}{3}-1} \;,
\end{equation}
\noindent
i.e. there are exactly two horizons with areal radius $\mathscr b(r_s^{(\pm)})$. Evaluating the areal radius $\mathscr b$ at these points gives

\begin{align}
\mathscr b\left(r_s^{(\pm)}\right)^2 =&\; \frac{M_{BH}}{2 M_{WH}} \left( \left(M_{BH} M_{WH}\right)^{\frac{1}{3}} \pm \sqrt{ \left(M_{BH} M_{WH}\right)^{\frac{2}{3}} - \lambda_k^{\frac{2}{3}}} \right)^3
\notag
\\
&\;+ \frac{M_{WH}}{2 M_{BH}}  \frac{\lambda_k^2}{\left( \left(M_{BH} M_{WH}\right)^{\frac{1}{3}} \pm \sqrt{ \left(M_{BH} M_{WH}\right)^{\frac{2}{3}} - \lambda_k^{\frac{2}{3}}} \right)^3} \;.
\end{align} 

\noindent
Indeed for $M_{BH} M_{WH} \gg \lambda_k$, we find 

\begin{align}
\mathscr b(r_s^{(+)})^2 &\simeq 4 M_{BH}^2 \left( 1 - \frac{3}{4} \left(\frac{\lambda_k}{M_{BH} M_{WH}}\right)^\frac{2}{3} - \mathcal{O}\left(\frac{\lambda_k^2}{M_{BH}^2 M_{WH}^2}\right)\right) \;, \label{eq:brspapprox}
	\\
\mathscr b(r_s^{(-)})^2 &\simeq 4 M_{WH}^2 \left( 1 - \frac{3}{4} \left(\frac{\lambda_k}{M_{BH} M_{WH}}\right)^\frac{2}{3} - \mathcal{O}\left(\frac{\lambda_k^2}{M_{BH}^2 M_{WH}^2}\right)\right) \;.
\end{align}
\noindent
From this, we see that for large black hole and white hole masses, the classical result is well approximated. Leading corrections are suppressed by powers of $\lambda_k$ independently of how $M_{BH}$ and $M_{WH}$ are chosen. The first order correction is negative.

Furthermore, the model predicts a transition surface where the minimal areal radius is reached and the interior region undergoes a transition from trapped to anti-trapped regions. The minimal value of $\mathscr b$ is reached when $\mathscr b' = 0$. As $\mathscr b \neq 0$ everywhere, this is also the case for $(\mathscr b^2)' = 0$, which simplifies the computations. Introducing the new coordinate 

\begin{equation}
z = \frac{\mathscr L_o r}{\lambda_j} + \sqrt{1 + \frac{\mathscr L_o^2 r^2}{\lambda_j^2}} \quad , \quad z \in (0,\infty)\;,
\end{equation}

\noindent
gives for Eq. \eqref{qmetriccoeff1}

$$
\mathscr b^2(z) = \frac{2 C^2 \lambda_k^2 \mathscr L_o^3 D}{\lambda_j^4} \; \frac{2 z}{z^2 +1} \;\frac{\frac{\lambda_j^6}{16 C^2 \lambda_k^2 \mathscr L_o^6} z^6 +1}{z^3} \;.
$$

\noindent
As furthermore $\dd z/\dd r \neq 0$, the condition for the transition surface becomes $\dd (\mathscr b^2)/\dd z = 0$. After some computations, we find

\begin{equation}
\frac{\dd \mathscr b^2}{\dd z} = 0 \quad \Leftrightarrow \quad - \frac{2 z^4 + z^2}{z^4 + z^2} \left( \frac{\lambda_j^6 z^6}{\mathscr L_o^6} + 16 C^2 \lambda_k^2 \right) + \frac{3 \lambda_j^6 z^6}{\mathscr L_o^6} = 0\;.
\end{equation}

\begin{figure}[t!]
	\centering\includegraphics[scale=0.5]{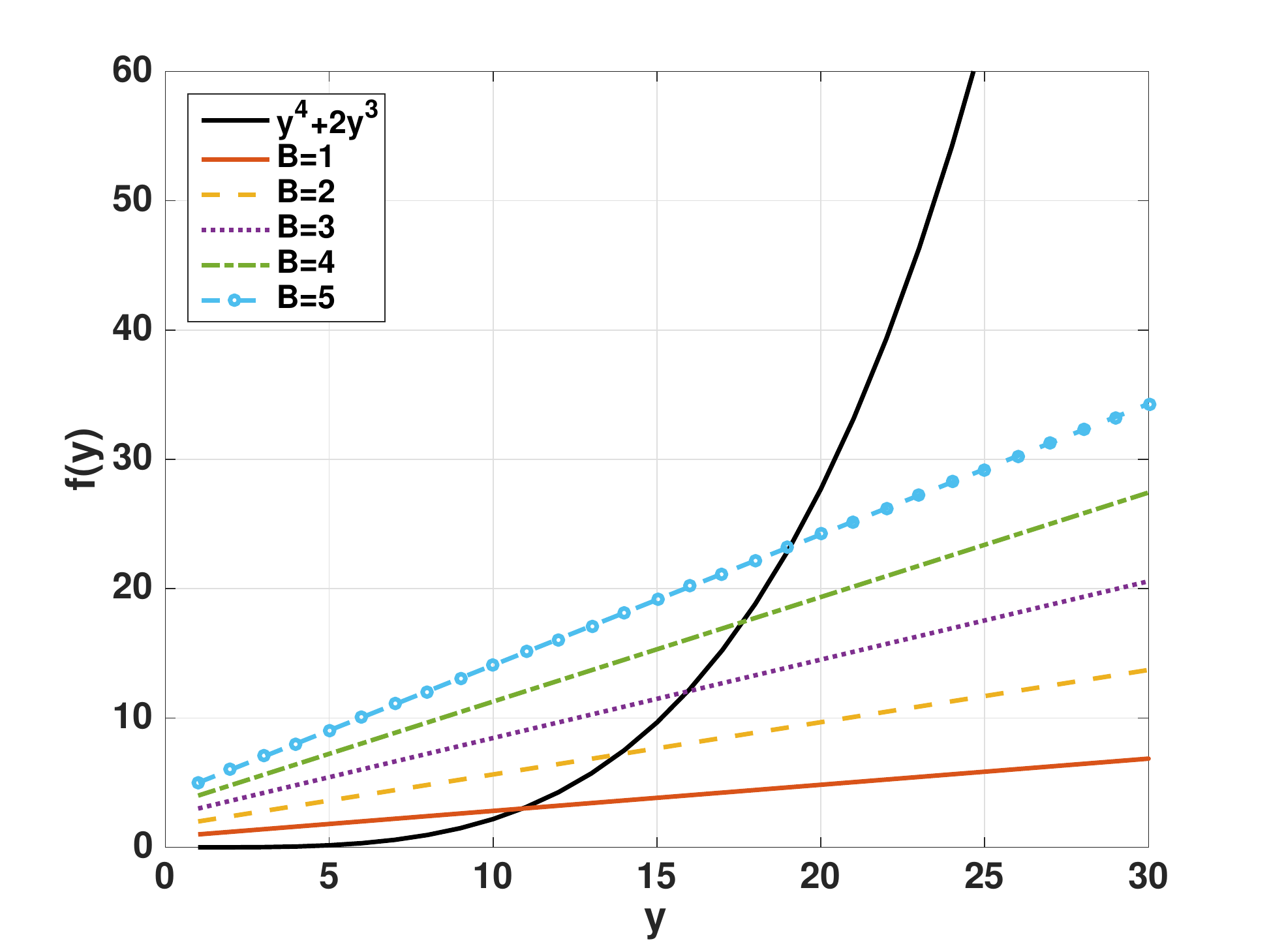}
	\caption{Graphical solutions of Eq. \eqref{eq:yTequ}. The black line corresponds to the left hand side of Eq. \eqref{eq:yTequ}, the coloured lines to the right hand side for different values of $B = M_{WH}^2/M_{BH}^2$. It is obvious, that there exists for $y >0$ always exactly one solution.}
	\label{fig:bT}
\end{figure}

\noindent
For $y = z^2$ and $y > 0$ this equation simplifies to (recall Eq. \eqref{eq:CDjk})

\begin{equation}\label{eq:yTequ}
y^4 + 2 y^3 = \frac{16 C^2 \lambda_k^2 \mathscr L_o^6}{\lambda_j^6} \left(2y+1\right) = \frac{M_{WH}^2}{M_{BH}^2}  \left(2y+1\right) \;,
\end{equation}

\noindent
which is a fourth order polynomial equation in $y$ for $y > 0$. Important at this point is that this equation has in $y > 0$ always one and only one solution, as one can easily convince oneself graphically (cfr. Fig. \ref{fig:bT}). Concluding, there exists always one unique minimal value of $\mathscr b$. This solution is given by

\begin{equation}\label{eq:yt}
y_{\mathcal{T}} = -\frac{1}{2}+\frac{1}{2} \sqrt{1+2^{2/3} \left(-B+B^2\right)^{1/3}}+\frac{1}{2} \sqrt{2-2^{2/3} \left(-B+B^2\right)^{1/3}+\frac{-8+16 B}{4 \sqrt{1+2^{2/3} \left(-B+B^2\right)^{1/3}}}} \;,
\end{equation}

\noindent
with $B = \frac{16 C^2 \lambda_k^2 \mathscr L_o^6}{\lambda_j^6} =M_{WH}^2/M_{BH}^2$. From this we can compute the value of the transition surface $\mathscr b_{\mathcal{T}} = \mathscr b(r_{\mathcal{T}})$. As this expression is complicated and not very insightful, we do not report it here.

This minimal value is then indeed a transition from trapped to anti-trapped regions. This can be easily checked by evaluating the expansions $\theta_{\pm}$ (cfr. e.g. \cite{HawkingTheLargeScale}) for $r = const.$ and $t = const.$ surfaces for $r_s^{(-)} < r < r_s^{(+)}$. For the future pointing unit null normals 

\be
u_{\pm}=u_{\pm}^a\frac{\partial}{\partial x^a}=\frac{1}{\sqrt{-2N}}\frac{\partial}{\partial r}\pm\frac{1}{\sqrt{-2a}}\frac{\partial}{\partial t}\;,
\ee

\noindent
this leads to
\be
\theta_{\pm}=S^{ab}\nabla_au^{\pm}_b=-\sqrt{-\frac{2}{N}}\frac{\mathscr b'}{\mathscr b}\;,
\ee
where $S^{ab}=g^{ab}+u_{+}^au_{-}^b+u_{-}^au_{+}^b$ is the projector on the metric 2-spheres (cfr. \cite{BodendorferEffectiveQuantumExtended}). Hence, in the interior both expansions are either negative or positive depending on the sign of $\mathscr b'$. As at the transition surface $\dot{\mathscr b}$ vanishes and changes from positive values on the black hole side to negative values on the white hole side, the minimal value of $\mathscr b$ characterises indeed a transition from trapped to anti-trapped regions, i.e. a transition from black hole to white hole interior.

Having done all this analysis we can now construct the Penrose diagram. For that we can redo all the steps explained in detail in \cite{BodendorferEffectiveQuantumExtended}. To be sure that this construction works, we need to check 1) that the asymptotic behaviour is Schwarzschild and 2) $a'(r_s^{(\pm)}) \neq 0$ and $\text{sign}(a'(r_s^{(\pm)}) = \pm 1$. The first one was already discussed in Sec. \ref{sec:CVeffectivemodel} and the beginning of this section. The second one can be verified easily by direct computation. As 1) and 2) are both true, we can draw the Penrose diagram which looks exactly as the one reported already in Fig. \ref{Penrosediag2}.

\section{Relation to connection variables}\label{sec:comparison}

We can relate these new curvature variables to the commonly used connection variables in the interior of the black hole. As discussed in Sec. \ref{sec:newvariables}, we have the relations (cfr. Eq. \eqref{vPtokj})

\be\label{eq:jktovP}
v_k=\left(\frac{3}{2}v_1\right)^{\frac{2}{3}}\frac{1}{P_2}  \quad,\quad  v_j=v_2-\frac{3v_1P_1}{2P_2} \quad,\quad k=\left(\frac{3}{2}v_1\right)^{\frac{1}{3}}P_1P_2 \quad,\quad j=P_2 \;.
\ee

\noindent
Inverting these relations yields

\begin{equation}
v_1 = \frac{2}{3} \left(v_k j\right)^{\frac{3}{2}} \quad, \quad v_2 = \frac{v_j j + v_k k}{j} \quad , \quad P_1 = \frac{k}{j \sqrt{v_k j}} \quad , \quad P_2 = j \;.
\end{equation}

\noindent
Using now the already known relations between $(v_1,P_1)$, $(v_2,P_2)$ with $(b,p_b)$, $(c,p_c)$ of Eqs. \eqref{connvaraibles1}, \eqref{connvaraibles2}, we get

 \begin{align}
p_b^2 = -\frac{8\left(v_k k + v_j j\right)}{j} \quad &, \quad |p_c| = 4 \; 2^{\frac{2}{3}} v_k j\;, 
\label{connvaraibleskj1}
\\
b = \text{sign}(p_b)\; \frac{\gamma}{4}\; \sqrt{-8\left(v_k k + v_j j\right) j} \quad &, \quad c= -\text{sign}(p_c)\; \frac{\gamma}{4}\; 2^{\frac{1}{3}} \frac{k}{j} \;.
\label{connvaraibleskj2}
\end{align}

\noindent
Having these relations it is possible to relate the polymerisation scales by demanding 

\begin{equation}
\lambda_j j \stackrel{!}{=} \lambda_2 P_2 \quad , \quad \lambda_k k \stackrel{!}{=} \lambda_1 P_1\;,
\end{equation}

\noindent
thus yielding

\begin{equation}
\lambda_2 = \lambda_j\quad , \quad \lambda_1 = \left(\frac{3v_1}{2}\right)^{\frac{1}{3}} P_2 \lambda_k \;.
\end{equation}

\noindent
Using furthermore the relations between the polymerisation scales $\lambda_1$, $\lambda_2$ and $\delta_b$, $\delta_c$ of Eqs. \eqref{eq:polyvP1}, \eqref{eq:polyvP12}, we find

\begin{equation}
\delta_b = \pm \frac{4 \lambda_j}{\gamma |p_b|} \quad , \quad \delta_c = \pm \frac{64\; 2^{\frac{1}{3}}}{\gamma^2} \frac{b}{p_b} \lambda_k \;.
\end{equation}

\noindent
From this we can read off that the scheme with constant $\lambda_j$, $\lambda_k$ is not of the common type. This scheme rather corresponds to a generalisation of a $\bar{\mu}$-scheme, where the polymerisation scales do not only depend on the triad components $p_c$, $p_b$ but also on the connection $c$, $b$ itself.
\footnote{Moreover, inserting the above transformations \eqref{connvaraibleskj1} and \eqref{connvaraibleskj2} into the Hamiltonian \eqref{eq:hampolyjk} does actually not lead to a ``polymerised Hamiltonian'' in connection variables.
The reason for this are connection-dependent terms in the $v_j$ transformation \eqref{eq:jktovP} (or \eqref{vPtokj}), leading to bare $b$ and $c$ in the final Hamiltonian.}

Having this relation we are in the position to ask if the plaquette argument of \cite{AshtekarQuantumExtensionof} (cfr. Eq. \eqref{eq:plaquette}) can be satisfied for mass independent $\lambda_k$ and $\lambda_j$. Our computations show that this is not true.

\section{Other possibilities: Non-scaling momenta}\label{sec:noscalingmodels}

For the sake of completeness, in this last section we would like to comment on some other possibilities in defining canonical phase space variables for Schwarzschild black holes. In particular, we will focus on the possibility of making also the canonical momentum $j$ independent of fiducial cell rescaling, while keeping the other momentum ($k$) to be the (square root of the) Kretchmann scalar. As we will see, in this case there is no second fiducial cell independent Dirac observable which can be related with the white hole mass and the relation between the masses is determined as an outcome of the effective dynamics.

Starting from the classical variables $(v_k,k,v_j,j)$ defined in Sec. \ref{sec:newvariables}, let us then consider the following transformation
\begin{align}\label{noscalvariables}
k\longmapsto K:=k\qquad&,\qquad v_k\longmapsto V_K:=v_k+\frac{(v_jj^2-2)}{jk}\left(1+k\frac{f'(k)}{f(k)}\right)\;,\nonumber\\
j\longmapsto J:=-\frac{k}{(v_jj^2-2)}f(k)\qquad&,\qquad v_j\longmapsto V_J:=\frac{(v_jj^2-2)^2}{jk}\frac{1}{f(k)}\;,
\end{align}

\noindent
where $f$ is a smooth function of $k$ only and $f'$ denotes its derivative w.r.t. $k$. As can be checked by direct computation, the transformation \eqref{noscalvariables} is canonical. In the gauge $\sqrt{n}=\frac{V_JJ^2}{Kf(K)}$, the Hamiltonian \eqref{newvarham} reads in terms of the new variables
\be\label{Hnoscalvariables1}
H_{\text{cl}}=3V_KK+V_JJ\left(3\frac{f'(K)}{f(K)}K+2\right)\:.
\ee

\noindent
Moreover, according to the behaviour \eqref{fidcellrescaling}, the above variables behave under rescaling of the fiducial cell as

\be
K\longmapsto K\quad,\quad V_K\longmapsto\alpha V_K\quad,\quad J\longmapsto J\quad,\quad V_J\longmapsto\alpha V_J\;,
\ee

\noindent
so that now both canonical momenta do not scale and their conjugate variables scale compatibly with having density weight 1 products $V_K K$, $V_J J$. The function $f(k)$ can be specified by means of the following argument. Looking at the definition of $J$ and recalling the on-shell values of the $(v_k,k,v_j,j)$ variables (cfr. Sec. \ref{sec:newvariables}), we see that $J\sim f(k)/b^2$ with $k=\frac{2M}{b^3}$ on-shell. Therefore, if the function $f$ is chosen in such a way that quantum effects due to polymerisation of $k$ are suppressed by inverse powers of $b$, then the quantum effects resulting from polymerisation of $J$ will always be subdominant w.r.t. those of the $k$-sector. A simple choice, which also make the resulting effective dynamics still analytically solvable, is provided by a power law of the kind $f(k)=k^\epsilon$ with $\epsilon\geq0$. With this choice, the Hamiltonian \eqref{Hnoscalvariables1} simplifies to

\be\label{Hnoscalvariables2}
H_{\text{cl}}=3V_KK+\left(3\epsilon+2\right)V_JJ\:,
\ee

\noindent
which is nothing but the generator of (anisotropic) dilatations in phase space. After polymerisation of the canonical momenta, the effective Hamiltonian is thus given by

\be\label{Heffnoscalvariables}
H_{\text{eff}}=3V_K\frac{\sin(\lambda_KK)}{\lambda_K}+\left(3\epsilon+2\right)V_J\frac{\sin(\lambda_JJ)}{\lambda_J}\;,
\ee

\noindent
where both polymerisation scales $\lambda_K$ and $\lambda_J$ now do not scale under rescaling of the fiducial cell, and they have dimension $L^2$ and $L^{2\epsilon+2}$, respectively. The equations of motion associated with the effective Hamiltonian \eqref{Heffnoscalvariables} are given by
\be
\begin{cases}
	K'=-3\frac{\sin(\lambda_KK)}{\lambda_K}\\
	V_K'=3V_K\cos(\lambda_KK)\\
	J'=-(3\epsilon+2)\frac{\sin(\lambda_JJ)}{\lambda_J}\\
	V_J'=(3\epsilon+2)V_J\cos(\lambda_JJ)\\
\end{cases}\;,
\ee

\noindent
which, by performing similar steps to those discussed in the previous sections, yield the solutions
\be\label{noscalsolutions}
\begin{cases}
	K(r)=\frac{2}{\lambda_K}\tan^{-1}\left(Ce^{-3r}\right)\\
	J(r)=\frac{2}{\lambda_J}\tan^{-1}\left(e^{-(3\epsilon+2)r}\right)\\
	V_J(r)=D\cosh\bigl((3\epsilon+2)r\bigr)\\
	V_K(r)=-\frac{D\lambda_K}{6C\lambda_J}(3\epsilon+2)\left(e^{3r}+C^2e^{-3r}\right)\\
\end{cases}
\ee

\noindent
where we get rid of one integration constant by using the Hamiltonian constraint for $V_K$, we use the gauge freedom in choosing the offset of the $r$-coordinate to set the integration constant entering the solution for $J$ to be 1, and we denote by $C$ and $D$ the remaining two integration constants.

Rephrasing now the classical expressions for the metric coefficients in terms of the new variables, we get

\be\label{bnoscalvariables1}
\mathscr b^2(r)=\frac{f(K)}{J}\left(1+\frac{f'(K)}{f(K)}K+\frac{V_KK}{V_JJ}\right)=\frac{K^\epsilon}{J}\left(1+\epsilon+\frac{V_KK}{V_JJ}\right)\;,
\ee

\noindent
and

\begin{align}
a(r)&=\frac{V_JJ^2}{2\mathscr b^2f(K)K}\left(V_KK+V_JJ\left(K\frac{f'(K)}{f(K)}+\frac{2 J}{K f(K)}\right)\right)\nonumber\\
&=\frac{1}{2\mathscr b^2}\frac{V_JJ^2}{K^{\epsilon+1}}\left(V_KK+V_JJ\left(\epsilon+\frac{2 J}{K^{\epsilon+1}}\right)\right)\;.\label{anoscalvariable}
\end{align}

\noindent
Taking the expression \eqref{bnoscalvariables1} for $\mathscr b^2$ and polymerising the occurring momenta yield
\begin{align}
&\mathscr b^2(r)=-\frac{\lambda_J}{\sin(\lambda_JJ)}\left(\frac{\sin(\lambda_KK)}{\lambda_K}\right)^\epsilon\left(1+\epsilon+\frac{V_K}{V_J}\frac{\lambda_J}{\sin(\lambda_JJ)}\frac{\sin(\lambda_KK)}{\lambda_K}\right)\nonumber\\
&\quad\;\overset{\text{on-shell}}{=}\frac{\lambda_J}{3}\left(\frac{2C}{\lambda_K(e^{-3r}+C^2e^{3r})}\right)^\epsilon\cosh\bigl((3\epsilon+2)r\bigr)\label{bnoscalvariables2}\;,
\end{align}

\noindent
from which we see that $\mathscr b$ is bounded from below and the bounce occurs when it reaches the minimal value. Similarly, the quantum-corrected expression for the coefficient $a(r)$ of the effective metric can be derived by polymerising the classical expression \eqref{anoscalvariable} and plugging in the solutions \eqref{noscalsolutions}. A straightforward calculation shows that the solutions \eqref{noscalsolutions} of the effective dynamics reproduce the correct classical behaviour in the $r\to\pm\infty$ limit, for which the metric coefficients correspondingly yield two asymptotic Schwarzschild spacetimes. We further observe that the integration constant $D$ appears only as a global factor in front of $a$. Hence, as in previous models it can be reabsorbed in a coordinate transformation $t \mapsto \tau$, showing that only $C$ has physical relevance, and only one fiducial cell independent Dirac observable can exist. In particular, looking at the asymptotic expressions for the areal radius \eqref{bnoscalvariables2}, we have

\begin{align}
\mathscr b^2_+&:=\mathscr b^2(r\to+\infty)=\frac{\lambda_J}{6}\left(\frac{2}{\lambda_KC}\right)^\epsilon\,e^{2r}\;,\label{bplus1}\\
\mathscr b^2_-&:=\mathscr b^2(r\to-\infty)=\frac{\lambda_J}{6}\left(\frac{2C}{\lambda_K}\right)^\epsilon\,e^{-2r}\;.\label{bminus1}
\end{align}

\noindent
Therefore, by requiring the on-shell value of the polymerised momentum $\frac{\sin(\lambda_KK)}{\lambda_K}$ to reduce in the $r\to\pm\infty$ limit to the classical value given by the square root of the Kretschmann scalar, we get the following conditions

\be
\frac{\sin(\lambda_KK)}{\lambda_K}\overset{\eqref{noscalsolutions}}{=}\frac{2C}{\lambda_K(e^{3r}+C^2e^{-3r})}=
\begin{cases}
	\overset{r\to+\infty}{\longrightarrow}\frac{2C}{\lambda_Ke^{3r}}\overset{\eqref{bplus1}}{=}\frac{2C}{\lambda_K}\left[\frac{\lambda_J}{6}\left(\frac{2}{C\lambda_K}\right)^\epsilon\right]^{\frac{3}{2}}\frac{1}{\mathscr b_+^3}\overset{!}{=}\frac{2M_{BH}}{\mathscr b_+^3}\\
	\\
	\overset{r\to-\infty}{\longrightarrow}\frac{2}{C\lambda_K}e^{3r}\overset{\eqref{bminus1}}{=}\frac{2}{C\lambda_K}\left[\frac{\lambda_J}{6}\left(\frac{2C}{\lambda_K}\right)^\epsilon\right]^{\frac{3}{2}}\frac{1}{\mathscr b_-^3}\overset{!}{=}\frac{2M_{WH}}{\mathscr b_-^3}
\end{cases}\;,
\ee

\noindent
from which it follows that the on-shell expressions for the mass Dirac observables are given by

\be\label{noscalDO}
2M_{BH}=\frac{2C}{\lambda_K}\left[\frac{\lambda_J}{6}\left(\frac{2}{C\lambda_K}\right)^\epsilon\right]^{\frac{3}{2}}\qquad,\qquad 2M_{WH}=\frac{2}{C\lambda_K}\left[\frac{\lambda_J}{6}\left(\frac{2C}{\lambda_K}\right)^\epsilon\right]^{\frac{3}{2}}\;.
\ee

\noindent
Note that, as anticipated at the beginning of this section, for the one-parameter family of models considered here in which both momenta (and hence polymerisation scales) do not scale under fiducial cell rescaling, there is only one relevant integration constant which appears in the expression for the masses. This means that for such class of models the freedom in fixing the initial conditions is completely encoded in specifying only one of the mass observables, while the relation between the two masses is determined by the effective dynamics itself. Indeed, using the above expressions \eqref{noscalDO}, we see that

\be
M_{BH}\cdot M_{WH}=\frac{1}{\lambda_K^2}\left(\frac{\lambda_J}{6}\right)^3\left(\frac{2}{\lambda_K}\right)^{3\epsilon}=const.\;,
\ee

\noindent
i.e., $M_{WH}\sim 1/M_{BH}$. Therefore, although for large black hole masses this would corresponds to a Planck mass regime on the white hole side which is then beyond the regime of applicability of a polymer-type effective description, this class of models can be considered as an explicit example to illustrate the relation between the scaling properties of the polymerisation scales and the Dirac observables.

In this respect, let us note that in the case of non-scaling momenta additional restrictions on the possible functional relation between the masses allowed by the dynamics come from symmetry arguments. Specifically, due to the invariance of the effective Hamiltonian under the replacement $K\mapsto\frac{\pi}{\lambda_K}-K$ and $J\mapsto\frac{\pi}{\lambda_J}-J$, which corresponds to a ``time'' reversal symmetry in following the spacetime evolution from the black hole classical regime to the white hole classical regime up to the same value of $\mathscr b$ in the Penrose diagram and consequently to $M_{BH}\mapsto M_{WH}$, $M_{WH}\mapsto M_{BH}$, the allowed functional relation $M_{WH}=\mathscr F(M_{BH})$ between the masses can only be of the kind $\mathscr F=\mathscr F^{-1}$. This selects two possibilities, namely $M_{WH}=M_{BH}$ or $M_{WH}=1/M_{BH}$. Which of such possibilities is determined by the dynamics depends on the details of the models. For the specific class of variables discussed in this section, a reciprocal mass relation is found. However, this does not exclude the possibility of uniquely selecting a symmetric bounce scenario by introducing suitable canonical variables and thus quantum corrections. The physical viability of the resulting model, which may or may not fulfil physical requirements on curvature upper bounds and onset of quantum effects, needs of course to be discussed on a case by case study. Note that this situation is similar to what was found in the Bianchi I setting \cite{WilsonEwingTheloopquantum}.

\section{Conclusions and outlook}\label{sec:conclusions}

In this paper, we discussed the appearance and role of Dirac observables in the static and spherically symmetric setting of black holes, whose interior region is modelled as a Kantowski-Sachs cosmological spacetime. As already mentioned in \cite{BodendorferEffectiveQuantumExtended}, classically there exist two Dirac observables, but only one of them is physically relevant. As discussed in the present paper, the appearance of two Dirac observables is immediate from the canonical analysis. At the classical level, we observed that one of them always has to be dependent on the fiducial length $L_o$ (or $\mathscr L_o$, respectively) and does not appear in the final solution for the metric. Hence, only one of them can be of physical relevance. The fact that only one of the canonical Dirac observables has physical relevance, can be explained by residual diffeomorphisms, which can not be treated within the symmetry reduced Hamiltonian framework. Precisely these allow to rescale the second observable away.

The situation changes in the effective quantum theory, where two fiducial cell independent Dirac observables can be constructed. We carefully discussed these observables in the case of already known polymer models of black holes \cite{CorichiLoopquantizationof,ModestoSemiclassicalLoopQuantum,AshtekarQuantumExtensionof,BodendorferEffectiveQuantumExtended} and found that in all of these models, two Dirac observables with physical relevance can be explicitly constructed. For models where the exterior spacetime is not available \cite{CorichiLoopquantizationof,AshtekarQuantumExtensionof}, these are the size of the black hole and white hole horizon, respectively. If the exterior region of the effective spacetime is also available and it is asymptotically flat, then it is possible to construct observables corresponding to the ADM masses. This provides us with important insights as initial conditions have to be fixed more carefully and can not simply compared to the classical setup, where one of the observables has no physical meaning. 
The appearance of two physically relevant observables can be explained by polymerisation scales scaling under fiducial cell changes. A Dirac observable which scales with the fiducial cell can then be multiplied by a suitable power of this also scaling polymerisation scale to arrive at a fiducial cell independent quantity. Both Dirac observables appear in the final metric and can not be removed by using residual diffeomorphisms.
These models have in common that a relation between black hole and white hole horizon/masses has to be fixed to meet criteria for physical viability based for instance on plaquette or Planckian curvature upper bound arguments.

We further introduced a class of models where both polymerisation scales $\lambda_K$, $\lambda_J$ are independent of the fiducial cell. In agreement with the above argument, we found that for this model only one physically relevant observable exist. A relation of the masses is then fixed as an outcome of the effective dynamics without restricting the initial conditions. By symmetry arguments, we further conclude that in these models the only possible relations between the masses are the symmetric bounce ($M_{BH} = M_{WH}$) or a reciprocal mass relation $M_{WH} \propto 1/M_{BH}$. This result in similar to the appearance of Kasner-transitions in anisotropic LQC \cite{WilsonEwingTheloopquantum}. The latter one is true for the presented class of variables. As this relation maps astrophysical black holes into sub-Planck size white holes, where the effective description can not be trusted any more, we do not give this model physical credibility.

In the second part of the paper, we focused instead on the possibility of surpassing previous limitations of initial conditions by constructing adapted canonical variables directly related to curvature invariants. To this aim, we discussed in details a new model recently introduced by the authors in a companion paper \cite{Bodendorferbvtypevariables} where one of the canonical momenta can off-shell be interpreted as $2 M_{\text{Misner-Sharp}}/b^3$ so that the corresponding on-shell value is exactly the square root of the Kretschmann scalar without any restriction on the initial conditions. In agreement with the general discussion of the first part of the paper, also for this new model the polymerisation scale $\lambda_j$ scales under a fiducial cell rescaling and we find two Dirac observables for the black hole and white hole masses. The main novel feature of this model is that the curvature is bounded from above without fixing a relation between the masses. 
We find a preference for masses which are symmetric and within the  range $1/8 < M_{BH}/M_{WH} < 8$, as then the overall dominant curvature scale is given by $3/(4\lambda_k^2)$ and both types of quantum corrections occur at large curvatures.

A further interesting property is that the final metric can be expressed in terms of the two masses and the polymerisation scale $\lambda_k$, which is related to the curvature, only. Thus, the fiducial cell dependent polymerisation scale $\lambda_j$ has no physical meaning, as the metric does not depend on it. The structure of the resulting quantum corrected effective spacetime of this model is qualitatively the same as in previous models \cite{BodendorferEffectiveQuantumExtended,ModestoSemiclassicalLoopQuantum}.

The Hamiltonian is formally the same as in \cite{BodendorferEffectiveQuantumExtended} and remarkably simple. It was shown that the quantum theory for \cite{BodendorferEffectiveQuantumExtended} can be constructed explicitly and analytically solved. As the Hamiltonians are formally equal, this is also true for the presented model. A detailed discussion of the quantum theory is left for future work. Of particular interest is then to understand the role of the Dirac observables in the quantum theory. 

In future work, further variables and polymerisation schemes should be explored. This includes more complicated schemes with different regularisation functions. The above discussed variable $k$ has a nice interpretation in terms of the Kretschmann scalar and leads to sensible curvature effects. Instead more suitable choices for the variable $j$ might be found. The analysis of Dirac observables and their physical role should be performed also for these new models, as well as for already existing ones.

An important remaining problem is then furthermore the relation of these symmetry reduced models to full LQG which, due to the similarity of the newly introduced variables with the adapted $(b,v)$-variables for LQC, we expect to be possible along the line of \cite{Bodendorfer2015}. 

\section*{Acknowledgements}

The authors were supported by an International Junior Research Group grant of the Elite Network of Bavaria. The authors would like to thank Andrea Dapor and Klaus Liegener for inspiring discussions.

\end{document}